\pdfoutput=1
\def\TechReportBuild{}
\documentclass[acmsmall]{acmart}
\PassOptionsToPackage{a4paper,margin=1in}{geometry}

\usepackage{xspace}
\usepackage{ifthen}
\usepackage{booktabs,tabularx,multirow}

\usepackage{geometry}
\usepackage{microtype}

\usepackage{graphicx}
\usepackage{caption}
\usepackage{enumitem}
\usepackage{xcolor}

\usepackage{mathtools,amssymb,amsfonts,bm}
\usepackage{amsthm}

\usepackage{algorithm}
\usepackage[noEnd=true, indLines=true]{algpseudocodex}

\usepackage{hyperref}
\usepackage{subcaption}

\newtheorem{theorem}{Theorem}

\newtheorem{definition}{Definition}
\newtheorem{example}{Example}

\makeatletter
\g@addto@macro\normalsize{%
  \setlength{\abovedisplayskip}{0pt}%
  \setlength{\abovedisplayshortskip}{0pt}%
  \setlength{\belowdisplayskip}{0pt}%
  \setlength{\belowdisplayshortskip}{0pt}%
}

\makeatother

\newcommand{\sig}{\rlap{\ensuremath{^{**}}}}

\newif\iftechreport
\ifdefined\TechReportBuild \techreporttrue \fi
\newcommand{\appref}[1]{\iftechreport Appendix~\ref{#1}\else the technical report~\cite{taxoreview_repo}\fi}

\makeatletter
\def\thm@space@setup{%
  \thm@preskip=2pt
  \thm@postskip=2pt
}
\makeatother

\newcommand{\kdd}{KDD20\xspace}
\newcommand{\neurips}{NeurIPS25\xspace}

\newcommand{\sys}{TaxoReview\xspace}

\DeclareMathOperator*{\argmax}{arg\,max}

\DeclareMathOperator{\desc}{desc}

\DeclareMathOperator{\prox}{prox}

\newcommand{\sysLLM}{TaxoReview\textsubscript{LLM}\xspace}

\newcommand{\sysHAC}{TaxoReview\textsubscript{HAC}\xspace}
\newcommand{\sysHDB}{TaxoReview\textsubscript{HDB}\xspace}
\newcommand{\sysUni}{TaxoReview\textsubscript{Uni}\xspace}
\newcommand{\sysCen}{TaxoReview\textsubscript{Cen}\xspace}

\newcommand{\sysAll}{TaxoReview\textsubscript{All}\xspace}
\newcommand{\sysRec}{TaxoReview\textsubscript{Rec}\xspace}
\newcommand{\sysRel}{TaxoReview\textsubscript{Rel}\xspace}
\newcommand{\sysRnd}{TaxoReview\textsubscript{Rnd}\xspace}
\newcommand{\sysvMF}{TaxoReview\textsubscript{vMF}\xspace}

\AtBeginDocument{%
  }

\setcopyright{cc}
\setcctype{by}
\acmJournal{PACMMOD}
\acmYear{2026} \acmVolume{4} \acmNumber{6 (SIGMOD)} \acmArticle{377}
\acmMonth{12} \acmDOI{10.1145/3856372}

\begin{document}
\setlength{\abovecaptionskip}{0pt}
\setlength{\belowcaptionskip}{0pt}

\title[Scholarly Data Taxonomy Management]{Scholarly Data Taxonomy Management: From Faithful Paper-Review Assignment to Scalable Panel Expansion}

\author{Daomin Ji}
\authornote{This work was done when Daomin Ji was a visiting student at The University of Queensland.}
\email{daomin.ji@student.rmit.edu.au}
\orcid{0009-0000-0037-3614}
\affiliation{%
  \institution{RMIT University}
  \city{Melbourne}
  \state{Victoria}
  \country{Australia}
}
\affiliation{%
  \institution{The University of Queensland}
  \city{Brisbane}
  \state{Queensland}
  \country{Australia}
}
\author{Zhifeng Bao}
\authornote{Zhifeng Bao is the corresponding author.}
\email{zhifeng.bao@uq.edu.au}
\orcid{0000-0003-2477-381X}
\affiliation{%
  \institution{The University of Queensland}
  \city{Brisbane}
  \state{Queensland}
  \country{Australia}
}
\author{Hui Luo}
\email{huil@uow.edu.au}
\orcid{0000-0002-7299-031X}
\affiliation{%
  \institution{University of Wollongong}
  \city{Wollongong}
  \state{New South Wales}
  \country{Australia}
}
\author{Zi Huang}
\email{huang@itee.uq.edu.au}
\orcid{0000-0002-9738-4949}
\affiliation{%
  \institution{The University of Queensland}
  \city{Brisbane}
  \state{Queensland}
  \country{Australia}
}
\author{Kai Qin}
\email{kqin@swin.edu.au}
\orcid{0000-0002-3612-2216}
\affiliation{%
  \institution{Swinburne University of Technology}
  \city{Melbourne}
  \state{Victoria}
  \country{Australia}
}

\renewcommand{\shortauthors}{Daomin Ji, Zhifeng Bao, Hui Luo, Zi Huang, \& Kai Qin}

\begin{abstract}
Paper--reviewer assignment is a cornerstone of peer review, yet achieving high assignment quality at modern conference scale, where leading venues now receive tens of thousands of submissions, remains challenging on two fronts. First, existing affinity-estimation methods operate in flat lexical or embedding spaces without modeling hierarchical relations between topics, and therefore struggle to distinguish fine-grained expertise from broad topical overlap. Second, existing platforms treat the reviewer panel as fixed and conference-local, leaving program chairs to manually recruit additional reviewers when coverage gaps arise. We present \sys, a scholarly data management system that addresses both limitations through a unified design principle: \emph{represent expertise and topics through coordinated taxonomy structures rather than flat embeddings.} \sys\ couples a \emph{Local Taxonomy Builder}, which induces a conference-specific weighted taxonomy and scores affinity via tree--Wasserstein distance over leaf distributions, with a \emph{Global Taxonomy Index}, a persistent, incrementally updated directed-acyclic taxonomy over canonical expertise atoms in a large expert lake, and a \emph{Global--Local Taxonomy Aligner} that bridges the two with a calibrated tree-based proxy, enabling scalable expert retrieval. Experiments on several benchmarks show that \sys\ improves affinity estimation over state-of-the-art baselines by up to $14.6\%$, produces a calibrated proxy that approximates the exact tree--Wasserstein score with Spearman correlation $0.91$, and supports panel expansion over a lake of $240$K experts at a cost comparable to flat indexing baselines.
\end{abstract}



\received{17 April 2026}
\received[revised]{19 August 2026}
\received[accepted]{12 September 2026}

\maketitle

\section{Introduction}
\label{sec:intro}

\begin{figure*}
    \centering
    \includegraphics[width=\linewidth]{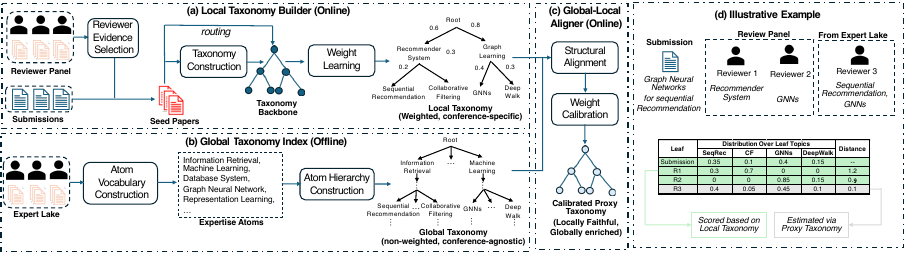}
    \caption{The architecture of \sys with an illustrative example.}
    \Description{Four-panel architecture diagram. (a) Local Taxonomy Builder (online): reviewer evidence selection and taxonomy construction over the submissions and selected seed papers, followed by weight learning, yield a weighted local taxonomy. (b) Global Taxonomy Index (offline): atom vocabulary construction and atom hierarchy construction over the expert lake yield a global taxonomy of expertise atoms. (c) Global-Local Aligner (online): structural alignment and weight calibration produce a calibrated proxy taxonomy. (d) Illustrative example: a submission on graph neural networks for sequential recommendation, two panel reviewers and one lake expert, with their distributions over leaf topics and their distances to the submission.}
    \label{fig:overview}
\end{figure*}

Assigning submissions to reviewers with relevant expertise is a foundational element of the peer-review process in scientific publishing~\cite{charlin2013tpms,mimno2007expertise,anjum2019pare,kou2015weighted,madden2006doubleblind,bhowmick2022connected}.
However, the scale of modern computer science conferences has made this task increasingly difficult. For example, NeurIPS~2025 received 21,575 submissions and engaged 21,921 reviewers~\cite{neurips2025factsheet}, underscoring the need for scalable and principled paper--reviewer assignment systems at this scale.

Achieving high global assignment quality hinges on two complementary factors.
\emph{First, faithful affinity estimation}: the matching system must faithfully quantify the expertise alignment between each submission and each committed reviewer, so that the downstream assignment objective reflects true topical fit rather than superficial lexical or semantic overlap.
\emph{Second, a review panel with sufficient topic coverage}: even with perfect affinity scores, no assignment can succeed if the committed panel lacks expertise in key submission topics.
Program chairs therefore routinely recruit additional reviewers~\cite{icml2025ac}, whether experts in emerging topics, replacement reviewers for late withdrawals, or seed reviewers for under-populated initial panels, drawing from a much larger external expert population. We refer to this recruitment process as \emph{panel expansion}.

Existing paper--reviewer assignment platforms fall short on both fronts. \emph{On affinity estimation}, systems ranging from classical lexical methods such as TPMS~\cite{charlin2013tpms} (adopted in CMT~\cite{microsoftcmt}) to state-of-the-art pretrained language model approaches~\cite{beltagy2019scibert,cohan2020specter,singh2023scirepeval,zhang2025chainoffactors} (adopted in OpenReview~\cite{openreview}) represent submissions and reviewers in \emph{flat lexical or embedding spaces}, including factor-specific representations~\cite{zhang2025chainoffactors}, and score affinity via pairwise similarity without modeling hierarchical relations between topics. Such flat representations introduce two structural shortcomings: multi-faceted submissions spanning several subfields are compressed into flat vectors that average out distinct concepts; and flat metric spaces treat all similarities as equivalent, failing to distinguish fine-grained, subtopic-level expertise from broad, high-level topical overlap. As illustrated in Example~\ref{ex:flat-limitations}, these representations cannot reliably separate a true multi-faceted expert from researchers whose relevance is partial or superficial.

\begin{example}
\label{ex:flat-limitations}
As shown in Fig.~\ref{fig:overview}(d), consider a submission titled ``Graph Neural Networks for Sequential Recommendation'', which spans two facets: Graph Neural Networks (GNNs) and Sequential Recommendation. The current panel contains a GNN specialist, covering only one facet, and a general recommendation researcher working on collaborative filtering, who shares high-level terminology with the submission but lacks depth in either facet. A flat embedding compresses both facets into a single vector and admits both candidates as false positives: the specialist scores misleadingly high because the one shared facet dominates the collapsed vector, while the generalist is rewarded for superficial vocabulary overlap that the flattened representation cannot distinguish from technical depth.
\end{example}

\emph{On panel expansion}, existing platforms such as CMT and OpenReview provide essentially no automated support. At the data level, \emph{experts in the wild are not organized at scale}: domain experts are scattered across heterogeneous resources with no unifying structure, so platforms treat the review panel as a fixed, conference-local input and delegate recruitment to program chairs, who identify candidates through personal networks, ad hoc searches, or browsing related papers, a process that is time-consuming, biased toward chairs' own communities, and inadequate at modern conference scale.
One might hope that a straightforward pipeline, assembling a large \emph{expert lake}, embedding each candidate, and scoring them against every submission, would suffice. Yet this naive solution inherits the flat-embedding pathology above and additionally breaks down at scale: exact affinity computation through a conference-specific pipeline is prohibitive, whereas a generic off-the-shelf embedding space induces a \emph{distribution shift} that leaves retrieved candidates poorly calibrated to the conference's topic distribution.

Both obstacles are, at their core, scholarly \emph{data management} problems. Faithful affinity estimation requires representing heterogeneous conference data, i.e., submissions and reviewer publication histories, in a structured form whose metric can be calibrated to each venue, rather than tuning a ranking heuristic. Panel expansion is a data discovery problem over a data lake: candidate experts scattered across heterogeneous open sources must be integrated under a unifying schema via entity canonicalization, indexed so that top-$k$ retrieval under a hierarchy-aware metric scores only a small candidate set, and maintained incrementally as new publications stream in. The required machinery, namely data integration and entity resolution, dynamic graph maintenance with reachability indexing, hierarchical index construction with amortized-cost guarantees, and calibrated query answering, is the subject matter of data management systems, and Section~\ref{sec:related} positions \sys\ against the corresponding lines of database work on data discovery, similarity indexing, and graph management.

We address both limitations through \sys, a scholarly data management system for paper--reviewer assignment grounded in a unified design principle: \emph{model expertise and topics through coordinated taxonomy structures rather than flat embedding vectors.} As illustrated in Fig.~\ref{fig:overview}, \sys\ instantiates this principle with three coordinated components: (i) a \emph{conference-specific local taxonomy} that models reviewer expertise and submission topics within the thematic scope of the current conference; (ii) a \emph{global taxonomy} that organizes expertise across the expert lake at scale and absorbs new publications and reviewers under continuous incremental maintenance; and (iii) a \emph{Global--Local Aligner} that compiles panel-expansion queries and calibrates global retrieval scores against the local taxonomy.

\noindent\textbf{(1) Taxonomy-driven affinity estimation.} To overcome the structural shortcomings of flat representations, \sys\ induces a conference-specific weighted taxonomy from the submission corpus augmented with selected reviewer evidence, projects each submission and reviewer onto its leaves as a probability distribution, and scores affinity via the weighted tree--Wasserstein distance~\cite{peyre2019computational,rubner2000emd,takezawa2021supervised}, a closed-form optimal-transport metric under which the cost of moving probability mass between two leaves equals the total edge weight along the connecting path. This addresses both shortcomings at once: multi-faceted submissions are preserved as distributions over multiple leaves rather than collapsed into a single centroid, and the cumulative edge-weight structure sharply penalizes broad, ancestor-only overlap while rewarding fine-grained leaf-level alignment, yielding a sharper separation between true experts and superficially related reviewers (Example~\ref{ex:flat-limitations}).

\noindent\textbf{(2) Principled and scalable panel expansion.} To reduce the manual recruitment burden on program chairs, \sys\ assembles an expert lake from open scholarly sources such as DBLP~\cite{ley2002dblp} and OpenAlex~\cite{priem2022openalex}, organized into a persistent, conference-agnostic global taxonomy. Two design choices keep this substrate practical at scale: the index is \emph{lightweight}, recording only \emph{canonical expertise atoms} rather than full paper texts, so the atom vocabulary grows far more slowly than the lake; and it supports \emph{incremental updates}, absorbing new atoms without rebuilding from scratch as the lake grows, keeping the index fresh across conference cycles at amortized cost. A tension remains: directly scoring candidates on the global taxonomy is efficient but distribution-shifted relative to the conference-specific affinity geometry, whereas rescoring every candidate on the local taxonomy is faithful but infeasible at large scale. \sys\ resolves this with a \emph{Global--Local Aligner} that aligns and calibrates the two taxonomies, yielding a calibrated proxy taxonomy globally enriched with expertise from the lake yet locally faithful to the conference-specific affinity geometry.

\begin{example}
Revisiting Example~\ref{ex:flat-limitations}, \sys\ induces a taxonomy with two relevant branches, \emph{Recommender Systems} and \emph{Graph Learning}, and routes the submission across the leaves of both rather than collapsing it into a single centroid. Under tree--Wasserstein affinity, both committed reviewers are correctly exposed as false positives. Since no suitable reviewer exists in the panel, the chairs can pose a paper-level query against the global taxonomy, and the Global--Local Aligner retrieves a promising reviewer from the expert lake whose expertise jointly spans both branches, closing the coverage gap without manual intervention.
\end{example}

The main technical contributions are summarized as follows:
\begin{itemize}[leftmargin=*,nosep]

\item We propose a novel taxonomy-driven affinity estimation framework that models submissions and reviewers as distributions over a learned weighted taxonomy and scores affinity via tree--Wasserstein distance (Section~\ref{sec:taxonomy}).

\item We design the Global Taxonomy Index, a hierarchical similarity index over canonical expertise atoms drawn from a continuously growing expert lake. It differs from existing index families in its item representation (weighted atom profiles over a subsumption directed acyclic graph (DAG) rather than tokens or flat vectors), its maintenance (a three-level cascade that avoids routine full rebuilds), and its query semantics (a build-once structure that serves per-conference calibrated metrics rather than a metric fixed at construction time) (Section~\ref{sec:index}).

\item We develop a Global--Local Aligner that fits a compact tree-based proxy on the global taxonomy to the conference-specific local tree--Wasserstein distances, enabling fast lake-scale retrieval while empirically preserving their rankings (Section~\ref{sec:completion}).

\item Across four benchmarks, \sys outperforms state-of-the-art baselines on affinity estimation by up to $14.6\%$. Its calibrated proxy taxonomy attains Spearman correlation $\rho = 0.91$ with the exact tree--Wasserstein score, while the Global Taxonomy Index sustains panel expansion over an expert lake of $240$K researchers at a cost comparable to flat indexing baselines (Section~\ref{sec:exp}).
\end{itemize}

\section{Problem Definition}
\label{sec:problem}

Paper--reviewer assignment involves three entities. The \emph{submission corpus} $\mathcal{P} = \{p_1, \ldots, p_M\}$ consists of $M$ submissions, each represented by its textual content $\mathrm{Text}(p)$ (e.g., title and abstract). The \emph{review panel} $\mathcal{R}_P = \{r_1, \ldots, r_N\}$ denotes the reviewers already committed to the current conference, each associated with a publication history $\mathcal{H}_r = \{q_{r,1}, \ldots, q_{r,|\mathcal{H}_r|}\}$ of prior publications with textual content $\mathrm{Text}(q)$. Beyond the committed panel, the \emph{expert lake} $\mathcal{R}_{\mathrm{W}}$ is a large-scale, continuously evolving external repository of domain experts maintained independently of any single conference; lake experts outside $\mathcal{R}_P$ are not yet part of the current review process but may be recruited to strengthen the panel.

The main goal is to assign reviewers to submissions so as to maximize the total expertise alignment. Let $x_{p,r} \in \{0,1\}$ indicate whether reviewer $r$ is assigned to submission $p$, and let $f(p, r)$ denote a real-valued affinity score (higher is better) measuring the expertise alignment between $p$ and $r$. The global paper--reviewer assignment is formulated as
\begin{equation}
\label{eq:assignment}
\max_{x \in \{0,1\}^{|\mathcal{P}| \times |\mathcal{R}_P|}}
\;\; \sum_{p \in \mathcal{P}} \sum_{r \in \mathcal{R}_P} f(p, r)\, x_{p,r},
\end{equation}
subject to (i) $\sum_{r} x_{p,r} = K$ for all $p \in \mathcal{P}$, (ii) $\sum_{p} x_{p,r} \leq L_r$ for all $r \in \mathcal{R}_P$, (iii) $x_{p,r} = 0$ for all $(p,r) \in \mathcal{C}$, and (iv) $x_{p,r} \in \{0,1\}$. Here, $K$ is the number of reviewers required per submission, $L_r$ is the maximum review load of reviewer $r$, and $\mathcal{C}$ encodes conflict-of-interest pairs, either derived from bibliographic metadata or declared by authors and reviewers (Section~\ref{sec:retrieval}).
The assignment quality is determined by two factors:
\begin{itemize}[leftmargin=*, noitemsep,topsep=0pt]
\item \textbf{Affinity fidelity.} The fidelity of the learned affinity scores $f(p, r)$ dictates whether the objective in Equation~\eqref{eq:assignment} faithfully reflects true expertise alignment.
\emph{Notably, once $f(p,r)$ is computed, Equation~\eqref{eq:assignment} reduces to a standard constrained integer program solvable via off-the-shelf combinatorial techniques, including min-cost flow~\cite{charlin2013tpms} and constraint-based optimization~\cite{tang2010expertise}}.
\item \textbf{Panel coverage.} The coverage of the panel $\mathcal{R}_P$ relative to the corpus $\mathcal{P}$ determines whether qualified reviewers exist for every submission topic in the first place: even an optimal solver cannot compensate for a panel whose collective expertise fails to span the submissions' thematic scope.
\end{itemize}

We therefore focus on two core tasks that jointly determine the achievable assignment quality: (1) improve the fidelity of pairwise affinity estimation over the committed panel, and (2) expand the panel by recruiting additional qualified reviewers from an external expert lake.

\begin{definition}[Paper--Reviewer Affinity Estimation]
\label{def:affinity}
Given a submission corpus $\mathcal{P}$ and a review panel $\mathcal{R}_P$, where each reviewer $r \in \mathcal{R}_P$ is associated with a publication history $\mathcal{H}_r$, the goal is to learn an affinity function $f : \mathcal{P} \times \mathcal{R}_P \to \mathbb{R}$ such that, for any pair $(p, r)$, the value $f(p, r)$ faithfully quantifies the semantic alignment between submission~$p$ and the expertise of reviewer~$r$.
\end{definition}

Consistent with established protocols~\cite{charlin2013tpms,zhang2025chainoffactors}, we assume the absence of ground-truth relevance labels (e.g., historical assignments or expert-annotated relevance scores) and explicit preference signals (e.g., bidding data), both of which are rarely available in practice.
Consequently, $f(p,r)$ must be derived exclusively from the unstructured text in $\mathrm{Text}(p)$ and $\{\mathrm{Text}(q)\}_{q \in \mathcal{H}_r}$.

\begin{definition}[Panel Expansion]
\label{def:completion}
Given a submission corpus $\mathcal{P}$, an initial review panel $\mathcal{R}_P$ (possibly empty), an expert lake $\mathcal{R}_{W}$, and a recruitment budget $B$, the goal is to identify an augmentation set $\mathcal{R}_A \subseteq \mathcal{R}_{W}$ with $|\mathcal{R}_A| \leq B$, such that solving the assignment in Equation~\eqref{eq:assignment} over the augmented panel $\mathcal{R}_P \cup \mathcal{R}_A$ maximizes the global assignment quality subject to the budget constraint.
\end{definition}

Unfortunately, solving Definition~\ref{def:completion} exactly is computationally intractable: the problem admits a reduction from Maximum Coverage and is therefore NP-hard (proof deferred to \appref{app:np-hard-proof}).

Given this intractability, we relax Definition~\ref{def:completion} to a tractable retrieval formulation: rather than jointly optimizing $\mathcal{R}_A$ and the resulting assignment, we decouple expert selection from the downstream combinatorial assignment and recast panel expansion as a top-$B$ retrieval problem over the expert lake.

\begin{definition}[Panel Expansion as Query-Anchored Top-$B$ Retrieval]
\label{def:topb-retrieval}
Given an expert lake $\mathcal{R}_{W}$, a query $q$ that specifies the additional expertise sought from the lake, a relevance function $\mathrm{rel}(q, r)$ scoring the alignment between $q$ and an expert $r \in \mathcal{R}_{W}$, and a budget $B$, the goal is to retrieve the augmentation set
$
\mathcal{R}_A^{\star} \;=\; \operatorname*{arg\,top\text{-}B}_{r \in \mathcal{R}_{W}} \; \mathrm{rel}(q, r),
$
i.e., the $B$ experts in $\mathcal{R}_{W}$ whose expertise is most relevant to $q$.
\end{definition}

The query $q$ is intentionally left abstract to accommodate diverse operational scenarios, from strengthening an entire panel to filling coverage gaps on specific topics or submissions; Section~\ref{sec:retrieval} compiles each scenario into a common weighted-submission representation served by a single retrieval primitive.

\section{System Overview}
\label{sec:overview}

We present \sys, a scholarly data management system that addresses the two tasks formalized in Section~\ref{sec:problem}. \sys\ anchors both tasks on two coordinated taxonomy structures, one local to each conference and one global over the expert lake, bridged dynamically on demand. As illustrated in Fig.~\ref{fig:overview}, \sys\ comprises three core components distributed across an offline preparation stage and an online execution stage. A summary of main notations is in \appref{app:notations}.

\noindent\textbf{Local Taxonomy Builder (Online Stage).}
The first component is invoked once per conference, at conference initiation, and resolves Task~1 (affinity estimation) over the committed panel $\mathcal{R}_P$; it runs in the online stage because its inputs, the submission corpus and the committed panel, only become available when the conference opens. It projects reviewers and submissions into a shared, weighted taxonomy tree, induced from the submission corpus augmented with a carefully selected subset of reviewer publications so that its leaf topics faithfully reflect the conference's thematic scope, and represents each as a distribution over the leaf nodes. The affinity $f(p,r)$ is then instantiated as the negative weighted tree--Wasserstein distance~\cite{takezawa2021supervised} between the two distributions, a closed-form optimal-transport metric whose edge weights encode the conference-specific cost of moving probability mass between topics. (Section~\ref{sec:taxonomy})

\noindent\textbf{Global Taxonomy Index (Offline Stage).}
The second component organizes the expertise within the large-scale expert lake $\mathcal{R}_W$ into a single, persistent index shared across all conferences served by \sys. Unlike prior taxonomies built from full paper texts, which are expensive to maintain and prone to drifting with textual trends rather than true expertise, this global taxonomy is constructed over canonical expertise atoms extracted from each expert's publication record. Anchoring the index on discrete atoms makes it venue-agnostic and lightweight to maintain: new publications trigger only local, incremental updates, and the one-time construction cost is amortized across every conference that queries it. (Section~\ref{sec:index})

\noindent\textbf{Global--Local Taxonomy Aligner (Online Stage).}
The final component executes online to reconcile the two taxonomies during panel expansion (Task~2). The challenge is a structural asymmetry: expert profiles are indexed on the global taxonomy $\mathcal{T}_G$, whereas the affinity metric $f(p,r)$ is defined on the local taxonomy $\mathcal{T}$. Projecting every candidate in $\mathcal{R}_W$ onto $\mathcal{T}$ is faithful but prohibitively expensive at scale; scoring candidates directly on $\mathcal{T}_G$ reuses the offline index efficiently, but its differing granularity and lack of conference-specific edge weights induce a \emph{distribution shift}, so its rankings no longer reflect the true local affinity. The \emph{Global--Local Aligner} combines the strengths of both through two sequential stages, \emph{structural alignment} and \emph{weight calibration}, producing a calibrated proxy taxonomy that is globally enriched with candidate reviewers from the lake while remaining locally faithful to the conference-specific affinity geometry. (Section~\ref{sec:completion})

\noindent\textbf{Remark.} \sys\ is fully automatic end-to-end: taxonomy induction, profile construction, and affinity estimation require no manual annotation. The only human input in the default pipeline is optional conflict declaration: conflicts not derivable from bibliographic metadata may be declared by reviewers or authors and enter the assignment problem in Eq.~\eqref{eq:assignment} as constraints (Section~\ref{sec:problem}). Optional human signals are nevertheless accommodated: self-declared expertise keywords and subject-area selections can be canonicalized against the atom catalog (Section~\ref{sec:atom-extraction}) and added to the expert's profile with configurable weights, participating in affinity estimation identically to publication-derived atoms, while manual corrections, such as excluding unsuitable reviewers or overriding individual assignments, apply downstream as constraints without altering the affinity model.

\section{Local Taxonomy Builder}
\label{sec:taxonomy}

\subsection{Taxonomy-Driven Affinity Estimation}
\label{sec:core}
We first present a novel taxonomy-based framework for paper--reviewer affinity estimation. Unlike existing taxonomy construction methods that produce \emph{hard assignments}, placing each paper at a single node, our framework requires a \emph{soft assignment} representing each submission and reviewer as a probability distribution over leaf topics, obtained by a post-hoc soft routing procedure on top of any hard-assignment backbone. Algorithm~\ref{alg:affinity} presents the five-stage pipeline; two remaining challenges, \emph{reviewer evidence selection} and \emph{edge-weight calibration}, are deferred to Sections~\ref{sec:evidence} and~\ref{sec:distance-learning}.

\begin{algorithm}[t]
\caption{Taxonomy-Based Affinity Estimation}
\label{alg:affinity}
\small
\begin{algorithmic}[1]
\Require Submission corpus $\mathcal{P}$, review panel $\mathcal{R}_P$ with publication histories $\{\mathcal{H}_r\}_{r \in \mathcal{R}_P}$
\Ensure Pairwise distance matrix $\mathbf{D} \in \mathbb{R}^{|\mathcal{P}| \times |\mathcal{R}_P|}$
\State \textit{// Stage 1: Reviewer evidence selection (\S\ref{sec:evidence})}
\State $\mathcal{E}^{\star} \gets \textsc{EvidenceSelect}(\mathcal{P}, \{\mathcal{H}_r\}_{r \in \mathcal{R}_P})$ \Comment{submodular selection}
\State \textit{// Stage 2: Taxonomy backbone construction}
\State $\mathcal{T}_0 \gets \textsc{InduceTaxonomy}(\mathcal{P} \cup \mathcal{E}^{\star})$ \Comment{hierarchical clustering over $\mathcal{D}_{\mathrm{seed}}$}
\State \textit{// Stage 3: Soft routing to leaf distributions (\S\ref{sec:core})}
\For{each submission $p \in \mathcal{P}$}
    \State $\phi_p \gets \textsc{SoftRoute}(p, \mathcal{T}_0)$ \Comment{Eq.~\eqref{eq:leaf-dist}}
\EndFor
\For{each panelist $r \in \mathcal{R}_P$}
    \For{each evidence paper $q \in \mathcal{E}_r^{\star}$}
        \State $\phi_q \gets \textsc{SoftRoute}(q, \mathcal{T}_0)$
    \EndFor
    \State $\phi_r \gets \frac{1}{|\mathcal{E}_r^{\star}|} \sum_{q \in \mathcal{E}_r^{\star}} \phi_q$ \Comment{Eq.~\eqref{eq:reviewer-dist}}
\EndFor
\State \textit{// Stage 4: Edge-weight calibration (\S\ref{sec:distance-learning})}
\State $\mathbf{w} \gets \textsc{CalibrateWeights}(\mathcal{T}_0, \{\phi_p\}_{p \in \mathcal{P}})$ \Comment{self-supervised NNLS}
\State $\mathcal{T} \gets (\mathcal{T}_0, \mathbf{w})$ \Comment{calibrated weighted taxonomy}
\State \textit{// Stage 5: Tree--Wasserstein affinity materialization}
\For{each $(p, r) \in \mathcal{P} \times \mathcal{R}_P$}
    \State $\mathbf{D}[p, r] \gets d_{\mathcal{T}}(\phi_p, \phi_r)$ \Comment{Eq.~\eqref{eq:tree-distance}}
\EndFor
\State \Return $\mathbf{D}$
\end{algorithmic}
\end{algorithm}

\noindent\textbf{Taxonomy Backbone Construction (Line~4).} On Line~4, an unweighted backbone $\mathcal{T}_0$ is induced over the augmented seed corpus $\mathcal{D}_{\mathrm{seed}} = \mathcal{P} \cup \mathcal{E}^{\star}$, where $\mathcal{E}^{\star}$ denotes the selected reviewer evidence produced by the selection procedure of Section~\ref{sec:evidence}, from which the calibrated taxonomy $\mathcal{T}$ is later derived. The framework is agnostic to the backbone construction: any hierarchical clustering algorithm (e.g., HDBSCAN~\cite{campello2015hdbscan}, hierarchical agglomerative clustering~\cite{murtagh2012algorithms}, or taxonomy induction guided by large language models (LLMs)~\cite{zeng2024chainoflayer,zhu2025context}) can serve as $\mathcal{T}_0$, provided it produces a rooted tree over $\mathcal{D}_{\mathrm{seed}}$. Unlike the global taxonomy of Section~\ref{sec:index}, whose nodes are discrete expertise atoms, the nodes of the local taxonomy are \emph{not} atoms: each internal node of $\mathcal{T}_0$ is a cluster of documents from $\mathcal{D}_{\text{seed}}$, its semantics defined implicitly by its member documents and their embedding centroid, so every node represents a conference-specific topic region rather than a canonical concept. This distinction is deliberate: the global taxonomy spans the entire expert lake and requires primitive, reusable units shared across authors and papers, whereas the local taxonomy needs only to partition a single conference corpus's thematic scope; the correspondence between the two node types is established by the Global--Local Aligner (Section~\ref{sec:structural-alignment}).

\noindent\textbf{Soft Routing to Leaf Distributions (Lines~6--11).} Once $\mathcal{T}_0$ is built, each paper is represented as a \emph{distribution} over all taxonomy leaves, since a hard single-leaf assignment cannot capture interdisciplinary submissions or reviewers spanning multiple subareas. At an internal node $u$ with children $u_1, \dots, u_{b_u}$, the routing probability of a paper $p$ to child $u_k$ is
\begin{equation*}\label{eq:routing}
  \rho_u(k \mid p) = \frac{\exp\!\big(\beta_u \cos(\mathbf{e}_p, \mathbf{c}_{u_k})\big)}
  {\sum_{\ell=1}^{b_u} \exp\!\big(\beta_u \cos(\mathbf{e}_p, \mathbf{c}_{u_\ell})\big)},
\end{equation*}
where $\mathbf{e}_p$ is the embedding of $p$, $\mathbf{c}_{u_k}$ is the centroid of child $u_k$'s submission set, and $\beta_u = 1 / (1 - \bar{s}_u + \epsilon)$ is an adaptive inverse temperature set by the average intra-child cosine similarity $\bar{s}_u$, with a small $\epsilon > 0$ for numerical stability. Cohesive children (high $\bar{s}_u$) yield large $\beta_u$ and sharp routing; loosely knit children yield small $\beta_u$ and diffuse mass, reflecting topical ambiguity. The leaf distribution of $p$ is then obtained by multiplying the routing probabilities along the unique root-to-leaf path $\pi(\ell)$,
\begin{equation}\label{eq:leaf-dist}
  \phi_p(\ell) = \!\!\prod_{(u,v) \in \pi(\ell)}\!\! \; \rho_u(v \mid p), \;\; \ell \in \mathcal{L},
\end{equation}
where $\mathcal{L}$ denotes the leaf set. Equation~\eqref{eq:leaf-dist} is applied differently to the two entity types:
\begin{itemize}[leftmargin=*,nosep]
  \item \emph{Submissions.} Each submission $p \in \mathcal{P}$ is routed directly through Equation~\eqref{eq:leaf-dist} to obtain its representation $\phi_p$.
  \item \emph{Reviewers.} Each reviewer $r \in \mathcal{R}_P$ is backed by a set of evidence papers $\mathcal{E}_r^{\star} \subseteq \mathcal{H}_r$. Each $q \in \mathcal{E}_r^{\star}$ is routed individually to produce $\phi_q$, and the reviewer's leaf distribution is the uniform mixture
  \begin{equation}\label{eq:reviewer-dist}
    \phi_r(\ell) = \frac{1}{|\mathcal{E}_r^{\star}|} \sum_{q \in \mathcal{E}_r^{\star}} \phi_q(\ell).
  \end{equation}
\end{itemize}
The mixture in Equation~\eqref{eq:reviewer-dist} is deliberately uniform, with no bibliometric scaling as in Section~\ref{sec:atom-extraction}. The global profile aggregates an expert's \emph{entire} history, so recency, authorship, and venue signals are needed to suppress outdated or peripheral entries; the local representation instead operates on the curated evidence set $\mathcal{E}^\star_r$, already restricted by the selection of Section~\ref{sec:evidence} to the reviewer's most representative and recent work. Applying the bibliometric factor $\alpha_{r,q}$ on top of this curated set is redundant and empirically harmful, skewing the leaf distribution away from the balanced topical coverage the selection establishes (ablation in \appref{app:bib-ablation}).

\noindent\textbf{Tree--Wasserstein Matching (Lines~16--18).}
With submissions and reviewers represented as leaf distributions, their affinity is measured by the tree--Wasserstein distance, an instance of optimal transport~\cite{rubner2000emd,peyre2019computational} on a weighted tree, where the transport cost between two leaves is the sum of edge weights along the path connecting them. This metric suits our setting for two reasons: it natively respects the hierarchy, charging each mismatch for the subtree boundaries it crosses so that the learned edge weights can penalize cross-branch mismatches according to their conference-specific semantic cost, a property neither point-wise divergences (Kullback--Leibler~\cite{kullback1951information}, Jensen--Shannon~\cite{lin1991jsd}) nor flat Wasserstein distances express; and it admits a closed form in terms of subtree masses, replacing the cubic-time program of general optimal transport with a single linear pass over the edges~\cite{takezawa2021supervised}.

Let $\mathbf{w} = (w_e)_{e \in E}$ denote the edge weight vector, computed by the calibration step on Lines~13--14 (Section~\ref{sec:distance-learning}). For each edge $e = (u, v)$ with $v$ the child endpoint, let $S_e = \desc(v)$ denote the leaves in the subtree rooted at $v$, and let $m_x(S_e) = \sum_{\ell \in S_e} x(\ell)$ be the mass that a leaf distribution $x$ places inside that subtree. The tree--Wasserstein distance between submission $p$ and reviewer $r$ is then
\begin{equation}\label{eq:tree-distance}
  d_{\mathcal{T}}(p, r) = \sum_{e \in E} w_e \,\big|\,m_{\phi_p}(S_e) - m_{\phi_r}(S_e)\,\big|,
\end{equation}
so that each edge contributes its own weight exactly once. The affinity function is then instantiated as $f(p, r) = -d_{\mathcal{T}}(p, r)$.

\subsection{Reviewer Evidence Selection}
\label{sec:evidence}
Within the framework, each reviewer is represented by routing a selected subset of their past publications through the taxonomy, so the fidelity of this representation is determined by which publications are selected. The selected evidence additionally enters the taxonomy construction itself: the submission corpus alone is often sparse in certain topic regions, and evidence reinforces these under-represented regions to stabilize the induced hierarchy; moreover, a taxonomy induced exclusively from submissions would situate reviewer expertise in poorly resolved regions built in its absence, degrading matching quality. We motivate the selection problem through three considerations.
\begin{itemize}[leftmargin=*,topsep=0pt]
  \item \textbf{Coverage.} The combined evidence of the selected set must span the thematic scope of the submission corpus. Topic regions that lack semantically related reviewer papers yield unreliable affinity scores regardless of the underlying reviewer expertise.
  \item \textbf{Proximity.} Coverage is operationalized as the cumulative semantic proximity of the selected evidence to each submission $p \in \mathcal{P}$, with the contribution of a candidate paper $q$ to $p$ quantified by the embedding cosine similarity $\cos(\mathbf{e}_p, \mathbf{e}_q)$.
  \item \textbf{Representativeness.} Not all publications in a reviewer's history $\mathcal{H}_r$ are equally informative, as sustained core research should outweigh peripheral, one-off projects. Each candidate $q$ is therefore weighted by a \emph{representativeness factor} $\gamma_q \in [0, 1]$, defined as the normalized average cosine similarity to its $k$-nearest neighbors within its author's history $\mathcal{H}_{r(q)}$. A high $\gamma_q$ indicates core expertise, whereas a low value flags peripheral work.
\end{itemize}

A further consequence of budgeted selection concerns experts with broad expertise: since each leaf distribution sums to one, an expert active in many topics could in principle spread mass thinly across leaves and be penalized relative to narrow experts. Two mechanisms bound this effect. First, the evidence set $\mathcal{E}^\star_r$ is capped at $B_r$ papers and selected for proximity and representativeness, so even a broad expert's mass concentrates on submission-relevant leaves. Second, the tree--Wasserstein distance of Equation~\eqref{eq:tree-distance} operates on subtree masses rather than individual leaves: mass spread within the same relevant branch still aggregates at that branch's edges, so within-branch breadth incurs little cost, while for multi-faceted submissions such breadth is precisely what lowers the transport distance (Example~\ref{ex:flat-limitations}).

\noindent\textbf{From Selection Criteria to an Objective.} The three criteria are translated into a single optimization problem in which each term corresponds directly to one criterion. Let $\mathcal{U} = \biguplus_{r \in \mathcal{R}_P} \mathcal{H}_r$ denote the pool of candidate papers across the panel, restricted to each reviewer's publications from the past five years; a publication co-authored by several panelists enters $\mathcal{U}$ once per panelist. Let $x_q \in \{0, 1\}$ indicate whether candidate $q$ is selected. The per-paper quality score
$
s(p, q) = \gamma_q \cdot \max\{0, \cos(\mathbf{e}_p, \mathbf{e}_q)\}
$
fuses \emph{proximity} (the cosine similarity, clipped at zero) and \emph{representativeness} ($\gamma_q$) into a single contribution of paper $q$ toward covering submission $p$. The cumulative quality-weighted proximity of submission $p$ under the current selection is then
\begin{equation}\label{eq:cum-prox}
  \prox(p; x) = \sum_{q \in \mathcal{U}} x_q \cdot s(p, q),
\end{equation}
and \emph{coverage} is enforced by passing this quantity through a concave aggregator $g(t) = \log(1 + t)$ before summing across submissions. The concavity is essential, as it imposes a diminishing-returns rule under which the marginal gain of adding further nearby papers to a well-covered submission drops sharply, redirecting the remaining selection budget toward under-served submissions. The full evidence selection problem is therefore
\begin{equation}\label{eq:evidence}
\begin{aligned}
  \max_{x} \quad & F(x) = \sum_{p \in \mathcal{P}} g\!\big(\prox(p; x)\big), \\
  \mathrm{s.t.} \quad & \sum_{q \in \mathcal{U}_r} x_q \leq B_r, \quad \forall\, r \in \mathcal{R}_P, \\
  & x_q \in \{0,1\}, \quad \forall\, q \in \mathcal{U},
\end{aligned}
\end{equation}
where $\mathcal{U}_r = \{q \in \mathcal{U} \mid r(q) = r\}$ partitions the candidates by authorship and $B_r$ is the per-reviewer evidence budget. The three criteria map cleanly onto Equation~\eqref{eq:evidence}: the outer sum over $p \in \mathcal{P}$ enforces \emph{coverage} of the submission corpus, the inner $\prox(p; x)$ aggregates \emph{proximity} contributions of the selected evidence to $p$, and the per-paper weight $\gamma_q$ inside $s(p, q)$ encodes \emph{representativeness}.

\noindent\textbf{Solution.} Given $g(t) = \log(1+t)$ and $s(p, q) \ge 0$, the objective $F(x)$ is a monotone submodular set function over $\mathcal{U}$, and, denoting by $\mathcal{E} \subseteq \mathcal{U}$ the incrementally selected evidence set, the per-reviewer budget constraints $|\mathcal{E} \cap \mathcal{U}_r| \leq B_r$ form a \emph{partition matroid}~\cite{oxley2011matroid}. Maximizing a monotone submodular function over a matroid is NP-hard in general, but the standard greedy procedure achieves a $\tfrac{1}{2}$-approximation~\cite{fisher1978analysis,nemhauser1978analysis}. Starting from $\mathcal{E} = \emptyset$, the algorithm iteratively selects the candidate maximizing the marginal gain subject to its reviewer's remaining budget,
\begin{equation}\label{eq:margin_gain}
  q^{\star} = \argmax_{\substack{q \in \mathcal{U} \setminus \mathcal{E} \,\land\, |\mathcal{E} \cap \mathcal{U}_{r(q)}| < B_{r(q)}}} \Delta F(q \mid \mathcal{E}),
\end{equation}
where $\Delta F(q \mid \mathcal{E}) = F(\mathcal{E} \cup \{q\}) - F(\mathcal{E})$ denotes the marginal gain from adding $q$ to the current selection. Proofs of submodularity and the approximation guarantee are deferred to \appref{app:proof-evidence}.

\subsection{Transport-Optimal Edge Weight Learning}
\label{sec:distance-learning}

Accurate matching through the taxonomy $\mathcal{T} = (V, E)$ requires an appropriate definition of the parent--child edge weights (i.e., semantic distances). Instantiating each weight as the geometric centroid distance, e.g., $w_{(u,v)} = \|\mathbf{c}_u - \mathbf{c}_v\|_2$, is uncalibrated and fails to adapt to a specific conference's affinity structure, while a supervised alternative is infeasible because neither matching labels nor bidding data are available (Section~\ref{sec:problem}). We therefore propose a self-supervised strategy that leverages pairwise semantic distances between submissions themselves as \emph{pseudo-labels}: edge weights are optimized so that tree-routed distances closely approximate the direct semantic distances in the submission embedding space.

\noindent\textbf{Problem Formulation.} Let $\mathbf{w} \in \mathbb{R}_{\ge 0}^{|E|}$ denote the vector of non-negative edge weights to be learned. For any pair of submissions $p_i, p_j \in \mathcal{P}$ with leaf distributions $\phi_{p_i}, \phi_{p_j}$, the tree-routed distance is the tree--Wasserstein distance of Equation~\eqref{eq:tree-distance},
\begin{equation}
    d_{\mathcal{T}}(p_i, p_j; \mathbf{w}) = \sum_{e \in E} w_e \,\big|m_{\phi_{p_i}}(S_e) - m_{\phi_{p_j}}(S_e)\big| = \mathbf{a}_{ij}^\top \mathbf{w}\,,
\end{equation}
where $[\mathbf{a}_{ij}]_e = \big|m_{\phi_{p_i}}(S_e) - m_{\phi_{p_j}}(S_e)\big| \in [0, 1]$ is the mass that edge $e$ separates; for point masses, $\mathbf{a}_{ij}$ reduces to the indicator of the edges on the path between the two leaves. Taking the pseudo-label distance as the embedding-space dissimilarity $\hat{d}_{ij} = 1 - \cos(\mathbf{e}_{p_i}, \mathbf{e}_{p_j})$, edge-weight learning is cast as a regularized least-squares problem over a sampled set of submission pairs $\mathcal{S}$,
\begin{equation}\label{eq:edge_weight_obj}
    \min_{\mathbf{w} \ge \mathbf{0}} \quad \mathcal{J}(\mathbf{w}) = \frac{1}{|\mathcal{S}|} \sum_{(p_i, p_j) \in \mathcal{S}} \Big( \mathbf{a}_{ij}^\top \mathbf{w} - \hat{d}_{ij} \Big)^2 + \lambda \|\mathbf{w}\|_2^2\,,
\end{equation}
where $\lambda > 0$ is an $L_2$ regularization coefficient that prevents overfitting and promotes structural stability.

\noindent\textbf{Solution.} The objective in Equation~\eqref{eq:edge_weight_obj} is strictly convex in $\mathbf{w}$ and therefore admits a unique global optimum (proof deferred to \appref{app:proof-edge-weight}). The optimum is obtained via Non-Negative Least Squares (NNLS)~\cite{lawson1995solving}, a classical active-set method for constrained least-squares problems, with modern implementations~\cite{bro1997fast} achieving substantial speedups over the original algorithm.

\section{Global Taxonomy Index}
\label{sec:index}

The expert lake $\mathcal{R}_{W}$ (Section~\ref{sec:problem}) comprises a large, continuously growing population of experts well beyond the committed panel of any single conference; we assemble it from open scholarly sources including DBLP~\cite{ley2002dblp} and OpenAlex~\cite{priem2022openalex} (construction details in \appref{app:lake}). Organizing a lake of this scale while serving diverse conferences imposes three requirements on the index: (i)~\emph{lightweight}, keeping construction and storage costs practical as the lake grows; (ii)~\emph{incrementally updatable}, absorbing new publications and reviewers without global rebuilds; and (iii)~\emph{structurally faithful}, so that index-derived distances admit efficient calibration to any conference-specific tree--Wasserstein affinity (Section~\ref{sec:taxonomy}). To satisfy these jointly, \sys\ constructs the global taxonomy over \emph{expertise atoms}, canonical and lightweight knowledge units extracted from publication histories rather than whole documents (paper titles and abstracts): the atom vocabulary is orders of magnitude smaller than the paper corpus, so only localized updates are needed as new concepts emerge, and the structure supports hierarchical retrieval calibrated by the Global--Local Aligner (Section~\ref{sec:completion}).

\subsection{Atom Vocabulary Construction}
\label{sec:atom-extraction}

We first construct the vocabulary that serves as the node set of the global taxonomy. Rather than clustering whole documents (titles and abstracts) as the local taxonomy does (Section~\ref{sec:taxonomy}), \sys\ compresses each expert's publication history into a weighted profile of \emph{expertise atoms}, which are keyphrases representing distinct technical concepts. We adopt keyphrases over richer compact representations (e.g., publication-level summaries) because the atoms of a growing global taxonomy must be \emph{primitive and reusable}: keyphrases denote canonical concepts (e.g., \textsf{approximate nearest neighbor search}) that recur across authors and papers, so expert profiles overlap on common atoms, and the DAG $\mathcal{T}_G$ can encode hypernym--hyponym subsumptions over them. Summaries, by contrast, entangle a concept with paper-specific claims; such near-unique units would grow linearly with the corpus, support little cross-author sharing, and blur the concept-granular correspondence the Global--Local Aligner (Section~\ref{sec:completion}) relies on. Atom granularity is likewise essential to \emph{incremental maintainability}: the atom vocabulary grows orders of magnitude more slowly than the paper corpus, so ingesting a new publication typically reduces to updating weights of existing atoms (Section~\ref{sec:maintenance}), whereas summary-level units would be created afresh for nearly every paper and require recurrent re-clustering. Summaries remain valuable as \emph{auxiliary context}: publication text informs keyphrase extraction and atom embeddings, so contextual signal is distilled into atoms rather than discarded.

\noindent\textbf{Extraction and Canonicalization.}
For each paper $q$ in the expert lake, candidate keyphrases are extracted using KeyBART~\cite{kulkarni2022keybart}, a BART-based model pretrained on scientific abstracts that jointly generates present and absent keyphrases in a single pass; we adopt it over statistical extractors (e.g., YAKE~\cite{campos2020yake}) and similarity-ranking extractors (e.g., KeyBERT~\cite{grootendorst2020keybert}) because its generative formulation handles compound technical terms more reliably. Since the same concept surfaces under varied forms (e.g., ``GNN'' versus ``graph neural network''), candidates are canonicalized through a standard blocking-and-matching entity resolution pipeline~\cite{christen2012er,papadakis2020blocking,elmagarmid2007duplicate,mudgal2018deepmatcher}.

\noindent\textbf{Filtering and Profiling.}
Many extracted candidates are uninformative: generic phrases such as ``related work'' or ``experimental results'' carry no expertise signal.
We prune these using an inverse document frequency (IDF) threshold, retaining only atoms that are specific enough to discriminate between research topics.
The survivors form the global atom vocabulary $\mathcal{A} = \{a_1, \ldots, a_L\}$, equipped with dense embeddings $\mathbf{e}_{a_i}$ and abstract-level co-occurrence vectors $\mathbf{f}_{a_i}$ used for the subsumption signals in Section~\ref{sec:global-taxonomy}.
For each expert $r \in \mathcal{R}_{W}$, an (unnormalized) atom profile $\mathbf{w}_r = (w_{r,1}, \ldots, w_{r,L})$ is then compiled, where $w_{r,i} = \sum_{q \in \mathcal{H}_r} \alpha_{r,q} \cdot \mathbb{1}[a_i \in q]$ quantifies the strength of expert $r$'s demonstrated engagement with atom $a_i$, aggregated over the publication history $\mathcal{H}_r$ and scaled by a bibliometric factor $\alpha_{r,q}$ so that the profile reflects the expert's current, primary, and refereed expertise rather than a raw publication count. The factor $\alpha_{r,q}$ incorporates three signals: the recency of $q$, the prominence of $r$'s authorship on $q$, and the quality of its publication venue. For a co-authored publication $q$, the extracted atoms are credited to every co-author $r$, but not uniformly: the authorship term $\omega_{\text{auth}}(r,q)$ applies harmonic position weighting over the author list, so that earlier-listed authors receive proportionally larger credit, and promotes corresponding authors to full weight when such metadata is available. Combined with the recency term, which discounts outdated evidence, and the venue term, which reflects publication quality, $\alpha_{r,q}$ yields a per-author, per-paper competence estimate rather than a uniform split of credit (see \appref{app:scaling-factor} for the full specification of all three components). The authorship term is field-configurable: for communities with other ordering conventions, e.g., alphabetically ordered fields, a uniform credit can be selected per publication from venue metadata (see \appref{app:scaling-factor}).

\subsection{Atom Hierarchy Construction}
\label{sec:global-taxonomy}

With the vocabulary established, the atoms are organized into a hierarchy encoding hypernym--hyponym subsumption. Whereas the local taxonomy $\mathcal{T}$ (Section~\ref{sec:taxonomy}) is a tree, admitting the closed-form tree--Wasserstein computation, the global taxonomy spans the full breadth of computer science and must accommodate concepts with multiple legitimate parents (``graph neural network'' descends from both ``deep learning'' and ``graph representation learning''), which a tree would arbitrarily collapse. We therefore adopt a directed acyclic graph (DAG) $\mathcal{T}_G = (V_G, E_G)$, where each edge $a_i \to a_j$ encodes that $a_j$ is a strict generalization of $a_i$; this choice does not propagate downstream, since the Global--Local Aligner (Section~\ref{sec:completion}) extracts a tree-structured projection of $\mathcal{T}_G$ prior to any tree--Wasserstein evaluation.

\noindent\textbf{Edge Signals and Construction.} Since our global index operates at the atom level and must be derived without supervision, existing methods that consume full paper texts (e.g., NetTaxo~\cite{shang2020nettaxo}) or presuppose a seed taxonomy (e.g., TaxoExpan~\cite{shen2020taxoexpan}) are inapplicable. We therefore construct $\mathcal{T}_G$ by combining three complementary distributional signals that capture both symmetric topical relatedness and asymmetric generalization (subsumption): \emph{(i) Cosine similarity} $\mathrm{sim}(a_i, a_j) = \cos(\mathbf{e}_{a_i}, \mathbf{e}_{a_j})$ identifies symmetric semantic neighborhoods. \emph{(ii) WeedsPrec} $\mathrm{WP}(a_i \to a_j)$~\cite{weeds2004characterising} measures asymmetric context subsetting, establishing $a_j$ as a parent since hyponyms appear in narrower contexts than their hypernyms. \emph{(iii) Pointwise Mutual Information} $\mathrm{PMI}(a_i, a_j)$ reinforces edges between atoms frequently co-mentioned in the same abstracts. Candidate directed edges are scored via a linear combination of these signals and retained if they exceed a threshold $\tau_{\mathrm{edge}}$; acyclicity is preserved by construction: an edge $a_i \to a_j$ is inserted only if the $O(1)$ reachability check on the indicator $z$ (below) confirms $a_i$ is not already an ancestor of $a_j$. A transitive reduction~\cite{aho1972transitive} is then applied to prune redundant shortcuts (e.g., a direct ``GNN'' $\to$ ``machine learning'' edge bypassing intermediate concepts), yielding the Hasse diagram of the induced partial order~\cite{davey2002lattices}. Finally, a Boolean indicator $z_{ij} = 1$ iff $a_j$ is reachable from $a_i$ is precomputed, enabling $O(1)$ ancestry queries that support fast incremental updates and the Global--Local Aligner~\cite{yu2010reachability,jin2009threehop}.

\noindent\textbf{Upward Expertise Propagation.} The base profile $\mathbf{w}_r$ (Section~\ref{sec:atom-extraction}) weights only explicitly published atoms, missing hierarchical inheritance (publishing heavily on ``graph attention networks'' implies expertise in the ancestor ``deep learning''). Direct weights are therefore propagated upward through $\mathcal{T}_G$: for each node $a_j$,
$
  w_r^{\uparrow}(a_j) = w_{r,j} + \sum_{a_k:\, z_{kj} = 1} w_{r,k},
$
so every descendant of $a_j$ contributes its direct weight once, even when it is reachable along several paths. The resulting profile supports multi-granularity retrieval, with leaf queries consuming direct weights and internal-node queries the aggregated mass of all specializations beneath. $\mathbf{w}_r^{\uparrow}$ is materialized in the node posting lists and serves only candidate lookup (Section~\ref{sec:retrieval}); transport scoring uses normalized direct weights (Section~\ref{sec:structural-alignment}).

\noindent\textbf{Auxiliary Information Layers.}
Beyond the structural core of $\mathcal{T}_G$ and its annotations $\{\mathbf{w}_r^{\uparrow}\}$, panel expansion requires several materialized views that amortize query-time computation: (1) a \emph{sparse atom catalog} mapping canonicalized surface variants to exact lexical postings~\cite{zobel2006inverted}; (2) a \emph{dense atom catalog} indexing the atom embeddings $\{\mathbf{e}_{a_i}\}$ for approximate nearest-neighbor retrieval~\cite{malkov2020hnsw,fu2019nsg,jegou2011pq}; (3) a \emph{node posting list} storing, for each $a_j \in V_G$, the pairs $(r, w_r^{\uparrow}(a_j))$ sorted by weight, enabling branch-level expert retrieval; and (4) an \emph{expert profile catalog} maintaining each expert's $\mathbf{w}_r$ and $\mathbf{w}_r^{\uparrow}$ with provenance metadata.

\subsection{Incremental Maintenance}
\label{sec:maintenance}

\begin{algorithm}[t]
\caption{Incremental Maintenance of Global Taxonomy Index}
\label{alg:maintenance}
\small
\begin{algorithmic}[1]
\Require New publication $q$ for expert $r$, vocabulary $\mathcal{A}$, DAG $\mathcal{T}_G$
\State Extract and canonicalize keyphrases from $q$ against $\mathcal{A}$
\State Partition into matched ($\mathcal{K}_{\mathrm{match}}$) and unmatched ($\mathcal{K}_{\mathrm{new}}$)
\State \textit{// Level 1: Atom-level update}
\For{each $a_i \in \mathcal{K}_{\mathrm{match}}$}
    \State Update profile $w_{r,i} \gets w_{r,i} + \alpha_{r,q}$ and co-occurrences $\mathbf{f}_{a_i}$
\EndFor
\For{each $k \in \mathcal{K}_{\mathrm{new}}$}
    \State Create $a_{\mathrm{new}}$ with embedding $\mathbf{e}_{a_{\mathrm{new}}}$; add to $\mathcal{A}$ if $\mathrm{IDF} \geq \tau_{\mathrm{idf}}$, with $w_{r,\mathrm{new}} \gets \alpha_{r,q}$
\EndFor
\State \textit{// Level 2: Taxonomy structure update}
\For{each new $a_{\mathrm{new}}$, and $a_i \in \mathcal{K}_{\mathrm{match}}$ with shifted $\mathbf{f}_{a_i}$}
    \State Score neighborhood candidate edges; insert/remove if crossing $\tau_{\mathrm{edge}}$
    \State Apply local transitive reduction and update the ancestry indicator $z$
\EndFor
\State \textit{// Level 3: Annotation and auxiliary-layer refresh}
\State Recompute propagated weights $w_r^{\uparrow}$ along affected DAG paths
\State Refresh sparse and dense atom catalogs for new or updated atoms
\State Refresh node posting lists for affected DAG nodes and their ancestors
\State Update expert profile catalog entry for $r$
\end{algorithmic}
\end{algorithm}

Rebuilding the global index for every new publication is prohibitive at large scale. \sys\ instead exploits \textit{update locality}: a typical publication introduces few (if any) new atoms and only slightly shifts local co-occurrence statistics, so the impact on auxiliary layers is restricted to the ancestral paths of the affected atoms. Algorithm~\ref{alg:maintenance} organizes maintenance as a \textbf{three-level cascade}, triggering higher-level structural updates only when lower-level changes make them necessary.

\noindent\textbf{Level 1: Atom-level update (Lines 4--7).} The cascade extracts and partitions the keyphrases of a new publication $q$ into existing ($\mathcal{K}_{\mathrm{match}}$) and novel ($\mathcal{K}_{\mathrm{new}}$) sets. Existing atoms receive a statistics-only (non-structural) update of profile weights and co-occurrence vectors (Lines 4--5); novel keyphrases exceeding the IDF threshold $\tau_{\mathrm{idf}}$ are inserted as new atoms credited to $r$ (Lines 6--7). This isolates the common case: when an expert publishes within their established area, no new atoms arise, so Level 2 is skipped unless co-occurrences drift, and only the affected annotations and auxiliary layers are refreshed at Level 3.

\noindent\textbf{Level 2: Taxonomy structure update (Lines 9--11).} Triggered only when Level 1 introduces new atoms or shifts a co-occurrence vector $\mathbf{f}_{a_i}$ beyond a drift threshold $\tau_{\mathrm{drift}}$, this level integrates new atoms by scoring candidate edges against their $k$-nearest embedding neighbors followed by a local transitive reduction that also updates the ancestry indicator $z$, and recomputes drifted atoms' outgoing edges to reflect shifting semantics (e.g., ``transformer'' evolving from an architecture term to a foundation-model concept), adding or removing edges crossing $\tau_{\mathrm{edge}}$ (Line 10).

\noindent\textbf{Level 3: Annotation and auxiliary-layer refresh (Lines 13--16).} The final stage synchronizes annotations and auxiliary layers: since hierarchical changes propagate strictly upward, the propagated weights $w_r^{\uparrow}$ and node posting lists are recomputed only for affected DAG nodes and their ancestors (Lines 13 and 15), the atom catalogs are refreshed for new or updated atoms (Line 14), and the expert profile catalog receives the revised weights (Line 16).

\noindent\textbf{Cost.} The amortized cost per maintenance event is $O(|\mathcal{A}_q| \cdot k + |\mathcal{A}_{\mathrm{new}}| \cdot k^2 + d^{\uparrow})$, where $|\mathcal{A}_q|$ denotes the number of keyphrases extracted from $q$, $|\mathcal{A}_{\mathrm{new}}|$ the number of newly created atoms, $k$ the embedding-neighborhood size, and $d^{\uparrow}$ the number of ancestors reached by upward propagation. A detailed per-level derivation is provided in \appref{app:maintenance-cost}.

\noindent\textbf{Remarks.} This bound presupposes canonical, reusable atoms: with publication-specific units (e.g., summaries) as atoms, nearly every event would create new atoms ($|\mathcal{A}_{\text{new}}| \approx |\mathcal{A}_q|$), re-activating the quadratic Level-2 term on every ingestion and requiring recurrent re-clustering, which is precisely what the keyphrase design avoids.

\section{Global--Local Taxonomy Aligner}
\label{sec:completion}

As introduced in Section~\ref{sec:overview}, panel expansion faces a structural asymmetry: expert profiles are indexed on $\mathcal{T}_G$, whereas the tree--Wasserstein affinity is defined on $\mathcal{T}$. Rescoring every candidate on $\mathcal{T}$ is faithful but infeasible at scale, while retrieving directly on $\mathcal{T}_G$ is efficient yet distribution-shifted, since $\mathcal{T}_G$ carries no metric aligned with $\mathcal{T}$. The central challenge is to approximate the conference-specific affinity at large scale without paying its full computational cost.

\noindent\textbf{A Cross-Taxonomy Alignment Lens.} Although $\mathcal{T}$ and $\mathcal{T}_G$ are constructed independently from different corpora, $\mathcal{T}_G$ is designed to span the breadth of computer science, so the topical region covered by any conference-specific $\mathcal{T}$ should correspond to a sub-region of $\mathcal{T}_G$. We exploit this containment by treating $\mathcal{T}$ as the structural and metric target against which a sub-structure of $\mathcal{T}_G$ is aligned: each conference node (a document-cluster topic region, Section~\ref{sec:core}) should align with a global atom, each conference edge with a surviving subsumption relation, and each conference edge weight with a calibrated weight on the aligned sub-structure. The resulting \emph{alignment problem} decomposes into two coupled sub-problems: (i) \emph{structural alignment} (Section~\ref{sec:structural-alignment}), recovering a compact tree-structured projection of $\mathcal{T}_G$ whose topology approximates that of $\mathcal{T}$; and (ii) \emph{weight calibration} (Section~\ref{sec:calibration}), fitting edge weights that render the projection metrically consistent with $\mathcal{T}$. Their solution yields a proxy taxonomy $\mathcal{T}_G^{\star}$ locally faithful to $\mathcal{T}$ yet globally enriched by the expert profiles indexed on $\mathcal{T}_G$. Algorithm~\ref{alg:aligner} summarizes the procedure.

\begin{algorithm}[t]
\caption{Global--Local Taxonomy Aligner}
\label{alg:aligner}
\small
\begin{algorithmic}[1]
\Require Global taxonomy $\mathcal{T}_G$; conference taxonomy $\mathcal{T}=(V,E)$ with node centroids $\{\mathbf{c}_v\}$; panel $\mathcal{R}_P$; submissions $\mathcal{P}$; $\lambda; w_0$
\Ensure Calibrated compressed taxonomy $\mathcal{T}_G^{\star}=(V^{\star}, E^{\star}, \mathbf{w}^{\star})$
\State \textit{// Structural Alignment: Anchor Selection (Sec.~\ref{sec:structural-alignment})}
\For{each $v \in V$}
    \State $\mathcal{A}_v^{\mathrm{cand}} \gets \mathrm{top\text{-}}k_c\{a_i : \cos(\mathbf{c}_v, \mathbf{e}_{a_i})\}$ over $\mathcal{T}_G$
\EndFor
\State $\mathcal{A}^{\mathrm{cand+}} \gets \mathrm{AncestorClosure}_{\mathcal{T}_G}\!\big(\bigcup_v \mathcal{A}_v^{\mathrm{cand}}\big)$
\State Solve the hierarchy-preserving ILP (Sec.~\ref{sec:structural-alignment}) with penalty $\lambda$ to obtain $\{y_{v,i}\}$
\State $\mathcal{A}^{\star} \gets \bigcup_v \{a_i : y_{v,i} = 1\}$
\State \textit{// Structural Alignment: Steiner Extraction \& Contraction (Sec.~\ref{sec:structural-alignment})}
\State Resolve DAG directionality on $\mathcal{A}^{\mathrm{cand+}}$ by retaining, per atom, the highest-scoring parent edge
\State $\mathcal{S} \gets \mathrm{SteinerTree}(\mathcal{T}_G, \mathcal{A}^{\star})$
\State $\mathcal{T}_G^{\star} \gets \mathrm{ContractNonBranchingChains}(\mathcal{S})$
\For{each $v^{\star} \in V^{\star}$}
    \State $w_p(v^{\star}) \gets \sum_{a_i \in \mathrm{atoms}(v^{\star})} w_{p,i}$;\quad $w_{r'}(v^{\star}) \gets \sum_{a_i \in \mathrm{atoms}(v^{\star})} w_{r',i}$
\EndFor
\State \textit{// Weight Calibration (Sec.~\ref{sec:calibration})}
\For{each $(p, r) \in \mathcal{P} \times \mathcal{R}_P$}
    \State $d_{\mathcal{T}}(p, r) \gets$ exact tree--Wasserstein on $\mathcal{T}$ (Sec.~\ref{sec:taxonomy})
    \State $[\mathbf{h}_{pr}]_e \gets \big|m_{\psi_p}(S_e) - m_{\psi_r}(S_e)\big|$ for each $e \in E^{\star}$
\EndFor
\State Solve the calibration LP in Eq.~\eqref{eq:calibration-lp} to obtain $\mathbf{w}^{\star}$
\State \Return $\mathcal{T}_G^{\star} = (V^{\star}, E^{\star}, \mathbf{w}^{\star})$
\end{algorithmic}
\end{algorithm}

\subsection{Structural Alignment}
\label{sec:structural-alignment}

Starting from $\mathcal{T}_G$, structural alignment extracts a compact tree-structured projection $\mathcal{T}_G^{\star}$ whose topology approximates that of $\mathcal{T}$, in three stages.

\noindent\textbf{Anchor Selection (Lines 2--6).}
For each conference node $v$ with centroid $\mathbf{c}_v$, a candidate pool $\mathcal{A}_v^{\mathrm{cand}}$ of the top-$k_c$ atoms is retrieved by centroid cosine similarity. Since the mapping must respect hierarchical structure in addition to semantic similarity, anchor selection is cast as a hierarchy-preserving optimization assigning exactly one global atom to each conference node, maximizing matched-pair embedding similarity while penalizing local parent--child edges that map to atoms violating the precomputed global ancestry (Section~\ref{sec:index}); the selected atoms form the anchor set $\mathcal{A}^{\star}$, one atom per conference node. The full integer linear program (ILP) formulation and solution are deferred to \appref{app:mapping-ilp}.

\noindent\textbf{Steiner Tree Extraction (Lines 8--9).}
The anchors in $\mathcal{A}^{\star}$ are connected through their common ancestors in $\mathcal{T}_G$, which must appear as internal vertices to preserve hypernymic structure. The expanded pool $\mathcal{A}^{\mathrm{cand+}}$, obtained by closing $\bigcup_v \mathcal{A}_v^{\mathrm{cand}}$ under DAG ancestry, supplies exactly these non-terminal vertices, so every anchor-to-ancestor path is already materialized and no second retrieval pass is needed. Directionality is resolved by retaining, for each atom in $\mathcal{A}^{\mathrm{cand+}}$, only the highest-scoring parent edge, reducing the induced subgraph to a tree on which the minimal subtree spanning $\mathcal{A}^{\star}$ (the Steiner tree) is computable in a single post-order pass.

\noindent\textbf{Chain Contraction (Lines 10--12).}
Any chain $a_i \to a_k \to a_j$ where $a_k$ has exactly one parent and one child in the Steiner tree is contracted into a direct edge $a_i \to a_j$, eliminating chains of generic ancestors (e.g., ``machine learning'' $\to$ ``learning algorithm'' $\to$ ``algorithm''). The compressed taxonomy $\mathcal{T}_G^{\star} = (V^{\star}, E^{\star})$ retains only the branching points where retrieval decisions are made; each compressed node $v^{\star}$ aggregates a set $\mathrm{atoms}(v^{\star})$ of original atoms. A submission or expert $x$ is represented on $\mathcal{T}_G^{\star}$ by normalizing its direct (unpropagated) profile weights, $\psi_x(v^{\star}) = w_x(v^{\star}) / \sum_{u^{\star} \in V^{\star}} w_x(u^{\star})$ with $w_x(v^{\star}) = \sum_{a_i \in \mathrm{atoms}(v^{\star})} w_{x,i}$; weights of atoms outside $\mathcal{T}_G^{\star}$ are dropped before normalization, and the subtree mass $m_{\psi_x}(S_e)$ sums $\psi_x$ over the nodes below edge $e$. No recomputation against $\mathcal{T}$ is required.

\subsection{Weight Calibration}
\label{sec:calibration}

Following structural alignment, the next step is to set the edge weights on the projected taxonomy. Two challenges arise from the absence of a direct supervision signal:
\begin{itemize}[leftmargin=*,nosep]
  \item \emph{(C1) Target distances are defined only on the committed panel.} The only quantity with a valid claim to being the conference-specific distance is $d_{\mathcal{T}}(p, r)$, obtained through the Section~\ref{sec:taxonomy} pipeline on the local taxonomy $\mathcal{T}$. However, the role of $\mathcal{T}_G^{\star}$ is to score lake candidates without passing them through $\mathcal{T}$, so the supervision signal is available on the committed panel but not on the lake.
  \item \emph{(C2) Edge weights are globally coupled.} The tree--Wasserstein proxy sums $w_e \cdot |m_{\psi_p}(S_e) - m_{\psi_r}(S_e)|$ over all edges, so modifying any single weight $w_e$ simultaneously perturbs the affinity of every submission--expert pair whose transport path crosses $e$. The calibration problem therefore does not decompose edge by edge, and greedy or subtree-local strategies are not applicable.
\end{itemize}

\noindent\textbf{Self-Supervised Calibration from the Panel (Lines 14--16).}
To address (C1), the edge weights of $\mathcal{T}_G^{\star}$ are fitted from a self-supervised signal on the committed panel, reusing the affinity scores $d_{\mathcal{T}}(p, r)$ already computed during matching. Both sides of every pair must carry a distribution on the global taxonomy: reviewers satisfy this automatically, since most reviewers in $\mathcal{R}_P$ also appear in the lake $\mathcal{R}_W$ and thus already carry an atom profile from the global index, and submissions are projected onto $\mathcal{T}_G^{\star}$ through the structural alignment to obtain a counterpart distribution $\psi_p$. The panel labels then serve as regression targets, producing a lightweight proxy that scores lake candidates on $\mathcal{T}_G^{\star}$ without passing them back through $\mathcal{T}$; label-generation cost scales with the panel rather than the lake. With an empty panel, Equation~\eqref{eq:calibration-lp} has no fitting term and returns the uniform prior $w_e^{\star} = w_0$.

\noindent\textbf{The Linearity Observation.}
To address (C2), we exploit the fact that, for any pair of distributions $\psi_p, \psi_r$ over the nodes of $\mathcal{T}_G^{\star}$, the tree--Wasserstein proxy distance is \emph{linear} in $\mathbf{w}^{\star}$,
\begin{equation}\label{eq:linearity}
  d_{\mathcal{T}_G^{\star}}(p, r; \mathbf{w}^{\star}) = \mathbf{h}_{pr}^{\top} \mathbf{w}^{\star},
  \quad [\mathbf{h}_{pr}]_e = \big|m_{\psi_p}(S_e) - m_{\psi_r}(S_e)\big|,
\end{equation}
where the activation vector $\mathbf{h}_{pr}$ depends only on the two distributions and the tree topology, and is therefore precomputed once per pair. Linearity turns the coupled calibration problem into a single convex regression in $\mathbf{w}^{\star}$: each pair $(p, r)$ becomes a training example with feature vector $\mathbf{h}_{pr}$ and target $d_{\mathcal{T}}(p, r)$, and all edges are fitted jointly from the resulting linear system with a guaranteed global optimum.

\noindent\textbf{The Calibration Objective (Line 17).}
Combining the linearity of Equation~\eqref{eq:linearity} with the self-supervision produced above, weight calibration reduces to a linear program,
\begin{equation}\label{eq:calibration-lp}
\begin{aligned}
\min_{\mathbf{w}^{\star},\, \boldsymbol{\delta}} \;\; & \sum_{p, r} \frac{\delta_{pr}}{\max\{d_{\mathcal{T}}(p, r), \eta\}} + \mu \sum_e \big|w_e^{\star} - w_0\big|, \\
\mathrm{s.t.} \;\; & \delta_{pr} \geq \pm\big(\mathbf{h}_{pr}^{\top} \mathbf{w}^{\star} - d_{\mathcal{T}}(p, r)\big), \; \forall\, p \in \mathcal{P},\, r \in \mathcal{R}_P, \\
& w_e^{\star} \geq \epsilon, \; \forall\, e \in E^{\star},
\end{aligned}
\end{equation}
where $w_0$ is a uniform scale prior, $\mu$ controls the trade-off between fitting the panel and regularizing toward the prior, and the fitting error is normalized by $\max\{d_{\mathcal{T}}(p, r), \eta\}$ with a small $\eta > 0$, which guards against zero targets, so that the LP optimizes relative rather than absolute discrepancy. The non-negativity constraint $w_e^{\star} \geq \epsilon$ preserves the validity of the resulting tree--Wasserstein metric. After linearizing the absolute values, the LP has $2|E^{\star}| + |\mathcal{P}| \cdot |\mathcal{R}_P|$ variables and is solved efficiently by standard interior-point methods~\cite{karmarkar1984new,wright1997primaldual} using off-the-shelf solvers such as Gurobi~\cite{gurobi} or CPLEX~\cite{cplex}.

\subsection{Two-Stage Retrieval for Online Queries}
\label{sec:retrieval}

With $\mathcal{T}_G^{\star}$ calibrated, panel-expansion queries are compiled and served at low cost through a single retrieval primitive, casting each operational scenario as a paper-weight function $\omega_q$ over the submission corpus that encodes which submissions require additional coverage.

\noindent\textbf{Query Types.}
Three query types cover the scenarios encountered in practice, each instantiating the abstract query $q$ of Definition~\ref{def:topb-retrieval}: (i) a \emph{panel-level query} places uniform weights over $\mathcal{P}$ to strengthen an inadequate committed panel across the whole corpus; (ii) a \emph{topic-level query} concentrates weight on submissions that fall within under-covered taxonomy branches, scaled by topic priority; and (iii) a \emph{paper-level query} assigns weight proportional to each submission's affinity gap relative to a target threshold, focusing retrieval on weakly assigned papers. Reducing all scenarios to a single weighted-submission abstraction allows the same two-stage pipeline to serve every case.

\noindent\textbf{Target Distribution.}
Regardless of how $\omega_q$ is constructed, the query is compiled into a target distribution $\psi_q$ over the nodes of $\mathcal{T}_G^{\star}$ by taking the $\omega_q$-weighted average of the submission distributions $\psi_p$. After this step, the retrieval pipeline is scenario-agnostic.

\noindent\textbf{Stage 1: Node-Level Candidate Location.}
The \emph{active node set} comprises the nodes of $\mathcal{T}_G^{\star}$ on which $\psi_q$ places mass above a threshold $\tau_{\mathrm{act}}$; probing their posting lists yields the candidate set $\mathcal{R}_q^{\mathrm{cand}}$ of experts with non-zero weight on at least one active node, shrinking the candidate space from the full lake in a single sparse-probe step.

\noindent\textbf{Stage 2: Calibrated Proxy Reranking.}
For each candidate $r' \in \mathcal{R}_q^{\mathrm{cand}}$, the calibrated proxy distance $\hat{d}_q(r')$ sums, over every non-root node, the calibrated weight of the edge to its parent scaled by the absolute difference between the candidate's and the query's subtree mass. Because $\mathcal{T}_G^{\star}$ is compact and subtree masses are precomputed, reranking is efficient even for thousands of candidates, yielding the top-$K_q$ shortlist $\mathcal{R}_q^{\mathrm{short}}$ for exact scoring on $\mathcal{T}$.

\noindent\textbf{Conflict Handling.} Conflicts of interest are handled uniformly for committed reviewers and lake experts, and independently of affinity estimation: they never modify affinity or proxy scores but are enforced downstream as hard constraints. Observable conflicts, i.e., co-authorship and, where metadata permits, shared institutional affiliation, are derived automatically from DBLP/OpenAlex records; those invisible to publication metadata, such as personal relationships or shared funding, are self-declared by authors or reviewers (cf.\ the remark in Section~\ref{sec:overview}). Conflicted pairs are excluded from the assignment problem of Eq.~\eqref{eq:assignment} by zeroing the corresponding $x_{p,r}$, and conflicted external experts are filtered from the shortlist $\mathcal{R}^{\text{short}}_q$ before any recommendation is returned.

\section{Experiments}
\label{sec:exp}

The evaluation of \sys proceeds along three dimensions corresponding to its three main contributions: (i) the accuracy of taxonomy-induced affinity estimation (Section~\ref{sec:exp:affinity}); (ii) the effectiveness of the calibrated proxy for panel expansion (Section~\ref{sec:exp:completion}); and (iii) the cost and efficiency of the Global Taxonomy Index (Section~\ref{sec:exp:index}). We further report an ablation study (Section~\ref{sec:exp:ablation}) that isolates the contribution of each component, a human evaluation (Section~\ref{sec:human-eval}) that validates both the assignment output and the expansion recommendations against blinded expert judgments, and a case study (Section~\ref{app:case}) that illustrates \sys's behavior on a representative multi-faceted submission.

\subsection{Experimental Setup}
\label{sec:exp:setup}

\noindent\textbf{Evaluation Data.}
The study uses two complementary data assets: conference datasets for affinity estimation and human evaluation, and an external expert lake for panel expansion.
\begin{itemize}[leftmargin=*,noitemsep,topsep=0pt]
\item \emph{Conference datasets.} We adopt four established benchmarks from prior work, namely NIPS (NIPS06~\cite{mimno2007expertise}), SIGIR (SIGIR07~\cite{karimzadehgan2008multiaspect}), SciRepEval~\cite{singh2023scirepeval}, and KDD (KDD20~\cite{zhang2025chainoffactors}). Each benchmark provides three-level human-annotated relevance judgments (strongly relevant, weakly relevant, and irrelevant) over at most a few hundred submissions and reviewers. In addition, we construct a real-world conference dataset, \neurips, from the publicly released NeurIPS 2025 acceptance and authorship records retrieved via the OpenReview API. It comprises $5{,}286$ accepted submissions and $11{,}235$ candidate reviewers and serves as the basis for our human evaluation (Section~\ref{sec:human-eval}). Table~\ref{tab:datasets} summarizes the dataset statistics.
\begin{table}[t]
\centering
\caption{Statistics of the benchmark datasets. NeurIPS25 lacks annotated pairs and is used for human evaluation.}
\label{tab:datasets}
\small
\begin{tabular}{l r r r}
\toprule
\textbf{Dataset} & \textbf{\#Papers} & \textbf{\#Reviewers} & \textbf{\#Annotated Pairs} \\
\midrule
NIPS06      & 34    & 190    & 393    \\
SciRepEval  & 107   & 661    & 1{,}729 \\
SIGIR07     & 73    & 189    & 13{,}797 \\
KDD20       & 174   & 737    & 3{,}480 \\
NeurIPS25   & 5{,}286 & 11{,}235 & -- \\
\bottomrule
\end{tabular}
\end{table}

\item \emph{Expert lake.} For panel expansion, we use an expert lake assembled from DBLP and OpenAlex covering researchers in computer science, comprising approximately $240$K experts and $1.3$M publications.
\end{itemize}

\noindent\textbf{Implementation Details.}
All experiments are conducted on a single NVIDIA RTX 5090 GPU (32\,GB) with 25 vCPUs (Intel Xeon Platinum 8470Q) and 90\,GB of RAM. Following the common paradigm of iterative LLM-based taxonomy construction~\cite{lan2026agenticscholar,kargupta2025taxoadapt}, we induce the local taxonomy backbone by prompting DeepSeek-V3.2 to iteratively partition and label topic nodes up to a maximum depth of $D = 4$. The reviewer evidence budget is set to $B_r = 15$ papers. To prevent temporal leakage, we exclude from each reviewer's profile any publication that appeared after the conference submission deadline. The source code, datasets, and experiment scripts are publicly available~\cite{taxoreview_repo}.

\subsection{Effectiveness on Affinity Estimation}
\label{sec:exp:affinity}

\noindent\textbf{Baselines.}
We compare \sys\ against six representative paper--reviewer matching methods spanning sparse retrieval, pretrained language model (PLM) encoders, and LLM-based relevance judgment. \textbf{TPMS}~\cite{charlin2013tpms} is a classical content-based baseline that scores affinity via TF--IDF cosine similarity. \textbf{SciBERT}~\cite{beltagy2019scibert} is a domain-specific BERT encoder pretrained on scientific corpora. \textbf{SPECTER}~\cite{cohan2020specter} is a document-level embedding model fine-tuned from SciBERT on the citation graph. \textbf{SPECTER2}~\cite{singh2023scirepeval} is its successor, for which we use the proximity-optimized variant reported to perform best for reviewer matching. \textbf{CoF}~\cite{zhang2025chainoffactors} is a state-of-the-art multi-factor framework that fuses semantic, citation, and topic signals within a unified language model. Finally, \textbf{LLM Judge} prompts DeepSeek-V3.2, the same LLM used by \sys, to rate the relevance of each submission to each reviewer's profile on a scale from 0 to 10. This baseline controls for the backbone LLM, allowing us to verify that the gains of \sys\ stem from its design rather than from the capability of the underlying LLM.

\noindent\textbf{Evaluation Metrics.}
For each submission, reviewers in the annotated candidate set are ranked by affinity score. Following~\cite{singh2023scirepeval,zhang2025chainoffactors}, ranking quality in the main comparison is measured by Precision@$5$ (P@5; reported as Soft and Hard precision, averaged over submissions): Soft admits reviewers labeled strongly or weakly relevant, while Hard admits only strongly relevant ones.

\noindent\textbf{Results}. Table~\ref{tab:effectiveness} reports effectiveness and end-to-end running time across the four established benchmarks (NIPS, SIGIR, SciRepEval, and KDD) along with their average.

\noindent\textit{Effectiveness.} \sys outperforms every baseline on all four benchmarks under both Soft and Hard settings, achieving the highest average Soft and Hard precision ($53.2\%$ and $26.5\%$), with all margins over the strongest baseline in each column, CoF on average (51.2\% and 24.7\%), significant at $p<0.01$. Among the baselines, CoF benefits from unifying topic, semantic, and citation signals, while TPMS remains competitive and outperforms the dense encoders on all four benchmarks under the Soft setting. LLM Judge, which shares the same backbone LLM (DeepSeek-V3.2) with \sys, attains an average precision of only $46.2\%$ (Soft) and $22.6\%$ (Hard), comparable to SPECTER2 but below both CoF and TPMS, and is competitive only on NIPS, the benchmark with the smallest candidate pool per submission. We attribute this gap to its pointwise scoring, which rates each submission--reviewer pair in isolation on a coarse 0--10 scale and thus lacks a shared reference frame for ranking reviewers. Since \sys surpasses LLM Judge by $15.2\%$ (Soft) and $17.3\%$ (Hard) on average, its gains stem from the taxonomy-driven affinity design rather than from the underlying LLM. Moreover, the relative gains of \sys over CoF are consistently larger on the Hard metrics than on the Soft metrics, reaching $14.6\%$ vs.\ $5.7\%$ on SIGIR, $10.8\%$ vs.\ $3.3\%$ on KDD, and $7.3\%$ vs.\ $3.9\%$ on average, indicating that hierarchical modeling preserves fine-grained topical boundaries that flat representations blur. This advantage is most pronounced on SIGIR and KDD, the benchmarks with the largest candidate pools per submission, where strongly and weakly relevant experts become densely packed in flat space and the tree--Wasserstein penalty on cross-branch transport sustains precision that flat cosine similarity cannot.

\noindent\textit{Efficiency.} \sys\ completes the four benchmarks in $27.4$\,min in total, faster than both CoF ($29.9$\,min) and LLM Judge ($43.2$\,min). CoF generates three separate embeddings and LLM-based factors per paper, while LLM Judge must query the LLM for every submission--reviewer pair. The single-encoder PLM baselines are faster ($4.7$--$5.7$\,min) because they require only one embedding pass. \sys\ additionally induces the local taxonomy, selects reviewer evidence, and calibrates edge weights, but this is a one-time cost per conference and small relative to an assignment cycle. In terms of monetary cost, \sys\ incurs only $5.3$ USD in LLM API fees, $28.4\%$ less than LLM Judge ($7.4$ USD). Both costs are negligible within a conference management budget, yet their growth differs fundamentally. The LLM cost of \sys\ grows with the number of documents used for taxonomy induction, whereas that of LLM Judge grows with the number of submission--reviewer pairs. This makes pairwise LLM judging prohibitive at modern conference scale.

\begin{table*}[t]
\centering
\caption{Precision@5 (\%) under Soft and Hard relevance and total running time on the four benchmarks. The best precision in each column is in bold and the strongest baseline is underlined ($^{**}$: improvement over the underlined baseline significant at $p<0.01$, pairwise bootstrap test).}
\label{tab:main_results}\label{tab:effectiveness}
\setlength{\tabcolsep}{4.5pt}
\resizebox{\ifdim\width>\textwidth\textwidth\else\width\fi}{!}{%
\begin{tabular}{lccccccccccr}
\toprule
 & \multicolumn{2}{c}{NIPS} & \multicolumn{2}{c}{SIGIR} & \multicolumn{2}{c}{SciRepEval} & \multicolumn{2}{c}{KDD} & \multicolumn{2}{c}{Average} & \multicolumn{1}{c}{Running} \\
\cmidrule(lr){2-3} \cmidrule(lr){4-5} \cmidrule(lr){6-7} \cmidrule(lr){8-9} \cmidrule(lr){10-11}
Method & Soft & Hard & Soft & Hard & Soft & Hard & Soft & Hard & Soft & Hard & \multicolumn{1}{c}{time} \\
\midrule
TPMS & 53.8 & 25.1 & 42.7 & \underline{21.2} & 64.3 & 32.8 & 33.1 & 14.6 & 48.5 & 23.4 & 39\,s \\
SciBERT & 48.0 & 20.9 & 35.7 & 16.5 & 59.3 & 26.7 & 24.4 & 10.7 & 41.9 & 18.7 & 4.7\,min \\
SPECTER & 52.6 & 23.3 & 37.9 & 13.7 & 61.1 & 29.6 & 31.3 & 13.8 & 45.7 & 20.1 & 5.3\,min \\
SPECTER2 & 53.6 & 26.8 & 38.7 & 15.6 & 63.1 & 31.4 & 30.6 & 13.6 & 46.5 & 21.9 & 5.7\,min \\
CoF & \underline{54.4} & 27.3 & \underline{45.8} & \underline{21.2} & \underline{67.9} & \underline{34.6} & \underline{36.8} & \underline{15.7} & \underline{51.2} & \underline{24.7} & 29.9\,min \\
LLM Judge & 51.8 & \underline{27.7} & 38.9 & 19.2 & 62.2 & 29.9 & 31.8 & 13.6 & 46.2 & 22.6 & 43.2\,min \\
\midrule
TaxoReview & \textbf{57.0}\sig & \textbf{28.7}\sig & \textbf{48.4}\sig & \textbf{24.3}\sig & \textbf{69.3}\sig & \textbf{35.6}\sig & \textbf{38.0}\sig & \textbf{17.4}\sig & \textbf{53.2}\sig & \textbf{26.5}\sig & 27.4\,min \\
\bottomrule
\end{tabular}%
}
\end{table*}

\subsection{Effectiveness on Panel Expansion}
\label{sec:exp:completion}

\noindent\textbf{Query Instantiation.} This experiment evaluates the panel expansion pipeline of \sys\ against the external expert lake. For each of the four benchmarks, we construct $10$ queries for each of the three query types defined in Section~\ref{sec:retrieval}, namely panel-level ($Q_\text{pan}$), topic-level ($Q_\text{top}$), and paper-level ($Q_\text{pap}$) queries, yielding $120$ queries in total. Since expansion is an on-demand operation with recruitment budget $B$, invoked at the chair's discretion or when taxonomy branches are under-covered or submissions fall short of the target affinity (Section~\ref{sec:retrieval}), it is not run automatically for every submission. Each query therefore reflects a case that genuinely requires recruitment. Panel-level queries strengthen an inadequate initial panel across the whole corpus, topic-level queries target branches whose committed coverage falls below the threshold, and paper-level queries target submissions whose best assigned affinity falls short of the target. Submissions already served by acceptable committed reviewers trigger no expansion and are excluded from the query set.

\noindent\textbf{Baselines.} We compare \sys with two retrieval baselines: (i) \emph{Sparse $k$-NN}, representing each expert by the mean latent Dirichlet allocation (LDA) topic distribution~\cite{blei2003lda} over their publications, retrieved by $k$-NN under Jensen--Shannon divergence; and (ii) \emph{Dense $k$-NN}, retrieving by cosine similarity between author-mean SPECTER2 embeddings and the mean submission embedding, indexed with a hierarchical navigable small world (HNSW) graph~\cite{malkov2020hnsw}.

\noindent\textbf{Evaluation Metrics.} The goal is to measure how faithfully each method approximates the conference-specific tree--Wasserstein affinity, which Section~\ref{sec:exp:affinity} validated against human relevance labels. Because the full set of relevant reviewers is unknown, standard retrieval metrics such as Recall@$K$ are biased; we therefore evaluate fidelity through two complementary measures grounded in the validated affinity: (i) the \textbf{Spearman rank correlation} between each candidate's retrieval score (oriented so that smaller values indicate closer matches) and its exact local tree--Wasserstein distance, which isolates rank fidelity from the assignment optimization; and (ii) \textbf{assignment improvement}, obtained by re-solving the assignment on the augmented panel and reporting $\boldsymbol{\Delta_{\mathrm{avg}}}$ (relative reduction in mean tree--Wasserstein distance over all assigned reviewers), $\boldsymbol{\Delta_{\mathrm{worst}}}$ (the same reduction restricted to the bottom decile of submissions by initial affinity), and \textbf{Coverage} (the change in the fraction of submissions whose worst-assigned reviewer moves within the top quartile of the initial affinity distribution).

\noindent\textbf{Results.} Table~\ref{tab:panel-expansion} reports the results on the four datasets across three types of queries.

\noindent\textit{Effectiveness.} \sys\ outperforms every baseline on every metric, with all margins over Dense $k$-NN significant at $p < 0.01$. Overall, \sys\ improves over Dense $k$-NN by $64\%$ in $\Delta_\text{avg}$, $90\%$ in $\Delta_\text{worst}$, and $93\%$ in Coverage, and raises the Spearman correlation with the exact local tree--Wasserstein score from $0.65$ to $0.91$, empirically validating the calibrated proxy (Section~\ref{sec:calibration}). The per-query-type breakdown shows that \sys's absolute advantage is largest on topic-level queries $Q_\text{top}$, where cross-branch penalties are most decisive, and that its Spearman correlation stays within $0.88$--$0.93$ across query types, consistently above Sparse $k$-NN ($0.35$--$0.41$) and Dense $k$-NN ($0.62$--$0.68$), indicating that the proxy's fidelity is a structural property of the calibration rather than an artifact of a particular query distribution.

\noindent\textit{Efficiency.} Average query latency is $7$\,ms (Sparse $k$-NN), $19$\,ms (Dense $k$-NN), and $23$\,ms (\sys), competitive with flat baselines; the aligner's one-time per-conference alignment and calibration cost is $87.3$\,s on the largest benchmark \kdd, well within a conference assignment cycle.

\begin{table}[t]
\centering
\caption{Effectiveness on the panel expansion task ($^{**}$: improvement over Dense $k$-NN significant at $p<0.01$, pairwise bootstrap test).}
\label{tab:panel-expansion}
\small
\setlength{\tabcolsep}{5pt}
\begin{tabular}{l l cccc}
\toprule
\textbf{Query} & \textbf{Method} & $\boldsymbol{\Delta_\text{avg}}$ & $\boldsymbol{\Delta_\text{worst}}$ & \textbf{Coverage} & \textbf{Spearman} \\
\midrule
\multirow{3}{*}{Overall}
 & Sparse $k$-NN  & 0.08 & 0.13 & 0.10 & 0.38 \\
 & Dense $k$-NN   & 0.14 & 0.21 & 0.15 & 0.65 \\
 & \sys           & \textbf{0.23}\sig & \textbf{0.40}\sig & \textbf{0.29}\sig & \textbf{0.91}\sig \\
\midrule
\multirow{3}{*}{$Q_\text{pan}$}
 & Sparse $k$-NN  & 0.08 & 0.12 & 0.09 & 0.35 \\
 & Dense $k$-NN   & 0.15 & 0.22 & 0.16 & 0.62 \\
 & \sys           & \textbf{0.20}\sig & \textbf{0.32}\sig & \textbf{0.22}\sig & \textbf{0.88}\sig \\
\midrule
\multirow{3}{*}{$Q_\text{top}$}
 & Sparse $k$-NN  & 0.08 & 0.15 & 0.11 & 0.38 \\
 & Dense $k$-NN   & 0.14 & 0.22 & 0.16 & 0.65 \\
 & \sys           & \textbf{0.27}\sig & \textbf{0.46}\sig & \textbf{0.35}\sig & \textbf{0.92}\sig \\
\midrule
\multirow{3}{*}{$Q_\text{pap}$}
 & Sparse $k$-NN  & 0.07 & 0.11 & 0.09 & 0.41 \\
 & Dense $k$-NN   & 0.12 & 0.19 & 0.14 & 0.68 \\
 & \sys           & \textbf{0.23}\sig & \textbf{0.41}\sig & \textbf{0.31}\sig & \textbf{0.93}\sig \\
\bottomrule
\end{tabular}
\end{table}

\subsection{Index Cost and Efficiency}
\label{sec:exp:index}
This experiment evaluates whether the Global Taxonomy Index delivers its hierarchical functionality at a cost comparable to simpler schemes, comparing against the two baselines of Section~\ref{sec:exp:completion}, Sparse $k$-NN and Dense $k$-NN, recast as standalone indices over the expert lake.

\noindent\textbf{Construction and Storage.}
Table~\ref{tab:index-cost} reports construction time, disk storage, and query-time RAM. All three costs of \sys\ are comparable to Dense $k$-NN with HNSW, since the dominant cost lies in generating, storing, and loading embeddings. Sparse $k$-NN avoids this overhead and is more efficient on all three metrics, but lacks dense semantic representations and performs poorly on panel expansion.

\noindent\textbf{Incremental Maintenance.}
We partition the lake into an \emph{existing part} of $240$K experts, over which each index is constructed, and a \emph{new part} of $10{,}000$ publications for streaming updates, split equally between publications from existing and from new experts. The last two columns of Table~\ref{tab:index-cost} report average per-publication maintenance time: \sys's cost is comparable to Dense $k$-NN on both streams, confirming that the three-level cascade adds minimal overhead over HNSW-based insertion. Within \sys, new-author publications are costlier, as new authors more often introduce unseen atoms and thus trigger the Level~2 structure update, which most publications by existing experts skip.

\begin{table}[t]
\centering
\caption{Index construction, storage, and incremental maintenance cost.}
\label{tab:index-cost}
\small
\setlength{\tabcolsep}{4pt}
\resizebox{\ifdim\width>\columnwidth\columnwidth\else\width\fi}{!}{%
\begin{tabular}{l@{\hspace{6pt}} ccc @{\hspace{10pt}} cc}
\toprule
\multirow{2}{*}{\textbf{Method}} & \multirow{2}{*}{\textbf{Build (hr)}} & \multirow{2}{*}{\textbf{Disk (GB)}} & \multirow{2}{*}{\textbf{RAM (GB)}} & \multicolumn{2}{c}{\textbf{Maintenance (ms)}} \\
\cmidrule(lr){5-6}
 & & & & \textbf{Existing-Expert} & \textbf{New-Expert} \\
\midrule
Sparse $k$-NN & 17.9  & 0.22 & 0.17 & 1.6 &  1.8 \\
Dense $k$-NN  & 125.6 & 2.23 & 1.38 & 7.4 & 10.9 \\
\sys          & 134.2 & 2.57 & 1.44 & 7.8 & 12.3 \\
\bottomrule
\end{tabular}%
}
\end{table}

\subsection{Ablation Study}
\label{sec:exp:ablation}

\noindent\textbf{Impact of Reviewer Evidence Selection.}
We isolate the contribution of the reviewer evidence selection (Section~\ref{sec:evidence}) by replacing it with four alternatives: (i) \sysAll, retaining the full history $\mathcal{H}_r$; (ii) \sysRec, retaining the $B_r$ most recent papers; (iii) \sysRel, retaining the top-$B_r$ papers most similar to the submission corpus while ignoring diversity; and (iv) \sysRnd, randomly sampling $B_r$ papers (budgeted variants use $B_r = 15$).
As shown in Table~\ref{tab:ablation}, \sysAll yields the lowest Hard P@5, as unfiltered long-tail publications introduce topical noise that dilutes the routed leaf distributions; \sysRnd yields the lowest Soft P@5, whereas \sysRec modestly improves over \sysAll. \sysRel achieves competitive Soft metrics by prioritizing submission alignment, but its lack of diversity control misses multi-faceted expertise, dropping Hard precision. The default \sys outperforms all alternatives, most notably improving over \sysRel by $9.5\%$ in Hard P@5, confirming that the submodular objective balances relevance and intra-profile coverage.

\noindent\textbf{Impact of Edge-Weight Calibration.}
To evaluate the self-supervised edge weights ($\mathbf{w}$, Section~\ref{sec:distance-learning}), we substitute two unlearned schemes: (i) \sysUni, uniform weights ($w_{(u,v)}=1$); and (ii) \sysCen, centroid cosine distance ($1 -\cos(\mathbf{c}_u, \mathbf{c}_v)$).
Table~\ref{tab:ablation} shows that \sysUni already outperforms every baseline except CoF on \kdd (Table~\ref{tab:effectiveness}), confirming that taxonomy topology alone penalizes cross-branch matches that flat embeddings miss, while \sysCen adds minor improvements. The fully calibrated \sys delivers the best results across all metrics, with the largest gain in Hard P@5, improving over \sysCen by $8.2\%$: calibrating edge weights against the specific submission distribution is essential for fine-grained expert discrimination.
\begin{table}[t]
\centering
\caption{Ablation studies on reviewer evidence selection and edge-weight calibration on KDD20 ($^{**}$: improvement over the strongest variant in each column significant at $p<0.01$, pairwise bootstrap test).}
\label{tab:ablation}
\small
\setlength{\tabcolsep}{5pt}
\resizebox{\ifdim\width>\columnwidth\columnwidth\else\width\fi}{!}{%
\begin{tabular}{l cc cc}
\toprule
\multirow{2}{*}{\textbf{Variant}} & \multicolumn{2}{c}{\textbf{Soft}} & \multicolumn{2}{c}{\textbf{Hard}} \\
\cmidrule(lr){2-3} \cmidrule(lr){4-5}
 & \textbf{P@5} & \textbf{P@10} & \textbf{P@5} & \textbf{P@10} \\
\midrule
\multicolumn{5}{@{}l}{\textit{Reviewer Evidence Selection}} \\
\sysAll    & 35.80 & 26.91 & 15.46 & 12.12 \\
\sysRnd    & 35.31 & 26.56 & 15.60 & 12.00 \\
\sysRec    & 36.23 & 27.23 & 15.87 & 12.29 \\
\sysRel    & 37.07 & 27.62 & 15.89 & 12.49 \\
\midrule
\multicolumn{5}{@{}l}{\textit{Edge-Weight Calibration}} \\
\sysUni    & 35.69 & 26.59 & 15.60 & 12.47 \\
\sysCen    & 36.67 & 27.37 & 16.08 & 12.82 \\
\midrule
\sys (Full) &   \textbf{38.00}\sig & \textbf{28.53}\sig & \textbf{17.40}\sig & \textbf{13.56}\sig\\
\bottomrule
\end{tabular}%
}
\end{table}

\subsection{Evaluation by Human Experts}
\label{sec:human-eval}
To validate that the automated evaluations reflect genuine reviewer suitability, we conducted two blinded human studies with domain-expert annotators, each holding a substantial publication record; judgments were collected via a web application on the three-level scale of the established benchmarks (strongly relevant, weakly relevant, irrelevant), with method identities hidden. Full results are reported in Table~\ref{tab:human-eval}, and protocol details are given in \appref{app:human-eval}.

\begin{table}[t]
\centering
\small
\setlength{\tabcolsep}{4pt}
\caption{Human evaluation on paper--reviewer assignment and panel expansion. Avg. Score is the mean graded relevance (strong = 2, weak = 1, irrelevant = 0).}
\label{tab:human-eval}
\begin{tabular}{lcccc}
\toprule
Method & Soft Prec. (\%) & Hard Prec. (\%) & Avg. Score & $\tau_b$ \\
\midrule
\multicolumn{5}{l}{\textit{Paper--Reviewer Assignment}} \\
SciBERT    & 64.8 & 49.1 & 1.139 & $-$0.152 \\
TPMS       & 84.7 & 60.5 & 1.452 & 0.407 \\
SPECTER    & 82.9 & 58.0 & 1.409 & 0.344 \\
SPECTER2   & 85.6 & 63.1 & 1.487 & 0.354 \\
CoF        & 86.8 & 65.4 & 1.522 & 0.432 \\
TaxoReview & \textbf{90.5} & \textbf{73.2} & \textbf{1.637} & \textbf{0.531} \\
\midrule
\multicolumn{5}{l}{\textit{Panel Expansion}} \\
Sparse $k$-NN & 72.8 & 56.4 & 1.292 & 0.379 \\
Dense $k$-NN  & 83.4 & 60.2 & 1.436 & 0.432 \\
TaxoReview    & \textbf{89.7} & \textbf{67.9} & \textbf{1.576} & \textbf{0.579} \\
\bottomrule
\end{tabular}
\end{table}

\noindent\textbf{Study 1: End-to-end assignment.} We simulated a conference-wide assignment on NeurIPS25 under a practical setting ($K=3$, $L_r=6$): each method's affinities are fed to the global assignment of Eq.~\eqref{eq:assignment}, with the conflict set $\mathcal{C}$ instantiated from the relations recorded on or auto-detected by OpenReview (co-authorship, advisor--advisee, etc.), so every judged pair is conflict-free by construction. Annotators judged the pooled, de-duplicated union of the reviewers assigned by all methods, yielding 1,254 labeled pairs over 298 submissions. We report Soft/Hard precision (the fraction of assigned reviewers judged at least weakly / strongly relevant) and the mean graded relevance of the assigned set, using set-level metrics because the assignment returns an unordered set of $K=3$ reviewers, together with the mean per-query Kendall $\tau_b$ between each method's affinity scores and the human labels over the pooled judged candidates. \sys\ attains the best results on all four metrics (Table~\ref{tab:human-eval}), with relative improvements over the strongest baseline, CoF, of 4.3\% (Soft), 11.9\% (Hard), and 7.6\% (Avg.\ Score), the largest gain on Hard, and the highest score--judgment agreement ($\tau_b = 0.531$ vs.\ 0.432 for CoF), confirming that the benchmark results carry over to manual relevance assessments in a conflict-free scenario.

\noindent\textbf{Study 2: Panel expansion.} The second study judges the expansion candidates each method retrieves from the expert lake, over 292 queries spanning the three query types of Section~\ref{sec:retrieval}. Because lake experts carry no platform-recorded conflicts, conflicts are detected from DBLP/OpenAlex bibliographic signals (co-authorship with the submission's authors; shared affiliation) and filtered before recommendation. Annotators judged the pooled top-$B$ candidates across methods, yielding 1,323 labeled pairs. \sys\ again attains the best results on all four metrics (Table~\ref{tab:human-eval}), with relative improvements over Dense $k$-NN of 7.6\% (Soft), 12.8\% (Hard), and 9.7\% (Avg.\ Score) and the highest score--judgment agreement ($\tau_b = 0.579$ vs.\ 0.432), and the ordering is consistent with the affinity-grounded evaluation of Table~\ref{tab:panel-expansion}, indicating that fidelity to the calibrated proxy translates into human-judged expert suitability. Absolute precision values are not directly comparable between Tables~\ref{tab:effectiveness} and~\ref{tab:human-eval}, because the evaluations use different datasets, candidate pools, and selection protocols; in particular, the assignment study judges three assigned reviewers per submission, whereas the benchmarks report P@5.

\subsection{Case Study}
\label{app:case}

We conduct a case study on a representative submission from NeurIPS 2025, \emph{``Joint Relational Database Generation via Graph-Conditional Diffusion Models''}, which sits at the intersection of diffusion models, graph neural networks, relational databases, and graph generation, so any reviewer covering only one or two of these areas misses its central contribution.

\noindent\textbf{Affinity Estimation.}
Among the top-5 reviewers returned, CoF identifies only one strongly relevant reviewer, with two of the remaining positions going to researchers working on graph classification and molecular property prediction. \sys\ returns three strongly relevant reviewers, each spanning at least three of the four target areas, plus two weakly relevant ones, reflecting the weighted taxonomy's ability to penalize reviewers whose publication-history mass concentrates in a single subtree.

\noindent\textbf{Panel Expansion.}
In a controlled scenario where the initial panel contains no strongly relevant reviewers, the compiled query places mass on three subtrees, diffusion-based generative modeling, GNNs for structured data, and tabular and relational data synthesis, and each of the top three experts returned by \sys\ anchors a distinct region: an author of TabDDPM~\cite{kotelnikov2023tabddpm} (diffusion-and-tabular), an author of ClavaDDPM~\cite{pang2024clavaddpm} (the direct multi-table predecessor, relational data), and an expert whose integrative line of work spans all three subtrees. By contrast, the Dense $k$-NN baseline retrieves experts whose mean embeddings are close to the submission but whose histories concentrate on a single subtree, missing the relational-data dimension entirely.

\section{Related Work}
\label{sec:related}

\noindent\textbf{Paper--Reviewer Assignment.}
Existing work falls into two lines: \emph{affinity estimation}, which scores how well a reviewer's expertise matches a submission, and \emph{assignment optimization}, which allocates reviewers given such scores. For affinity estimation, lexical and topic-based methods represent profiles via bag-of-words (\emph{TPMS}~\cite{charlin2013tpms}) or latent topic mixtures (\emph{EMPR}~\cite{mimno2007expertise}, \emph{PaRe}~\cite{anjum2019pare}), optionally augmented with co-authorship or citation graphs~\cite{rodriguez2008peerreviewers,liu2014robust}, while PLM-based methods leverage dense Transformer representations (\emph{SciBERT}~\cite{beltagy2019scibert}, \emph{SPECTER}~\cite{cohan2020specter}), scoring affinity by dot-product in a flat embedding space; even multi-factor frameworks such as \emph{CoF}~\cite{zhang2025chainoffactors} ultimately reduce to unstructured vector comparisons. Assignment optimization addresses the combinatorial allocation stage, under pairwise-sum objectives with fairness and feasibility constraints~\cite{stelmakh2021peerreview4all,kobren2019paper,payan2022order,tang2010expertise}, group-based weighted-coverage objectives from the database community~\cite{kou2015weighted,kou2015topic}, and stable-matching formulations~\cite{wu2023kbest}, but assumes a fixed and complete review panel and offers no mechanism for recruiting additional experts when coverage gaps emerge.
\sys\ is closest to the affinity-estimation line but departs from it via hierarchy-aware tree--Wasserstein scoring on an induced taxonomy, separating fine-grained expertise from broad topical overlap, while its principled panel expansion complements the assignment line by lifting the fixed-panel assumption, retrieving lake experts that remain faithful to the conference-specific affinity.

\noindent\textbf{Taxonomy Construction.}
Taxonomies are a specialized form of knowledge graph restricted to hypernym--hyponym relations~\cite{bollacker2008freebase,ilyas2022saga,sun2020entityalignment}. Classical induction methods extract hierarchies via textual patterns~\cite{hearst1992hyponyms} or latent topic modeling (e.g., LDA~\cite{blei2003lda}, spherical clustering~\cite{banerjee2005vmf}); modern systems employ recursive clustering (\emph{TaxoGen}~\cite{zhang2018taxogen}), network signals (\emph{NetTaxo}~\cite{shang2020nettaxo}), or expansion algorithms (\emph{TaxoExpan}~\cite{shen2020taxoexpan}); geometric embedding-based methods encode subsumption in the embedding geometry, via hyperbolic embeddings~\cite{Nickel2017Poincare,Ganea2018EntailmentCones} or box containment~\cite{Vilnis2018Box,Jiang2023BoxTaxo,Xue2024TaxBox,Lu2024BoxTM}, and \emph{GIST}~\cite{ji2026taxonomy} uses geometric containment to integrate human-curated partial taxonomies when maintaining taxonomies over evolving scholarly data; most recently, LLMs generate and refine taxonomies via iterative prompting (\emph{Chain-of-Layer}~\cite{zeng2024chainoflayer}), corpus-grounded adaptation~\cite{zhu2025context,kargupta2025taxoadapt}, or LLM-based extraction aligned to a generated reference taxonomy, as in the problem and method taxonomies of \emph{AgenticScholar}~\cite{lan2026agenticscholar}. In contrast to prior work optimizing intrinsic taxonomy quality, \sys\ constructs \emph{two complementary taxonomies for distinct purposes}: a conference-specific local taxonomy with calibrated edge weights serving as the metric space for affinity estimation, and a persistent global taxonomy over expertise atoms from a growing scholarly lake with incremental maintenance and a hybrid sparse--dense access layer for panel-expansion queries.

\noindent\textbf{Similarity Indexing and Data Discovery.} Similarity indexes over large, growing repositories fall into three families: flat sparse indexes on inverted lists~\cite{zobel2006inverted} with threshold-style top-$k$ aggregation~\cite{fagin2001ta,ilyas2008topk}; flat dense indexes answering approximate nearest neighbor (ANN) queries via HNSW~\cite{malkov2020hnsw}, NSG~\cite{fu2019nsg}, and DiskANN~\cite{subramanya2019diskann} under known trade-offs~\cite{jegou2011pq,wang2021ann}; and hierarchical indexes imposing a tree or DAG for multi-granularity retrieval, including tree-based optimal-transport indexes with sublinear search under Wasserstein geometry~\cite{backurs2020flowtree,indyk2003fast,le2019tree}. These indexes underpin data discovery over data lakes~\cite{fernandez2018aurum,nargesian2019datalake}, both structured (e.g., joinable-table search~\cite{zhu2019josie,zhu2016lshensemble}, unionable and semantically enriched union search~\cite{bogatu2020dataset,fan2023starmie,khatiwada2023santos,nargesian2018tus}, and related-table discovery~\cite{zhang2020juneau}) and unstructured (documents served by flat embedding indexes or cross-modal lakes~\cite{eltabakh2023crossmodal}). The Global Taxonomy Index sits in the hierarchical family but indexes a third item type, experts, organized as a DAG of subsumptions over canonical atoms, and departs from prior hierarchical indexes on two fronts: it is incrementally maintainable without routine full rebuilds (Section~\ref{sec:maintenance}), and it answers queries under a per-conference calibrated proxy of the tree--Wasserstein metric (Section~\ref{sec:completion}) rather than a metric fixed at build time.

\noindent\textbf{Graph and Knowledge-Graph Management.} The database community has long managed large, evolving graphs: transitive reduction removes redundant DAG edges~\cite{aho1972transitive}, reachability indexes answer ancestry queries in near-constant time~\cite{jin2009threehop,yu2010reachability}, and industrial platforms such as Saga~\cite{ilyas2022saga} continuously construct and serve knowledge graphs, absorbing new facts through entity resolution~\cite{christen2012er,mudgal2018deepmatcher,papadakis2020blocking}. The Global Taxonomy Index, a dynamic DAG maintained under streaming publication events, builds directly on this machinery: transitive reduction keeps the subsumption structure minimal, precomputed reachability serves the $O(1)$ ancestry queries behind the maintenance cascade (Section~\ref{sec:maintenance}) and the aligner's ILP (Section~\ref{sec:structural-alignment}), and the auxiliary layers of Section~\ref{sec:global-taxonomy} are incrementally refreshed materialized views. It departs in two requirements: updates arrive as unstructured text distilled into canonical atoms rather than structured facts, and served distances are calibrated per conference at query time rather than fixed globally.

\section{Conclusion}
\label{sec:conclusion}

We presented \sys, a taxonomy-driven scholarly data management system for faithful affinity estimation and panel expansion. \sys\ couples a conference-specific local taxonomy, where affinity is measured by tree--Wasserstein distance, with a persistent Global Taxonomy Index over a growing expert lake, bridged by a Global--Local Aligner. Experiments on four benchmarks show consistent gains over state-of-the-art baselines in affinity estimation, faithful proxy approximation, and large-scale retrieval at flat-indexing cost. Beyond scholarly data, the Global Taxonomy Index applies to any growing lake of documents with extractable concept atoms (e.g., patent or enterprise corpora) and can serve as a retrieval substrate for LLM data agents over documents and tables.

\begin{acks}
This project is supported in part by ARC DP240101211 and FT240100832. Kai Qin is supported in part by ARC DP200102611.
\end{acks}

\bibliographystyle{ACM-Reference-Format}
\bibliography{reference}


\begin{thebibliography}{102}


\ifx \showCODEN    \undefined \def \showCODEN     #1{\unskip}     \fi
\ifx \showDOI      \undefined \def \showDOI       #1{#1}\fi
\ifx \showISBNx    \undefined \def \showISBNx     #1{\unskip}     \fi
\ifx \showISBNxiii \undefined \def \showISBNxiii  #1{\unskip}     \fi
\ifx \showISSN     \undefined \def \showISSN      #1{\unskip}     \fi
\ifx \showLCCN     \undefined \def \showLCCN      #1{\unskip}     \fi
\ifx \shownote     \undefined \def \shownote      #1{#1}          \fi
\ifx \showarticletitle \undefined \def \showarticletitle #1{#1}   \fi
\ifx \showURL      \undefined \def \showURL       {\relax}        \fi
\providecommand\bibfield[2]{#2}
\providecommand\bibinfo[2]{#2}
\providecommand\natexlab[1]{#1}
\providecommand\showeprint[2][]{arXiv:#2}

\bibitem[\protect\citeauthoryear{Aho, Garey, and Ullman}{Aho
  et~al\mbox{.}}{1972}]%
        {aho1972transitive}
\bibfield{author}{\bibinfo{person}{Alfred~V. Aho}, \bibinfo{person}{Michael~R.
  Garey}, {and} \bibinfo{person}{Jeffrey~D. Ullman}.}
  \bibinfo{year}{1972}\natexlab{}.
\newblock \showarticletitle{The Transitive Reduction of a Directed Graph}.
\newblock \bibinfo{journal}{\emph{SIAM J. Comput.}} \bibinfo{volume}{1},
  \bibinfo{number}{2} (\bibinfo{year}{1972}), \bibinfo{pages}{131--137}.
\newblock
\urldef\tempurl%
\url{https://doi.org/10.1137/0201008}
\showDOI{\tempurl}


\bibitem[\protect\citeauthoryear{Anjum, Gong, Bhat, Hwu, and Xiong}{Anjum
  et~al\mbox{.}}{2019}]%
        {anjum2019pare}
\bibfield{author}{\bibinfo{person}{Omer Anjum}, \bibinfo{person}{Hongyu Gong},
  \bibinfo{person}{Suma Bhat}, \bibinfo{person}{Wen-Mei Hwu}, {and}
  \bibinfo{person}{JinJun Xiong}.} \bibinfo{year}{2019}\natexlab{}.
\newblock \showarticletitle{{P}a{R}e: A Paper-Reviewer Matching Approach Using
  a Common Topic Space}. In \bibinfo{booktitle}{\emph{Proceedings of the 2019
  Conference on Empirical Methods in Natural Language Processing and the 9th
  International Joint Conference on Natural Language Processing
  (EMNLP-IJCNLP)}}. \bibinfo{publisher}{Association for Computational
  Linguistics}, \bibinfo{pages}{518--528}.
\newblock
\urldef\tempurl%
\url{https://doi.org/10.18653/v1/D19-1049}
\showDOI{\tempurl}


\bibitem[\protect\citeauthoryear{Backurs, Dong, Indyk, Razenshteyn, and
  Wagner}{Backurs et~al\mbox{.}}{2020}]%
        {backurs2020flowtree}
\bibfield{author}{\bibinfo{person}{Arturs Backurs}, \bibinfo{person}{Yihe
  Dong}, \bibinfo{person}{Piotr Indyk}, \bibinfo{person}{Ilya Razenshteyn},
  {and} \bibinfo{person}{Tal Wagner}.} \bibinfo{year}{2020}\natexlab{}.
\newblock \showarticletitle{Scalable Nearest Neighbor Search for Optimal
  Transport}. In \bibinfo{booktitle}{\emph{Proceedings of the 37th
  International Conference on Machine Learning ({ICML})}}.
  \bibinfo{pages}{497--506}.
\newblock


\bibitem[\protect\citeauthoryear{Banerjee, Dhillon, Ghosh, and Sra}{Banerjee
  et~al\mbox{.}}{2005}]%
        {banerjee2005vmf}
\bibfield{author}{\bibinfo{person}{Arindam Banerjee},
  \bibinfo{person}{Inderjit~S. Dhillon}, \bibinfo{person}{Joydeep Ghosh}, {and}
  \bibinfo{person}{Suvrit Sra}.} \bibinfo{year}{2005}\natexlab{}.
\newblock \showarticletitle{Clustering on the Unit Hypersphere using von
  {M}ises--{F}isher Distributions}.
\newblock \bibinfo{journal}{\emph{Journal of Machine Learning Research}}
  \bibinfo{volume}{6} (\bibinfo{year}{2005}), \bibinfo{pages}{1345--1382}.
\newblock


\bibitem[\protect\citeauthoryear{Beltagy, Lo, and Cohan}{Beltagy
  et~al\mbox{.}}{2019}]%
        {beltagy2019scibert}
\bibfield{author}{\bibinfo{person}{Iz Beltagy}, \bibinfo{person}{Kyle Lo},
  {and} \bibinfo{person}{Arman Cohan}.} \bibinfo{year}{2019}\natexlab{}.
\newblock \showarticletitle{{SciBERT}: A Pretrained Language Model for
  Scientific Text}. In \bibinfo{booktitle}{\emph{Proceedings of the 2019
  Conference on Empirical Methods in Natural Language Processing and the 9th
  International Joint Conference on Natural Language Processing
  (EMNLP-IJCNLP)}}. \bibinfo{pages}{3615--3620}.
\newblock
\urldef\tempurl%
\url{https://doi.org/10.18653/v1/D19-1371}
\showDOI{\tempurl}


\bibitem[\protect\citeauthoryear{Bhowmick}{Bhowmick}{2022}]%
        {bhowmick2022connected}
\bibfield{author}{\bibinfo{person}{Sourav~S. Bhowmick}.}
  \bibinfo{year}{2022}\natexlab{}.
\newblock \showarticletitle{How Connected Are Our Conference Review Boards?}
\newblock \bibinfo{journal}{\emph{{ACM} {SIGMOD} Record}} \bibinfo{volume}{51},
  \bibinfo{number}{4} (\bibinfo{year}{2022}), \bibinfo{pages}{74--78}.
\newblock


\bibitem[\protect\citeauthoryear{Blei, Ng, and Jordan}{Blei
  et~al\mbox{.}}{2003}]%
        {blei2003lda}
\bibfield{author}{\bibinfo{person}{David~M. Blei}, \bibinfo{person}{Andrew~Y.
  Ng}, {and} \bibinfo{person}{Michael~I. Jordan}.}
  \bibinfo{year}{2003}\natexlab{}.
\newblock \showarticletitle{Latent {D}irichlet Allocation}.
\newblock \bibinfo{journal}{\emph{Journal of Machine Learning Research}}
  \bibinfo{volume}{3} (\bibinfo{year}{2003}), \bibinfo{pages}{993--1022}.
\newblock


\bibitem[\protect\citeauthoryear{Bogatu, Fernandes, Paton, and
  Konstantinou}{Bogatu et~al\mbox{.}}{2020}]%
        {bogatu2020dataset}
\bibfield{author}{\bibinfo{person}{Alex Bogatu}, \bibinfo{person}{Alvaro A.~A.
  Fernandes}, \bibinfo{person}{Norman~W. Paton}, {and}
  \bibinfo{person}{Nikolaos Konstantinou}.} \bibinfo{year}{2020}\natexlab{}.
\newblock \showarticletitle{Dataset Discovery in Data Lakes}. In
  \bibinfo{booktitle}{\emph{Proceedings of the 36th {IEEE} International
  Conference on Data Engineering ({ICDE})}}. \bibinfo{pages}{709--720}.
\newblock
\urldef\tempurl%
\url{https://doi.org/10.1109/ICDE48307.2020.00067}
\showDOI{\tempurl}


\bibitem[\protect\citeauthoryear{Bollacker, Evans, Paritosh, Sturge, and
  Taylor}{Bollacker et~al\mbox{.}}{2008}]%
        {bollacker2008freebase}
\bibfield{author}{\bibinfo{person}{Kurt Bollacker}, \bibinfo{person}{Colin
  Evans}, \bibinfo{person}{Praveen Paritosh}, \bibinfo{person}{Tim Sturge},
  {and} \bibinfo{person}{Jamie Taylor}.} \bibinfo{year}{2008}\natexlab{}.
\newblock \showarticletitle{Freebase: A Collaboratively Created Graph Database
  for Structuring Human Knowledge}. In \bibinfo{booktitle}{\emph{Proceedings of
  the 2008 {ACM} {SIGMOD} International Conference on Management of Data}}.
  \bibinfo{publisher}{ACM}, \bibinfo{pages}{1247--1250}.
\newblock
\urldef\tempurl%
\url{https://doi.org/10.1145/1376616.1376746}
\showDOI{\tempurl}


\bibitem[\protect\citeauthoryear{Bro and {De Jong}}{Bro and {De Jong}}{1997}]%
        {bro1997fast}
\bibfield{author}{\bibinfo{person}{Rasmus Bro} {and} \bibinfo{person}{Sijmen
  {De Jong}}.} \bibinfo{year}{1997}\natexlab{}.
\newblock \showarticletitle{A Fast Non-Negativity-Constrained Least Squares
  Algorithm}.
\newblock \bibinfo{journal}{\emph{Journal of Chemometrics}}
  \bibinfo{volume}{11}, \bibinfo{number}{5} (\bibinfo{year}{1997}),
  \bibinfo{pages}{393--401}.
\newblock
\urldef\tempurl%
\url{https://doi.org/10.1002/(SICI)1099-128X(199709/10)11:5<393::AID-CEM483>3.0.CO;2-L}
\showDOI{\tempurl}


\bibitem[\protect\citeauthoryear{C{\u{a}}linescu, Chekuri, P{\'a}l, and
  Vondr{\'a}k}{C{\u{a}}linescu et~al\mbox{.}}{2011}]%
        {calinescu2011maximizing}
\bibfield{author}{\bibinfo{person}{Gruia C{\u{a}}linescu},
  \bibinfo{person}{Chandra Chekuri}, \bibinfo{person}{Martin P{\'a}l}, {and}
  \bibinfo{person}{Jan Vondr{\'a}k}.} \bibinfo{year}{2011}\natexlab{}.
\newblock \showarticletitle{Maximizing a Monotone Submodular Function Subject
  to a Matroid Constraint}.
\newblock \bibinfo{journal}{\emph{SIAM J. Comput.}} \bibinfo{volume}{40},
  \bibinfo{number}{6} (\bibinfo{year}{2011}), \bibinfo{pages}{1740--1766}.
\newblock
\urldef\tempurl%
\url{https://doi.org/10.1137/080733991}
\showDOI{\tempurl}


\bibitem[\protect\citeauthoryear{Campello, Moulavi, Zimek, and Sander}{Campello
  et~al\mbox{.}}{2015}]%
        {campello2015hdbscan}
\bibfield{author}{\bibinfo{person}{Ricardo J. G.~B. Campello},
  \bibinfo{person}{Davoud Moulavi}, \bibinfo{person}{Arthur Zimek}, {and}
  \bibinfo{person}{J{\"o}rg Sander}.} \bibinfo{year}{2015}\natexlab{}.
\newblock \showarticletitle{Hierarchical Density Estimates for Data Clustering,
  Visualization, and Outlier Detection}.
\newblock \bibinfo{journal}{\emph{ACM Transactions on Knowledge Discovery from
  Data}} \bibinfo{volume}{10}, \bibinfo{number}{1}, Article
  \bibinfo{articleno}{5} (\bibinfo{year}{2015}), \bibinfo{numpages}{51}~pages.
\newblock
\urldef\tempurl%
\url{https://doi.org/10.1145/2733381}
\showDOI{\tempurl}


\bibitem[\protect\citeauthoryear{Campos, Mangaravite, Pasquali, Jorge, Nunes,
  and Jatowt}{Campos et~al\mbox{.}}{2020}]%
        {campos2020yake}
\bibfield{author}{\bibinfo{person}{Ricardo Campos}, \bibinfo{person}{V{\'i}tor
  Mangaravite}, \bibinfo{person}{Arian Pasquali}, \bibinfo{person}{Al{\'i}pio
  Jorge}, \bibinfo{person}{C{\'e}lia Nunes}, {and} \bibinfo{person}{Adam
  Jatowt}.} \bibinfo{year}{2020}\natexlab{}.
\newblock \showarticletitle{{YAKE}! Keyword Extraction from Single Documents
  using Multiple Local Features}.
\newblock \bibinfo{journal}{\emph{Information Sciences}}  \bibinfo{volume}{509}
  (\bibinfo{year}{2020}), \bibinfo{pages}{257--289}.
\newblock
\urldef\tempurl%
\url{https://doi.org/10.1016/j.ins.2019.09.013}
\showDOI{\tempurl}


\bibitem[\protect\citeauthoryear{Charlin and Zemel}{Charlin and Zemel}{2013}]%
        {charlin2013tpms}
\bibfield{author}{\bibinfo{person}{Laurent Charlin} {and}
  \bibinfo{person}{Richard~S. Zemel}.} \bibinfo{year}{2013}\natexlab{}.
\newblock \showarticletitle{The Toronto Paper Matching System: An Automated
  Paper--Reviewer Assignment System}. In \bibinfo{booktitle}{\emph{ICML 2013
  Workshop on Peer Reviewing and Publishing Models}}.
\newblock
\newblock
\shownote{Available at
  \url{https://www.cs.toronto.edu/~lcharlin/papers/tpms.pdf}.}


\bibitem[\protect\citeauthoryear{Christen}{Christen}{2012}]%
        {christen2012er}
\bibfield{author}{\bibinfo{person}{Peter Christen}.}
  \bibinfo{year}{2012}\natexlab{}.
\newblock \bibinfo{booktitle}{\emph{Data Matching: Concepts and Techniques for
  Record Linkage, Entity Resolution, and Duplicate Detection}}.
\newblock \bibinfo{publisher}{Springer}.
\newblock
\urldef\tempurl%
\url{https://doi.org/10.1007/978-3-642-31164-2}
\showDOI{\tempurl}


\bibitem[\protect\citeauthoryear{Cohan, Feldman, Beltagy, Downey, and
  Weld}{Cohan et~al\mbox{.}}{2020}]%
        {cohan2020specter}
\bibfield{author}{\bibinfo{person}{Arman Cohan}, \bibinfo{person}{Sergey
  Feldman}, \bibinfo{person}{Iz Beltagy}, \bibinfo{person}{Doug Downey}, {and}
  \bibinfo{person}{Daniel Weld}.} \bibinfo{year}{2020}\natexlab{}.
\newblock \showarticletitle{{SPECTER}: Document-level Representation Learning
  using Citation-informed Transformers}. In
  \bibinfo{booktitle}{\emph{Proceedings of the 58th Annual Meeting of the
  Association for Computational Linguistics}}. \bibinfo{publisher}{Association
  for Computational Linguistics}, \bibinfo{address}{Online},
  \bibinfo{pages}{2270--2282}.
\newblock
\urldef\tempurl%
\url{https://doi.org/10.18653/v1/2020.acl-main.207}
\showDOI{\tempurl}


\bibitem[\protect\citeauthoryear{Davey and Priestley}{Davey and
  Priestley}{2002}]%
        {davey2002lattices}
\bibfield{author}{\bibinfo{person}{Brian~A. Davey} {and}
  \bibinfo{person}{Hilary~A. Priestley}.} \bibinfo{year}{2002}\natexlab{}.
\newblock \bibinfo{booktitle}{\emph{Introduction to Lattices and Order}
  (\bibinfo{edition}{2nd} ed.)}.
\newblock \bibinfo{publisher}{Cambridge University Press}.
\newblock
\showISBNx{978-0-521-78451-1}


\bibitem[\protect\citeauthoryear{Elmagarmid, Ipeirotis, and
  Verykios}{Elmagarmid et~al\mbox{.}}{2007}]%
        {elmagarmid2007duplicate}
\bibfield{author}{\bibinfo{person}{Ahmed~K. Elmagarmid},
  \bibinfo{person}{Panagiotis~G. Ipeirotis}, {and}
  \bibinfo{person}{Vassilios~S. Verykios}.} \bibinfo{year}{2007}\natexlab{}.
\newblock \showarticletitle{Duplicate Record Detection: A Survey}.
\newblock \bibinfo{journal}{\emph{{IEEE} Transactions on Knowledge and Data
  Engineering}} \bibinfo{volume}{19}, \bibinfo{number}{1}
  (\bibinfo{year}{2007}), \bibinfo{pages}{1--16}.
\newblock
\urldef\tempurl%
\url{https://doi.org/10.1109/TKDE.2007.250581}
\showDOI{\tempurl}


\bibitem[\protect\citeauthoryear{Eltabakh, Kunjir, Elmagarmid, and
  Ahmad}{Eltabakh et~al\mbox{.}}{2023}]%
        {eltabakh2023crossmodal}
\bibfield{author}{\bibinfo{person}{Mohamed~Y. Eltabakh},
  \bibinfo{person}{Mayuresh Kunjir}, \bibinfo{person}{Ahmed Elmagarmid}, {and}
  \bibinfo{person}{Mohammad~Shahmeer Ahmad}.} \bibinfo{year}{2023}\natexlab{}.
\newblock \showarticletitle{Cross Modal Data Discovery over Structured and
  Unstructured Data Lakes}.
\newblock \bibinfo{journal}{\emph{Proceedings of the {VLDB} Endowment}}
  \bibinfo{volume}{16}, \bibinfo{number}{11} (\bibinfo{year}{2023}),
  \bibinfo{pages}{3377--3390}.
\newblock
\urldef\tempurl%
\url{https://doi.org/10.14778/3611479.3611533}
\showDOI{\tempurl}


\bibitem[\protect\citeauthoryear{Fagin, Lotem, and Naor}{Fagin
  et~al\mbox{.}}{2001}]%
        {fagin2001ta}
\bibfield{author}{\bibinfo{person}{Ronald Fagin}, \bibinfo{person}{Amnon
  Lotem}, {and} \bibinfo{person}{Moni Naor}.} \bibinfo{year}{2001}\natexlab{}.
\newblock \showarticletitle{Optimal Aggregation Algorithms for Middleware}. In
  \bibinfo{booktitle}{\emph{Proceedings of the Twentieth {ACM}
  {SIGMOD}-{SIGACT}-{SIGART} Symposium on Principles of Database Systems
  ({PODS})}}. \bibinfo{pages}{102--113}.
\newblock
\urldef\tempurl%
\url{https://doi.org/10.1145/375551.375567}
\showDOI{\tempurl}


\bibitem[\protect\citeauthoryear{Fan, Wang, Li, Zhang, and Miller}{Fan
  et~al\mbox{.}}{2023}]%
        {fan2023starmie}
\bibfield{author}{\bibinfo{person}{Grace Fan}, \bibinfo{person}{Jin Wang},
  \bibinfo{person}{Yuliang Li}, \bibinfo{person}{Dan Zhang}, {and}
  \bibinfo{person}{Ren{\'e}e~J. Miller}.} \bibinfo{year}{2023}\natexlab{}.
\newblock \showarticletitle{Semantics-Aware Dataset Discovery from Data Lakes
  with Contextualized Column-Based Representation Learning}.
\newblock \bibinfo{journal}{\emph{Proceedings of the {VLDB} Endowment}}
  \bibinfo{volume}{16}, \bibinfo{number}{7} (\bibinfo{year}{2023}),
  \bibinfo{pages}{1726--1739}.
\newblock
\urldef\tempurl%
\url{https://doi.org/10.14778/3587136.3587146}
\showDOI{\tempurl}


\bibitem[\protect\citeauthoryear{Fernandez, Abedjan, Koko, Yuan, Madden, and
  Stonebraker}{Fernandez et~al\mbox{.}}{2018}]%
        {fernandez2018aurum}
\bibfield{author}{\bibinfo{person}{Raul~Castro Fernandez},
  \bibinfo{person}{Ziawasch Abedjan}, \bibinfo{person}{Famien Koko},
  \bibinfo{person}{Gina Yuan}, \bibinfo{person}{Samuel Madden}, {and}
  \bibinfo{person}{Michael Stonebraker}.} \bibinfo{year}{2018}\natexlab{}.
\newblock \showarticletitle{Aurum: A Data Discovery System}. In
  \bibinfo{booktitle}{\emph{Proceedings of the 34th {IEEE} International
  Conference on Data Engineering ({ICDE})}}. \bibinfo{pages}{1001--1012}.
\newblock
\urldef\tempurl%
\url{https://doi.org/10.1109/ICDE.2018.00094}
\showDOI{\tempurl}


\bibitem[\protect\citeauthoryear{Fisher, Nemhauser, and Wolsey}{Fisher
  et~al\mbox{.}}{1978}]%
        {fisher1978analysis}
\bibfield{author}{\bibinfo{person}{Marshall~L. Fisher},
  \bibinfo{person}{George~L. Nemhauser}, {and} \bibinfo{person}{Laurence~A.
  Wolsey}.} \bibinfo{year}{1978}\natexlab{}.
\newblock \showarticletitle{An Analysis of Approximations for Maximizing
  Submodular Set Functions---{II}}.
\newblock \bibinfo{journal}{\emph{Mathematical Programming Studies}}
  \bibinfo{volume}{8} (\bibinfo{year}{1978}), \bibinfo{pages}{73--87}.
\newblock
\urldef\tempurl%
\url{https://doi.org/10.1007/BFb0121195}
\showDOI{\tempurl}


\bibitem[\protect\citeauthoryear{Fu, Xiang, Wang, and Cai}{Fu
  et~al\mbox{.}}{2019}]%
        {fu2019nsg}
\bibfield{author}{\bibinfo{person}{Cong Fu}, \bibinfo{person}{Chao Xiang},
  \bibinfo{person}{Changxu Wang}, {and} \bibinfo{person}{Deng Cai}.}
  \bibinfo{year}{2019}\natexlab{}.
\newblock \showarticletitle{Fast Approximate Nearest Neighbor Search With The
  Navigating Spreading-Out Graph}.
\newblock \bibinfo{journal}{\emph{Proceedings of the {VLDB} Endowment}}
  \bibinfo{volume}{12}, \bibinfo{number}{5} (\bibinfo{year}{2019}),
  \bibinfo{pages}{461--474}.
\newblock
\urldef\tempurl%
\url{https://doi.org/10.14778/3303753.3303754}
\showDOI{\tempurl}


\bibitem[\protect\citeauthoryear{Ganea, B{\'e}cigneul, and Hofmann}{Ganea
  et~al\mbox{.}}{2018}]%
        {Ganea2018EntailmentCones}
\bibfield{author}{\bibinfo{person}{Octavian{-}Eugen Ganea},
  \bibinfo{person}{Gary B{\'e}cigneul}, {and} \bibinfo{person}{Thomas
  Hofmann}.} \bibinfo{year}{2018}\natexlab{}.
\newblock \showarticletitle{Hyperbolic Entailment Cones for Learning
  Hierarchical Embeddings}. In \bibinfo{booktitle}{\emph{Proceedings of the
  35th International Conference on Machine Learning (ICML)}}.
  \bibinfo{publisher}{PMLR}, \bibinfo{pages}{1646--1655}.
\newblock
\urldef\tempurl%
\url{https://proceedings.mlr.press/v80/ganea18a.html}
\showURL{%
\tempurl}


\bibitem[\protect\citeauthoryear{Grootendorst}{Grootendorst}{2020}]%
        {grootendorst2020keybert}
\bibfield{author}{\bibinfo{person}{Maarten Grootendorst}.}
  \bibinfo{year}{2020}\natexlab{}.
\newblock \bibinfo{title}{{KeyBERT}: Minimal Keyword Extraction with {BERT}}.
\newblock
\newblock
\urldef\tempurl%
\url{https://doi.org/10.5281/zenodo.4461265}
\showDOI{\tempurl}


\bibitem[\protect\citeauthoryear{{Gurobi Optimization, LLC}}{{Gurobi
  Optimization, LLC}}{2024}]%
        {gurobi}
\bibfield{author}{\bibinfo{person}{{Gurobi Optimization, LLC}}.}
  \bibinfo{year}{2024}\natexlab{}.
\newblock \bibinfo{title}{Gurobi Optimizer Reference Manual}.
\newblock \bibinfo{howpublished}{\url{https://www.gurobi.com}}.
\newblock


\bibitem[\protect\citeauthoryear{Hagen}{Hagen}{2008}]%
        {hagen2008harmonic}
\bibfield{author}{\bibinfo{person}{Nils~T. Hagen}.}
  \bibinfo{year}{2008}\natexlab{}.
\newblock \showarticletitle{Harmonic Allocation of Authorship Credit:
  Source-Level Correction of Bibliometric Bias Assures Accurate Publication and
  Citation Analysis}.
\newblock \bibinfo{journal}{\emph{PLOS ONE}} \bibinfo{volume}{3},
  \bibinfo{number}{12} (\bibinfo{year}{2008}), \bibinfo{pages}{e4021}.
\newblock
\urldef\tempurl%
\url{https://doi.org/10.1371/journal.pone.0004021}
\showDOI{\tempurl}


\bibitem[\protect\citeauthoryear{Hearst}{Hearst}{1992}]%
        {hearst1992hyponyms}
\bibfield{author}{\bibinfo{person}{Marti~A. Hearst}.}
  \bibinfo{year}{1992}\natexlab{}.
\newblock \showarticletitle{Automatic Acquisition of Hyponyms from Large Text
  Corpora}. In \bibinfo{booktitle}{\emph{Proceedings of COLING 1992}}.
  \bibinfo{pages}{539--545}.
\newblock


\bibitem[\protect\citeauthoryear{Hochbaum}{Hochbaum}{1997}]%
        {hochbaum1997approximation}
\bibfield{author}{\bibinfo{person}{Dorit~S. Hochbaum}.}
  \bibinfo{year}{1997}\natexlab{}.
\newblock \showarticletitle{Approximation Algorithms for {NP}-Hard Problems}.
\newblock In \bibinfo{booktitle}{\emph{Approximation Algorithms for {NP}-Hard
  Problems}}, \bibfield{editor}{\bibinfo{person}{Dorit~S. Hochbaum}} (Ed.).
  \bibinfo{publisher}{PWS Publishing Co.}, \bibinfo{address}{Boston, MA, USA},
  Chapter~3, \bibinfo{pages}{94--143}.
\newblock
\showISBNx{978-0-534-94968-6}
\newblock
\shownote{Chapter on covering problems, including Maximum Coverage.}


\bibitem[\protect\citeauthoryear{Hodge and Greenberg}{Hodge and
  Greenberg}{1981}]%
        {hodge1981credit}
\bibfield{author}{\bibinfo{person}{Scott~E. Hodge} {and}
  \bibinfo{person}{David~A. Greenberg}.} \bibinfo{year}{1981}\natexlab{}.
\newblock \showarticletitle{Publication Credit}.
\newblock \bibinfo{journal}{\emph{Science}} \bibinfo{volume}{213},
  \bibinfo{number}{4511} (\bibinfo{year}{1981}), \bibinfo{pages}{950}.
\newblock
\urldef\tempurl%
\url{https://doi.org/10.1126/science.213.4511.950.b}
\showDOI{\tempurl}


\bibitem[\protect\citeauthoryear{{IBM}}{{IBM}}{2024}]%
        {cplex}
\bibfield{author}{\bibinfo{person}{{IBM}}.} \bibinfo{year}{2024}\natexlab{}.
\newblock \bibinfo{title}{{IBM ILOG CPLEX Optimization Studio}}.
\newblock
  \bibinfo{howpublished}{\url{https://www.ibm.com/products/ilog-cplex-optimization-studio}}.
\newblock


\bibitem[\protect\citeauthoryear{{ICML Program Chairs}}{{ICML Program
  Chairs}}{2025}]%
        {icml2025ac}
\bibfield{author}{\bibinfo{person}{{ICML Program Chairs}}.}
  \bibinfo{year}{2025}\natexlab{}.
\newblock \bibinfo{title}{{ICML} 2025 Area Chair Instructions}.
\newblock
  \bibinfo{howpublished}{\url{https://icml.cc/Conferences/2025/AreaChairInstructions}}.
\newblock


\bibitem[\protect\citeauthoryear{Ilyas, Beskales, and Soliman}{Ilyas
  et~al\mbox{.}}{2008}]%
        {ilyas2008topk}
\bibfield{author}{\bibinfo{person}{Ihab~F. Ilyas}, \bibinfo{person}{George
  Beskales}, {and} \bibinfo{person}{Mohamed~A. Soliman}.}
  \bibinfo{year}{2008}\natexlab{}.
\newblock \showarticletitle{A Survey of Top-$k$ Query Processing Techniques in
  Relational Database Systems}.
\newblock \bibinfo{journal}{\emph{Comput. Surveys}} \bibinfo{volume}{40},
  \bibinfo{number}{4} (\bibinfo{year}{2008}), \bibinfo{pages}{11:1--11:58}.
\newblock
\urldef\tempurl%
\url{https://doi.org/10.1145/1391729.1391730}
\showDOI{\tempurl}


\bibitem[\protect\citeauthoryear{Ilyas, Rekatsinas, Konda, Pound, Qi, and
  Soliman}{Ilyas et~al\mbox{.}}{2022}]%
        {ilyas2022saga}
\bibfield{author}{\bibinfo{person}{Ihab~F. Ilyas}, \bibinfo{person}{Theodoros
  Rekatsinas}, \bibinfo{person}{Vishnu Konda}, \bibinfo{person}{Jeffrey Pound},
  \bibinfo{person}{Xiaoguang Qi}, {and} \bibinfo{person}{Mohamed Soliman}.}
  \bibinfo{year}{2022}\natexlab{}.
\newblock \showarticletitle{Saga: A Platform for Continuous Construction and
  Serving of Knowledge at Scale}. In \bibinfo{booktitle}{\emph{Proceedings of
  the 2022 International Conference on Management of Data ({SIGMOD})}}.
  \bibinfo{publisher}{ACM}, \bibinfo{pages}{2259--2272}.
\newblock
\urldef\tempurl%
\url{https://doi.org/10.1145/3514221.3526049}
\showDOI{\tempurl}


\bibitem[\protect\citeauthoryear{Indyk and Thaper}{Indyk and Thaper}{2003}]%
        {indyk2003fast}
\bibfield{author}{\bibinfo{person}{Piotr Indyk} {and} \bibinfo{person}{Nitin
  Thaper}.} \bibinfo{year}{2003}\natexlab{}.
\newblock \showarticletitle{Fast Image Retrieval via Embeddings}. In
  \bibinfo{booktitle}{\emph{Third Workshop on Statistical and Computational
  Theories of Vision ({ICCV})}}.
\newblock


\bibitem[\protect\citeauthoryear{J{\'e}gou, Douze, and Schmid}{J{\'e}gou
  et~al\mbox{.}}{2011}]%
        {jegou2011pq}
\bibfield{author}{\bibinfo{person}{Herv{\'e} J{\'e}gou},
  \bibinfo{person}{Matthijs Douze}, {and} \bibinfo{person}{Cordelia Schmid}.}
  \bibinfo{year}{2011}\natexlab{}.
\newblock \showarticletitle{Product Quantization for Nearest Neighbor Search}.
\newblock \bibinfo{journal}{\emph{{IEEE} Transactions on Pattern Analysis and
  Machine Intelligence}} \bibinfo{volume}{33}, \bibinfo{number}{1}
  (\bibinfo{year}{2011}), \bibinfo{pages}{117--128}.
\newblock
\urldef\tempurl%
\url{https://doi.org/10.1109/TPAMI.2010.57}
\showDOI{\tempurl}


\bibitem[\protect\citeauthoryear{Ji, Luo, Bao, Gan, and Huang}{Ji
  et~al\mbox{.}}{2026}]%
        {ji2026taxonomy}
\bibfield{author}{\bibinfo{person}{Daomin Ji}, \bibinfo{person}{Hui Luo},
  \bibinfo{person}{Zhifeng Bao}, \bibinfo{person}{Junhao Gan}, {and}
  \bibinfo{person}{Zi Huang}.} \bibinfo{year}{2026}\natexlab{}.
\newblock \showarticletitle{Taxonomy Maintenance in the Wild over Evolving
  Scholarly Data: Reliability, Efficiency, and Cost-Effectiveness}.
\newblock \bibinfo{journal}{\emph{Proceedings of the ACM on Management of
  Data}} \bibinfo{volume}{4}, \bibinfo{number}{4 (SIGMOD)}, Article
  \bibinfo{articleno}{289} (\bibinfo{date}{Sept.} \bibinfo{year}{2026}).
\newblock
\urldef\tempurl%
\url{https://doi.org/10.1145/3837127}
\showDOI{\tempurl}


\bibitem[\protect\citeauthoryear{Jiang, Yao, Wang, and Sun}{Jiang
  et~al\mbox{.}}{2023}]%
        {Jiang2023BoxTaxo}
\bibfield{author}{\bibinfo{person}{Song Jiang}, \bibinfo{person}{Qiyue Yao},
  \bibinfo{person}{Qifan Wang}, {and} \bibinfo{person}{Yizhou Sun}.}
  \bibinfo{year}{2023}\natexlab{}.
\newblock \showarticletitle{A Single Vector Is Not Enough: Taxonomy Expansion
  via Box Embeddings}. In \bibinfo{booktitle}{\emph{Proceedings of the ACM Web
  Conference (The Web Conference)}}. \bibinfo{pages}{2467--2476}.
\newblock
\urldef\tempurl%
\url{https://doi.org/10.1145/3543507.3583310}
\showDOI{\tempurl}


\bibitem[\protect\citeauthoryear{Jin, Xiang, Ruan, and Fuhry}{Jin
  et~al\mbox{.}}{2009}]%
        {jin2009threehop}
\bibfield{author}{\bibinfo{person}{Ruoming Jin}, \bibinfo{person}{Yang Xiang},
  \bibinfo{person}{Ning Ruan}, {and} \bibinfo{person}{David Fuhry}.}
  \bibinfo{year}{2009}\natexlab{}.
\newblock \showarticletitle{3-{HOP}: A High-Compression Indexing Scheme for
  Reachability Query}. In \bibinfo{booktitle}{\emph{Proceedings of the 2009
  {ACM} {SIGMOD} International Conference on Management of Data}}.
  \bibinfo{pages}{813--826}.
\newblock
\urldef\tempurl%
\url{https://doi.org/10.1145/1559845.1559930}
\showDOI{\tempurl}


\bibitem[\protect\citeauthoryear{Kargupta, Zhang, Zhang, Zhang, Mitra, and
  Han}{Kargupta et~al\mbox{.}}{2025}]%
        {kargupta2025taxoadapt}
\bibfield{author}{\bibinfo{person}{Priyanka Kargupta}, \bibinfo{person}{Nan
  Zhang}, \bibinfo{person}{Yunyi Zhang}, \bibinfo{person}{Rui Zhang},
  \bibinfo{person}{Prasenjit Mitra}, {and} \bibinfo{person}{Jiawei Han}.}
  \bibinfo{year}{2025}\natexlab{}.
\newblock \showarticletitle{TaxoAdapt: Aligning LLM-Based Multidimensional
  Taxonomy Construction to Evolving Research Corpora}. In
  \bibinfo{booktitle}{\emph{Proceedings of the 63rd Annual Meeting of the
  Association for Computational Linguistics (Volume 1: Long Papers)}}.
  \bibinfo{pages}{29834--29850}.
\newblock
\urldef\tempurl%
\url{https://aclanthology.org/2025.acl-long.1442/}
\showURL{%
\tempurl}


\bibitem[\protect\citeauthoryear{Karimzadehgan, Zhai, and
  Belford}{Karimzadehgan et~al\mbox{.}}{2008}]%
        {karimzadehgan2008multiaspect}
\bibfield{author}{\bibinfo{person}{Maryam Karimzadehgan},
  \bibinfo{person}{ChengXiang Zhai}, {and} \bibinfo{person}{Geneva Belford}.}
  \bibinfo{year}{2008}\natexlab{}.
\newblock \showarticletitle{Multi-Aspect Expertise Matching for Review
  Assignment}. In \bibinfo{booktitle}{\emph{Proceedings of the 17th {ACM}
  Conference on Information and Knowledge Management ({CIKM})}}.
  \bibinfo{pages}{1113--1122}.
\newblock
\urldef\tempurl%
\url{https://doi.org/10.1145/1458082.1458230}
\showDOI{\tempurl}


\bibitem[\protect\citeauthoryear{Karmarkar}{Karmarkar}{1984}]%
        {karmarkar1984new}
\bibfield{author}{\bibinfo{person}{Narendra Karmarkar}.}
  \bibinfo{year}{1984}\natexlab{}.
\newblock \showarticletitle{A New Polynomial-Time Algorithm for Linear
  Programming}.
\newblock \bibinfo{journal}{\emph{Combinatorica}} \bibinfo{volume}{4},
  \bibinfo{number}{4} (\bibinfo{year}{1984}), \bibinfo{pages}{373--395}.
\newblock
\urldef\tempurl%
\url{https://doi.org/10.1007/BF02579150}
\showDOI{\tempurl}


\bibitem[\protect\citeauthoryear{Khatiwada, Fan, Shraga, Chen, Gatterbauer,
  Miller, and Riedewald}{Khatiwada et~al\mbox{.}}{2023}]%
        {khatiwada2023santos}
\bibfield{author}{\bibinfo{person}{Aamod Khatiwada}, \bibinfo{person}{Grace
  Fan}, \bibinfo{person}{Roee Shraga}, \bibinfo{person}{Zixuan Chen},
  \bibinfo{person}{Wolfgang Gatterbauer}, \bibinfo{person}{Ren{\'e}e~J.
  Miller}, {and} \bibinfo{person}{Mirek Riedewald}.}
  \bibinfo{year}{2023}\natexlab{}.
\newblock \showarticletitle{{SANTOS}: Relationship-Based Semantic Table Union
  Search}.
\newblock \bibinfo{journal}{\emph{Proceedings of the {ACM} on Management of
  Data}} \bibinfo{volume}{1}, \bibinfo{number}{1} (\bibinfo{year}{2023}),
  \bibinfo{pages}{9:1--9:25}.
\newblock
\urldef\tempurl%
\url{https://doi.org/10.1145/3588689}
\showDOI{\tempurl}


\bibitem[\protect\citeauthoryear{Kobren, Saha, and McCallum}{Kobren
  et~al\mbox{.}}{2019}]%
        {kobren2019paper}
\bibfield{author}{\bibinfo{person}{Ari Kobren}, \bibinfo{person}{Barna Saha},
  {and} \bibinfo{person}{Andrew McCallum}.} \bibinfo{year}{2019}\natexlab{}.
\newblock \showarticletitle{Paper Matching with Local Fairness Constraints}. In
  \bibinfo{booktitle}{\emph{Proceedings of the 25th ACM SIGKDD International
  Conference on Knowledge Discovery \& Data Mining (KDD)}}.
\newblock


\bibitem[\protect\citeauthoryear{Kotelnikov, Baranchuk, Rubachev, and
  Babenko}{Kotelnikov et~al\mbox{.}}{2023}]%
        {kotelnikov2023tabddpm}
\bibfield{author}{\bibinfo{person}{Akim Kotelnikov}, \bibinfo{person}{Dmitry
  Baranchuk}, \bibinfo{person}{Ivan Rubachev}, {and} \bibinfo{person}{Artem
  Babenko}.} \bibinfo{year}{2023}\natexlab{}.
\newblock \showarticletitle{{TabDDPM}: Modelling Tabular Data with Diffusion
  Models}. In \bibinfo{booktitle}{\emph{Proceedings of the 40th International
  Conference on Machine Learning (ICML)}}.
\newblock
\urldef\tempurl%
\url{https://arxiv.org/abs/2209.15421}
\showURL{%
\tempurl}


\bibitem[\protect\citeauthoryear{Kou, U, Mamoulis, and Gong}{Kou
  et~al\mbox{.}}{2015a}]%
        {kou2015weighted}
\bibfield{author}{\bibinfo{person}{Ngai~Meng Kou}, \bibinfo{person}{Leong~Hou
  U}, \bibinfo{person}{Nikos Mamoulis}, {and} \bibinfo{person}{Zhiguo Gong}.}
  \bibinfo{year}{2015}\natexlab{a}.
\newblock \showarticletitle{Weighted Coverage based Reviewer Assignment}. In
  \bibinfo{booktitle}{\emph{Proceedings of the 2015 {ACM} {SIGMOD}
  International Conference on Management of Data}}.
  \bibinfo{pages}{2031--2046}.
\newblock
\urldef\tempurl%
\url{https://doi.org/10.1145/2723372.2723727}
\showDOI{\tempurl}


\bibitem[\protect\citeauthoryear{Kou, U, Mamoulis, Li, Li, and Gong}{Kou
  et~al\mbox{.}}{2015b}]%
        {kou2015topic}
\bibfield{author}{\bibinfo{person}{Ngai~Meng Kou}, \bibinfo{person}{Leong~Hou
  U}, \bibinfo{person}{Nikos Mamoulis}, \bibinfo{person}{Yong Li},
  \bibinfo{person}{Ye Li}, {and} \bibinfo{person}{Zhiguo Gong}.}
  \bibinfo{year}{2015}\natexlab{b}.
\newblock \showarticletitle{A Topic-based Reviewer Assignment System}.
\newblock \bibinfo{journal}{\emph{Proceedings of the {VLDB} Endowment}}
  \bibinfo{volume}{8}, \bibinfo{number}{12} (\bibinfo{year}{2015}),
  \bibinfo{pages}{1852--1855}.
\newblock
\urldef\tempurl%
\url{https://doi.org/10.14778/2824032.2824084}
\showDOI{\tempurl}


\bibitem[\protect\citeauthoryear{Kulkarni, Mahata, Arora, and Bhowmik}{Kulkarni
  et~al\mbox{.}}{2022}]%
        {kulkarni2022keybart}
\bibfield{author}{\bibinfo{person}{Mayank Kulkarni}, \bibinfo{person}{Debanjan
  Mahata}, \bibinfo{person}{Ravneet Arora}, {and} \bibinfo{person}{Rajarshi
  Bhowmik}.} \bibinfo{year}{2022}\natexlab{}.
\newblock \showarticletitle{Learning Rich Representation of Keyphrases from
  Text}. In \bibinfo{booktitle}{\emph{Findings of the Association for
  Computational Linguistics: NAACL 2022}}. \bibinfo{pages}{891--906}.
\newblock
\urldef\tempurl%
\url{https://doi.org/10.18653/v1/2022.findings-naacl.67}
\showDOI{\tempurl}


\bibitem[\protect\citeauthoryear{Kullback and Leibler}{Kullback and
  Leibler}{1951}]%
        {kullback1951information}
\bibfield{author}{\bibinfo{person}{Solomon Kullback} {and}
  \bibinfo{person}{Richard~A. Leibler}.} \bibinfo{year}{1951}\natexlab{}.
\newblock \showarticletitle{On Information and Sufficiency}.
\newblock \bibinfo{journal}{\emph{The Annals of Mathematical Statistics}}
  \bibinfo{volume}{22}, \bibinfo{number}{1} (\bibinfo{year}{1951}),
  \bibinfo{pages}{79--86}.
\newblock
\urldef\tempurl%
\url{https://doi.org/10.1214/aoms/1177729694}
\showDOI{\tempurl}


\bibitem[\protect\citeauthoryear{Lan, Wang, Bao, Li, Ji, Lee, Luo, Huang, Qiu,
  and Hua}{Lan et~al\mbox{.}}{2026}]%
        {lan2026agenticscholar}
\bibfield{author}{\bibinfo{person}{Hai Lan}, \bibinfo{person}{Tingting Wang},
  \bibinfo{person}{Zhifeng Bao}, \bibinfo{person}{Guoliang Li},
  \bibinfo{person}{Daomin Ji}, \bibinfo{person}{Ge Lee}, \bibinfo{person}{Feng
  Luo}, \bibinfo{person}{Zi Huang}, \bibinfo{person}{Hailang Qiu}, {and}
  \bibinfo{person}{Gang Hua}.} \bibinfo{year}{2026}\natexlab{}.
\newblock \showarticletitle{{AgenticScholar}: Agentic Data Management with
  Pipeline Orchestration for Scholarly Corpora}.
\newblock \bibinfo{journal}{\emph{arXiv preprint arXiv:2603.13774}}
  (\bibinfo{year}{2026}).
\newblock
\urldef\tempurl%
\url{https://arxiv.org/abs/2603.13774}
\showURL{%
\tempurl}


\bibitem[\protect\citeauthoryear{Lawson and Hanson}{Lawson and Hanson}{1995}]%
        {lawson1995solving}
\bibfield{author}{\bibinfo{person}{Charles~L. Lawson} {and}
  \bibinfo{person}{Richard~J. Hanson}.} \bibinfo{year}{1995}\natexlab{}.
\newblock \bibinfo{booktitle}{\emph{Solving Least Squares Problems}}.
  \bibinfo{series}{Classics in Applied Mathematics}, Vol.~\bibinfo{volume}{15}.
\newblock \bibinfo{publisher}{Society for Industrial and Applied Mathematics
  (SIAM)}, \bibinfo{address}{Philadelphia, PA}.
\newblock
\urldef\tempurl%
\url{https://doi.org/10.1137/1.9781611971217}
\showDOI{\tempurl}
\newblock
\shownote{Originally published by Prentice-Hall, 1974.}


\bibitem[\protect\citeauthoryear{Le, Yamada, Fukumizu, and Cuturi}{Le
  et~al\mbox{.}}{2019}]%
        {le2019tree}
\bibfield{author}{\bibinfo{person}{Tam Le}, \bibinfo{person}{Makoto Yamada},
  \bibinfo{person}{Kenji Fukumizu}, {and} \bibinfo{person}{Marco Cuturi}.}
  \bibinfo{year}{2019}\natexlab{}.
\newblock \showarticletitle{Tree-Sliced Variants of {W}asserstein Distances}.
  In \bibinfo{booktitle}{\emph{Advances in Neural Information Processing
  Systems 32 ({NeurIPS})}}. \bibinfo{pages}{12283--12294}.
\newblock


\bibitem[\protect\citeauthoryear{Ley}{Ley}{2002}]%
        {ley2002dblp}
\bibfield{author}{\bibinfo{person}{Michael Ley}.}
  \bibinfo{year}{2002}\natexlab{}.
\newblock \showarticletitle{The {DBLP} Computer Science Bibliography:
  Evolution, Research Issues, Perspectives}. In
  \bibinfo{booktitle}{\emph{String Processing and Information Retrieval
  ({SPIRE})}}. \bibinfo{pages}{1--10}.
\newblock
\urldef\tempurl%
\url{https://doi.org/10.1007/3-540-45735-6\_1}
\showDOI{\tempurl}


\bibitem[\protect\citeauthoryear{Lin and Bilmes}{Lin and Bilmes}{2011}]%
        {lin2011submodular}
\bibfield{author}{\bibinfo{person}{Hui Lin} {and} \bibinfo{person}{Jeff
  Bilmes}.} \bibinfo{year}{2011}\natexlab{}.
\newblock \showarticletitle{A Class of Submodular Functions for Document
  Summarization}. In \bibinfo{booktitle}{\emph{Proceedings of the 49th Annual
  Meeting of the Association for Computational Linguistics: Human Language
  Technologies}}, \bibfield{editor}{\bibinfo{person}{Dekang Lin},
  \bibinfo{person}{Yuji Matsumoto}, {and} \bibinfo{person}{Rada Mihalcea}}
  (Eds.). \bibinfo{publisher}{Association for Computational Linguistics},
  \bibinfo{address}{Portland, Oregon, USA}, \bibinfo{pages}{510--520}.
\newblock
\urldef\tempurl%
\url{https://aclanthology.org/P11-1052/}
\showURL{%
\tempurl}


\bibitem[\protect\citeauthoryear{Lin}{Lin}{1991}]%
        {lin1991jsd}
\bibfield{author}{\bibinfo{person}{Jianhua Lin}.}
  \bibinfo{year}{1991}\natexlab{}.
\newblock \showarticletitle{Divergence Measures Based on the {S}hannon
  Entropy}.
\newblock \bibinfo{journal}{\emph{IEEE Transactions on Information Theory}}
  \bibinfo{volume}{37}, \bibinfo{number}{1} (\bibinfo{year}{1991}),
  \bibinfo{pages}{145--151}.
\newblock
\urldef\tempurl%
\url{https://doi.org/10.1109/18.61115}
\showDOI{\tempurl}


\bibitem[\protect\citeauthoryear{Liu, Suel, and Memon}{Liu
  et~al\mbox{.}}{2014}]%
        {liu2014robust}
\bibfield{author}{\bibinfo{person}{Xiang Liu}, \bibinfo{person}{Torsten Suel},
  {and} \bibinfo{person}{Nasir~D. Memon}.} \bibinfo{year}{2014}\natexlab{}.
\newblock \showarticletitle{A Robust Model for Paper Reviewer Assignment}. In
  \bibinfo{booktitle}{\emph{Proceedings of the 8th ACM Conference on
  Recommender Systems (RecSys '14)}}. \bibinfo{publisher}{ACM},
  \bibinfo{pages}{25--32}.
\newblock
\urldef\tempurl%
\url{https://doi.org/10.1145/2645710.2645749}
\showDOI{\tempurl}


\bibitem[\protect\citeauthoryear{Lu, Chen, Mao, Rao, Xie, Wang, and Li}{Lu
  et~al\mbox{.}}{2024}]%
        {Lu2024BoxTM}
\bibfield{author}{\bibinfo{person}{Yuyin Lu}, \bibinfo{person}{Hegang Chen},
  \bibinfo{person}{Pengbo Mao}, \bibinfo{person}{Yanghui Rao},
  \bibinfo{person}{Haoran Xie}, \bibinfo{person}{Fu~Lee Wang}, {and}
  \bibinfo{person}{Qing Li}.} \bibinfo{year}{2024}\natexlab{}.
\newblock \showarticletitle{Self-supervised Topic Taxonomy Discovery in the Box
  Embedding Space}.
\newblock \bibinfo{journal}{\emph{Transactions of the Association for
  Computational Linguistics (TACL)}}  \bibinfo{volume}{12}
  (\bibinfo{year}{2024}), \bibinfo{pages}{1401--1416}.
\newblock
\urldef\tempurl%
\url{https://doi.org/10.1162/tacl_a_00712}
\showDOI{\tempurl}


\bibitem[\protect\citeauthoryear{Madden and DeWitt}{Madden and DeWitt}{2006}]%
        {madden2006doubleblind}
\bibfield{author}{\bibinfo{person}{Samuel Madden} {and}
  \bibinfo{person}{David~J. DeWitt}.} \bibinfo{year}{2006}\natexlab{}.
\newblock \showarticletitle{Impact of Double-Blind Reviewing on {SIGMOD}
  Publication Rates}.
\newblock \bibinfo{journal}{\emph{{ACM} {SIGMOD} Record}} \bibinfo{volume}{35},
  \bibinfo{number}{2} (\bibinfo{year}{2006}), \bibinfo{pages}{29--32}.
\newblock


\bibitem[\protect\citeauthoryear{Malkov and Yashunin}{Malkov and
  Yashunin}{2020}]%
        {malkov2020hnsw}
\bibfield{author}{\bibinfo{person}{Yury~A. Malkov} {and}
  \bibinfo{person}{Dmitry~A. Yashunin}.} \bibinfo{year}{2020}\natexlab{}.
\newblock \showarticletitle{Efficient and Robust Approximate Nearest Neighbor
  Search Using Hierarchical Navigable Small World Graphs}.
\newblock \bibinfo{journal}{\emph{{IEEE} Transactions on Pattern Analysis and
  Machine Intelligence}} \bibinfo{volume}{42}, \bibinfo{number}{4}
  (\bibinfo{year}{2020}), \bibinfo{pages}{824--836}.
\newblock
\urldef\tempurl%
\url{https://doi.org/10.1109/TPAMI.2018.2889473}
\showDOI{\tempurl}


\bibitem[\protect\citeauthoryear{{Microsoft Research}}{{Microsoft
  Research}}{2026}]%
        {microsoftcmt}
\bibfield{author}{\bibinfo{person}{{Microsoft Research}}.}
  \bibinfo{year}{2026}\natexlab{}.
\newblock \bibinfo{title}{Microsoft CMT: Conference Management Toolkit}.
\newblock \bibinfo{howpublished}{\url{https://cmt3.research.microsoft.com/}}.
\newblock
\newblock
\shownote{Accessed: 2026-02-10.}


\bibitem[\protect\citeauthoryear{Mimno and McCallum}{Mimno and
  McCallum}{2007}]%
        {mimno2007expertise}
\bibfield{author}{\bibinfo{person}{David Mimno} {and} \bibinfo{person}{Andrew
  McCallum}.} \bibinfo{year}{2007}\natexlab{}.
\newblock \showarticletitle{Expertise Modeling for Matching Papers with
  Reviewers}. In \bibinfo{booktitle}{\emph{Proceedings of the 13th ACM SIGKDD
  International Conference on Knowledge Discovery and Data Mining (KDD '07)}}.
  \bibinfo{publisher}{ACM}, \bibinfo{pages}{500--509}.
\newblock
\urldef\tempurl%
\url{https://doi.org/10.1145/1281192.1281247}
\showDOI{\tempurl}


\bibitem[\protect\citeauthoryear{Mudgal, Li, Rekatsinas, Doan, Park, Krishnan,
  Deep, Arcaute, and Raghavendra}{Mudgal et~al\mbox{.}}{2018}]%
        {mudgal2018deepmatcher}
\bibfield{author}{\bibinfo{person}{Sidharth Mudgal}, \bibinfo{person}{Han Li},
  \bibinfo{person}{Theodoros Rekatsinas}, \bibinfo{person}{AnHai Doan},
  \bibinfo{person}{Youngchoon Park}, \bibinfo{person}{Ganesh Krishnan},
  \bibinfo{person}{Rohit Deep}, \bibinfo{person}{Esteban Arcaute}, {and}
  \bibinfo{person}{Vijay Raghavendra}.} \bibinfo{year}{2018}\natexlab{}.
\newblock \showarticletitle{Deep Learning for Entity Matching: A Design Space
  Exploration}. In \bibinfo{booktitle}{\emph{Proceedings of the 2018
  International Conference on Management of Data ({SIGMOD})}}.
  \bibinfo{pages}{19--34}.
\newblock
\urldef\tempurl%
\url{https://doi.org/10.1145/3183713.3196926}
\showDOI{\tempurl}


\bibitem[\protect\citeauthoryear{Murtagh and Contreras}{Murtagh and
  Contreras}{2012}]%
        {murtagh2012algorithms}
\bibfield{author}{\bibinfo{person}{Fionn Murtagh} {and} \bibinfo{person}{Pedro
  Contreras}.} \bibinfo{year}{2012}\natexlab{}.
\newblock \showarticletitle{Algorithms for Hierarchical Clustering: An
  Overview}.
\newblock \bibinfo{journal}{\emph{Wiley Interdisciplinary Reviews: Data Mining
  and Knowledge Discovery}} \bibinfo{volume}{2}, \bibinfo{number}{1}
  (\bibinfo{year}{2012}), \bibinfo{pages}{86--97}.
\newblock


\bibitem[\protect\citeauthoryear{Nargesian, Zhu, Miller, Pu, and
  Arocena}{Nargesian et~al\mbox{.}}{2019}]%
        {nargesian2019datalake}
\bibfield{author}{\bibinfo{person}{Fatemeh Nargesian}, \bibinfo{person}{Erkang
  Zhu}, \bibinfo{person}{Ren{\'e}e~J. Miller}, \bibinfo{person}{Ken~Q. Pu},
  {and} \bibinfo{person}{Patricia~C. Arocena}.}
  \bibinfo{year}{2019}\natexlab{}.
\newblock \showarticletitle{Data Lake Management: Challenges and
  Opportunities}.
\newblock \bibinfo{journal}{\emph{Proceedings of the {VLDB} Endowment}}
  \bibinfo{volume}{12}, \bibinfo{number}{12} (\bibinfo{year}{2019}),
  \bibinfo{pages}{1986--1989}.
\newblock
\urldef\tempurl%
\url{https://doi.org/10.14778/3352063.3352116}
\showDOI{\tempurl}


\bibitem[\protect\citeauthoryear{Nargesian, Zhu, Pu, and Miller}{Nargesian
  et~al\mbox{.}}{2018}]%
        {nargesian2018tus}
\bibfield{author}{\bibinfo{person}{Fatemeh Nargesian}, \bibinfo{person}{Erkang
  Zhu}, \bibinfo{person}{Ken~Q. Pu}, {and} \bibinfo{person}{Ren{\'e}e~J.
  Miller}.} \bibinfo{year}{2018}\natexlab{}.
\newblock \showarticletitle{Table Union Search on Open Data}.
\newblock \bibinfo{journal}{\emph{Proceedings of the {VLDB} Endowment}}
  \bibinfo{volume}{11}, \bibinfo{number}{7} (\bibinfo{year}{2018}),
  \bibinfo{pages}{813--825}.
\newblock
\urldef\tempurl%
\url{https://doi.org/10.14778/3192965.3192973}
\showDOI{\tempurl}


\bibitem[\protect\citeauthoryear{Nemhauser, Wolsey, and Fisher}{Nemhauser
  et~al\mbox{.}}{1978}]%
        {nemhauser1978analysis}
\bibfield{author}{\bibinfo{person}{George~L. Nemhauser},
  \bibinfo{person}{Laurence~A. Wolsey}, {and} \bibinfo{person}{Marshall~L.
  Fisher}.} \bibinfo{year}{1978}\natexlab{}.
\newblock \showarticletitle{An Analysis of Approximations for Maximizing
  Submodular Set Functions---{I}}.
\newblock \bibinfo{journal}{\emph{Mathematical Programming}}
  \bibinfo{volume}{14}, \bibinfo{number}{1} (\bibinfo{year}{1978}),
  \bibinfo{pages}{265--294}.
\newblock


\bibitem[\protect\citeauthoryear{{NeurIPS Foundation}}{{NeurIPS
  Foundation}}{2025}]%
        {neurips2025factsheet}
\bibfield{author}{\bibinfo{person}{{NeurIPS Foundation}}.}
  \bibinfo{year}{2025}\natexlab{}.
\newblock \bibinfo{title}{NeurIPS 2025 Fact Sheet}.
\newblock
  \bibinfo{howpublished}{\url{https://media.neurips.cc/Conferences/NeurIPS2025/press/NeurIPS2025-Fact_Sheet.pdf}}.
\newblock
\newblock
\shownote{Accessed: Jan 2026.}


\bibitem[\protect\citeauthoryear{Nickel and Kiela}{Nickel and Kiela}{2017}]%
        {Nickel2017Poincare}
\bibfield{author}{\bibinfo{person}{Maximilian Nickel} {and}
  \bibinfo{person}{Douwe Kiela}.} \bibinfo{year}{2017}\natexlab{}.
\newblock \showarticletitle{Poincar{\'e} Embeddings for Learning Hierarchical
  Representations}. In \bibinfo{booktitle}{\emph{Advances in Neural Information
  Processing Systems (NeurIPS)}}. \bibinfo{pages}{6338--6347}.
\newblock


\bibitem[\protect\citeauthoryear{{OpenReview}}{{OpenReview}}{2026}]%
        {openreview}
\bibfield{author}{\bibinfo{person}{{OpenReview}}.}
  \bibinfo{year}{2026}\natexlab{}.
\newblock \bibinfo{title}{OpenReview}.
\newblock \bibinfo{howpublished}{\url{https://openreview.net/}}.
\newblock
\newblock
\shownote{Accessed: 2026-02-10.}


\bibitem[\protect\citeauthoryear{Oxley}{Oxley}{2011}]%
        {oxley2011matroid}
\bibfield{author}{\bibinfo{person}{James~G. Oxley}.}
  \bibinfo{year}{2011}\natexlab{}.
\newblock \bibinfo{booktitle}{\emph{Matroid Theory} (\bibinfo{edition}{2nd}
  ed.)}. \bibinfo{series}{Oxford Graduate Texts in Mathematics},
  Vol.~\bibinfo{volume}{21}.
\newblock \bibinfo{publisher}{Oxford University Press},
  \bibinfo{address}{Oxford}.
\newblock


\bibitem[\protect\citeauthoryear{Pang, Shafieinejad, Liu, Hazlewood, and
  He}{Pang et~al\mbox{.}}{2024}]%
        {pang2024clavaddpm}
\bibfield{author}{\bibinfo{person}{Wei Pang}, \bibinfo{person}{Masoumeh
  Shafieinejad}, \bibinfo{person}{Lucy Liu}, \bibinfo{person}{Stephanie
  Hazlewood}, {and} \bibinfo{person}{Xi He}.} \bibinfo{year}{2024}\natexlab{}.
\newblock \showarticletitle{{ClavaDDPM}: Multi-relational Data Synthesis with
  Cluster-guided Diffusion Models}. In \bibinfo{booktitle}{\emph{Advances in
  Neural Information Processing Systems (NeurIPS)}}.
\newblock
\urldef\tempurl%
\url{https://arxiv.org/abs/2405.17724}
\showURL{%
\tempurl}


\bibitem[\protect\citeauthoryear{Papadakis, Skoutas, Thanos, and
  Palpanas}{Papadakis et~al\mbox{.}}{2020}]%
        {papadakis2020blocking}
\bibfield{author}{\bibinfo{person}{George Papadakis},
  \bibinfo{person}{Dimitrios Skoutas}, \bibinfo{person}{Emmanouil Thanos},
  {and} \bibinfo{person}{Themis Palpanas}.} \bibinfo{year}{2020}\natexlab{}.
\newblock \showarticletitle{Blocking and Filtering Techniques for Entity
  Resolution: A Survey}.
\newblock \bibinfo{journal}{\emph{Comput. Surveys}} \bibinfo{volume}{53},
  \bibinfo{number}{2} (\bibinfo{year}{2020}), \bibinfo{pages}{31:1--31:42}.
\newblock
\urldef\tempurl%
\url{https://doi.org/10.1145/3377455}
\showDOI{\tempurl}


\bibitem[\protect\citeauthoryear{Payan and Zick}{Payan and Zick}{2022}]%
        {payan2022order}
\bibfield{author}{\bibinfo{person}{Justin Payan} {and} \bibinfo{person}{Yair
  Zick}.} \bibinfo{year}{2022}\natexlab{}.
\newblock \showarticletitle{I Will Have Order! Optimizing Orders for Fair
  Reviewer Assignment}. In \bibinfo{booktitle}{\emph{Proceedings of the 31st
  International Joint Conference on Artificial Intelligence (IJCAI)}}.
  \bibinfo{pages}{440--446}.
\newblock


\bibitem[\protect\citeauthoryear{Peyr{\'e} and Cuturi}{Peyr{\'e} and
  Cuturi}{2019}]%
        {peyre2019computational}
\bibfield{author}{\bibinfo{person}{Gabriel Peyr{\'e}} {and}
  \bibinfo{person}{Marco Cuturi}.} \bibinfo{year}{2019}\natexlab{}.
\newblock \showarticletitle{Computational Optimal Transport}.
\newblock \bibinfo{journal}{\emph{Foundations and Trends in Machine Learning}}
  \bibinfo{volume}{11}, \bibinfo{number}{5--6} (\bibinfo{year}{2019}),
  \bibinfo{pages}{355--607}.
\newblock
\urldef\tempurl%
\url{https://doi.org/10.1561/2200000073}
\showDOI{\tempurl}


\bibitem[\protect\citeauthoryear{Priem, Piwowar, and Orr}{Priem
  et~al\mbox{.}}{2022}]%
        {priem2022openalex}
\bibfield{author}{\bibinfo{person}{Jason Priem}, \bibinfo{person}{Heather
  Piwowar}, {and} \bibinfo{person}{Richard Orr}.}
  \bibinfo{year}{2022}\natexlab{}.
\newblock \bibinfo{title}{{OpenAlex}: A fully-open index of scholarly works,
  authors, venues, institutions, and concepts}.
\newblock \bibinfo{howpublished}{\url{https://arxiv.org/abs/2205.01833}}.
\newblock
\showeprint[arxiv]{2205.01833}~[cs.DL]


\bibitem[\protect\citeauthoryear{Rodriguez and Bollen}{Rodriguez and
  Bollen}{2008}]%
        {rodriguez2008peerreviewers}
\bibfield{author}{\bibinfo{person}{Marko~A. Rodriguez} {and}
  \bibinfo{person}{Johan Bollen}.} \bibinfo{year}{2008}\natexlab{}.
\newblock \showarticletitle{An Algorithm to Determine Peer-Reviewers}. In
  \bibinfo{booktitle}{\emph{Proceedings of the 17th ACM Conference on
  Information and Knowledge Management (CIKM '08)}}. \bibinfo{publisher}{ACM},
  \bibinfo{pages}{319--328}.
\newblock
\urldef\tempurl%
\url{https://doi.org/10.1145/1458082.1458127}
\showDOI{\tempurl}


\bibitem[\protect\citeauthoryear{Rubner, Tomasi, and Guibas}{Rubner
  et~al\mbox{.}}{2000}]%
        {rubner2000emd}
\bibfield{author}{\bibinfo{person}{Yossi Rubner}, \bibinfo{person}{Carlo
  Tomasi}, {and} \bibinfo{person}{Leonidas~J. Guibas}.}
  \bibinfo{year}{2000}\natexlab{}.
\newblock \showarticletitle{The Earth Mover's Distance as a Metric for Image
  Retrieval}.
\newblock \bibinfo{journal}{\emph{International Journal of Computer Vision}}
  \bibinfo{volume}{40}, \bibinfo{number}{2} (\bibinfo{year}{2000}),
  \bibinfo{pages}{99--121}.
\newblock
\urldef\tempurl%
\url{https://doi.org/10.1023/A:1026543900054}
\showDOI{\tempurl}


\bibitem[\protect\citeauthoryear{Shang, Zhang, Liu, Li, and Han}{Shang
  et~al\mbox{.}}{2020}]%
        {shang2020nettaxo}
\bibfield{author}{\bibinfo{person}{Jingbo Shang}, \bibinfo{person}{Xinyang
  Zhang}, \bibinfo{person}{Liyuan Liu}, \bibinfo{person}{Sha Li}, {and}
  \bibinfo{person}{Jiawei Han}.} \bibinfo{year}{2020}\natexlab{}.
\newblock \showarticletitle{NetTaxo: Automated Topic Taxonomy Construction from
  Text-Rich Network}. In \bibinfo{booktitle}{\emph{Proceedings of The Web
  Conference (WWW)}}. \bibinfo{address}{Taipei, Taiwan}.
\newblock
\urldef\tempurl%
\url{https://doi.org/10.1145/3366423.3380259}
\showDOI{\tempurl}


\bibitem[\protect\citeauthoryear{Shen, Shen, Xiong, Wang, Wang, and Han}{Shen
  et~al\mbox{.}}{2020}]%
        {shen2020taxoexpan}
\bibfield{author}{\bibinfo{person}{Jiaming Shen}, \bibinfo{person}{Zhihong
  Shen}, \bibinfo{person}{Chenyan Xiong}, \bibinfo{person}{Chi Wang},
  \bibinfo{person}{Kuansan Wang}, {and} \bibinfo{person}{Jiawei Han}.}
  \bibinfo{year}{2020}\natexlab{}.
\newblock \showarticletitle{TaxoExpan: Self-supervised Taxonomy Expansion with
  Position-Enhanced Graph Neural Network}. In
  \bibinfo{booktitle}{\emph{Proceedings of The Web Conference (WWW)}}.
\newblock
\urldef\tempurl%
\url{https://doi.org/10.1145/3366423.3380132}
\showDOI{\tempurl}


\bibitem[\protect\citeauthoryear{Singh, D’Arcy, Cohan, Downey, and
  Feldman}{Singh et~al\mbox{.}}{2023}]%
        {singh2023scirepeval}
\bibfield{author}{\bibinfo{person}{Amanpreet Singh}, \bibinfo{person}{Mike
  D’Arcy}, \bibinfo{person}{Arman Cohan}, \bibinfo{person}{Doug Downey},
  {and} \bibinfo{person}{Sergey Feldman}.} \bibinfo{year}{2023}\natexlab{}.
\newblock \showarticletitle{{SciRepEval}: A Multi-Format Benchmark for
  Scientific Document Representations}. In
  \bibinfo{booktitle}{\emph{Proceedings of the 2023 Conference on Empirical
  Methods in Natural Language Processing}}. \bibinfo{pages}{5548--5566}.
\newblock


\bibitem[\protect\citeauthoryear{Stelmakh, Shah, and Singh}{Stelmakh
  et~al\mbox{.}}{2021}]%
        {stelmakh2021peerreview4all}
\bibfield{author}{\bibinfo{person}{Ivan Stelmakh}, \bibinfo{person}{Nihar~B.
  Shah}, {and} \bibinfo{person}{Aarti Singh}.} \bibinfo{year}{2021}\natexlab{}.
\newblock \showarticletitle{PeerReview4All: Fair and Accurate Reviewer
  Assignment in Peer Review}.
\newblock \bibinfo{journal}{\emph{Journal of Machine Learning Research}}
  \bibinfo{volume}{22}, \bibinfo{number}{163} (\bibinfo{year}{2021}),
  \bibinfo{pages}{1--66}.
\newblock
\urldef\tempurl%
\url{https://www.jmlr.org/papers/v22/20-190.html}
\showURL{%
\tempurl}


\bibitem[\protect\citeauthoryear{Subramanya, Devvrit, Kadekodi, Krishnaswamy,
  and Simhadri}{Subramanya et~al\mbox{.}}{2019}]%
        {subramanya2019diskann}
\bibfield{author}{\bibinfo{person}{Suhas~Jayaram Subramanya},
  \bibinfo{person}{Devvrit}, \bibinfo{person}{Rohan Kadekodi},
  \bibinfo{person}{Ravishankar Krishnaswamy}, {and}
  \bibinfo{person}{Harsha~Vardhan Simhadri}.} \bibinfo{year}{2019}\natexlab{}.
\newblock \showarticletitle{{DiskANN}: Fast Accurate Billion-Point Nearest
  Neighbor Search on a Single Node}. In \bibinfo{booktitle}{\emph{Advances in
  Neural Information Processing Systems 32 ({NeurIPS})}}.
  \bibinfo{pages}{13748--13758}.
\newblock


\bibitem[\protect\citeauthoryear{Sun, Zhang, Hu, Wang, Chen, Akrami, and
  Li}{Sun et~al\mbox{.}}{2020}]%
        {sun2020entityalignment}
\bibfield{author}{\bibinfo{person}{Zequn Sun}, \bibinfo{person}{Qingheng
  Zhang}, \bibinfo{person}{Wei Hu}, \bibinfo{person}{Chengming Wang},
  \bibinfo{person}{Muhao Chen}, \bibinfo{person}{Farahnaz Akrami}, {and}
  \bibinfo{person}{Chengkai Li}.} \bibinfo{year}{2020}\natexlab{}.
\newblock \showarticletitle{A Benchmarking Study of Embedding-based Entity
  Alignment for Knowledge Graphs}.
\newblock \bibinfo{journal}{\emph{Proceedings of the {VLDB} Endowment}}
  \bibinfo{volume}{13}, \bibinfo{number}{12} (\bibinfo{year}{2020}),
  \bibinfo{pages}{2326--2340}.
\newblock
\urldef\tempurl%
\url{https://doi.org/10.14778/3407790.3407828}
\showDOI{\tempurl}


\bibitem[\protect\citeauthoryear{Takezawa, Sato, and Yamada}{Takezawa
  et~al\mbox{.}}{2021}]%
        {takezawa2021supervised}
\bibfield{author}{\bibinfo{person}{Yuki Takezawa}, \bibinfo{person}{Ryoma
  Sato}, {and} \bibinfo{person}{Makoto Yamada}.}
  \bibinfo{year}{2021}\natexlab{}.
\newblock \showarticletitle{Supervised Tree-{W}asserstein Distance}. In
  \bibinfo{booktitle}{\emph{Proceedings of the 38th International Conference on
  Machine Learning (ICML)}} \emph{(\bibinfo{series}{Proceedings of Machine
  Learning Research})}, Vol.~\bibinfo{volume}{139}. \bibinfo{publisher}{PMLR},
  \bibinfo{pages}{10086--10095}.
\newblock


\bibitem[\protect\citeauthoryear{Tang, Tang, and Tan}{Tang
  et~al\mbox{.}}{2010}]%
        {tang2010expertise}
\bibfield{author}{\bibinfo{person}{Wenbin Tang}, \bibinfo{person}{Jie Tang},
  {and} \bibinfo{person}{Chenhao Tan}.} \bibinfo{year}{2010}\natexlab{}.
\newblock \showarticletitle{Expertise Matching via Constraint-Based
  Optimization}. In \bibinfo{booktitle}{\emph{IEEE/WIC/ACM International
  Conference on Web Intelligence and Intelligent Agent Technology}}.
  \bibinfo{pages}{34--41}.
\newblock


\bibitem[\protect\citeauthoryear{TaxoReview}{TaxoReview}{2026}]%
        {taxoreview_repo}
TaxoReview \bibinfo{year}{2026}\natexlab{}.
\newblock \bibinfo{title}{{TaxoReview}: Source Code and Technical Report}.
\newblock
  \bibinfo{howpublished}{\url{https://github.com/DataAutonomyLab/TaxoReview}}.
\newblock
\newblock
\shownote{Accessed: 2026-10-19.}


\bibitem[\protect\citeauthoryear{Vilnis, Li, Murty, and McCallum}{Vilnis
  et~al\mbox{.}}{2018}]%
        {Vilnis2018Box}
\bibfield{author}{\bibinfo{person}{Luke Vilnis}, \bibinfo{person}{Xiang Li},
  \bibinfo{person}{Shikhar Murty}, {and} \bibinfo{person}{Andrew McCallum}.}
  \bibinfo{year}{2018}\natexlab{}.
\newblock \showarticletitle{Probabilistic Embedding of Knowledge Graphs with
  Box Lattice Measures}. In \bibinfo{booktitle}{\emph{Proceedings of the 56th
  Annual Meeting of the Association for Computational Linguistics (ACL)}}.
  \bibinfo{pages}{263--272}.
\newblock
\urldef\tempurl%
\url{https://doi.org/10.18653/v1/P18-1025}
\showDOI{\tempurl}


\bibitem[\protect\citeauthoryear{Wang, Xu, Yue, and Wang}{Wang
  et~al\mbox{.}}{2021}]%
        {wang2021ann}
\bibfield{author}{\bibinfo{person}{Mengzhao Wang}, \bibinfo{person}{Xiaoliang
  Xu}, \bibinfo{person}{Qiang Yue}, {and} \bibinfo{person}{Yuxiang Wang}.}
  \bibinfo{year}{2021}\natexlab{}.
\newblock \showarticletitle{A Comprehensive Survey and Experimental Comparison
  of Graph-Based Approximate Nearest Neighbor Search}.
\newblock \bibinfo{journal}{\emph{Proceedings of the {VLDB} Endowment}}
  \bibinfo{volume}{14}, \bibinfo{number}{11} (\bibinfo{year}{2021}),
  \bibinfo{pages}{1964--1978}.
\newblock
\urldef\tempurl%
\url{https://doi.org/10.14778/3476249.3476255}
\showDOI{\tempurl}


\bibitem[\protect\citeauthoryear{Weeds, Weir, and McCarthy}{Weeds
  et~al\mbox{.}}{2004}]%
        {weeds2004characterising}
\bibfield{author}{\bibinfo{person}{Julie Weeds}, \bibinfo{person}{David Weir},
  {and} \bibinfo{person}{Diana McCarthy}.} \bibinfo{year}{2004}\natexlab{}.
\newblock \showarticletitle{Characterising Measures of Lexical Distributional
  Similarity}. In \bibinfo{booktitle}{\emph{Proceedings of the 20th
  International Conference on Computational Linguistics ({COLING} 2004)}}.
  \bibinfo{publisher}{COLING}, \bibinfo{address}{Geneva, Switzerland},
  \bibinfo{pages}{1015--1021}.
\newblock
\urldef\tempurl%
\url{https://aclanthology.org/C04-1146/}
\showURL{%
\tempurl}


\bibitem[\protect\citeauthoryear{Wright}{Wright}{1997}]%
        {wright1997primaldual}
\bibfield{author}{\bibinfo{person}{Stephen~J. Wright}.}
  \bibinfo{year}{1997}\natexlab{}.
\newblock \bibinfo{booktitle}{\emph{Primal-Dual Interior-Point Methods}}.
\newblock \bibinfo{publisher}{Society for Industrial and Applied Mathematics
  (SIAM)}, \bibinfo{address}{Philadelphia, PA}.
\newblock
\urldef\tempurl%
\url{https://doi.org/10.1137/1.9781611971453}
\showDOI{\tempurl}


\bibitem[\protect\citeauthoryear{Wu, U, and Karras}{Wu et~al\mbox{.}}{2023}]%
        {wu2023kbest}
\bibfield{author}{\bibinfo{person}{Siyuan Wu}, \bibinfo{person}{Leong~Hou U},
  {and} \bibinfo{person}{Panagiotis Karras}.} \bibinfo{year}{2023}\natexlab{}.
\newblock \showarticletitle{k-Best Egalitarian Stable Marriages for Task
  Assignment}.
\newblock \bibinfo{journal}{\emph{Proceedings of the {VLDB} Endowment}}
  \bibinfo{volume}{16}, \bibinfo{number}{11} (\bibinfo{year}{2023}),
  \bibinfo{pages}{3240--3252}.
\newblock
\urldef\tempurl%
\url{https://doi.org/10.14778/3611479.3611522}
\showDOI{\tempurl}


\bibitem[\protect\citeauthoryear{Xue, Shen, Ren, Guo, Pu, and Lu}{Xue
  et~al\mbox{.}}{2024}]%
        {Xue2024TaxBox}
\bibfield{author}{\bibinfo{person}{Wei Xue}, \bibinfo{person}{Yongliang Shen},
  \bibinfo{person}{Wenqi Ren}, \bibinfo{person}{Jietian Guo},
  \bibinfo{person}{Shiliang Pu}, {and} \bibinfo{person}{Weiming Lu}.}
  \bibinfo{year}{2024}\natexlab{}.
\newblock \showarticletitle{Insert or Attach: Taxonomy Completion via Box
  Embedding}. In \bibinfo{booktitle}{\emph{Proceedings of the 62nd Annual
  Meeting of the Association for Computational Linguistics (ACL)}}.
  \bibinfo{pages}{3851--3863}.
\newblock
\urldef\tempurl%
\url{https://doi.org/10.18653/v1/2024.acl-long.212}
\showDOI{\tempurl}


\bibitem[\protect\citeauthoryear{Yu and Cheng}{Yu and Cheng}{2010}]%
        {yu2010reachability}
\bibfield{author}{\bibinfo{person}{Jeffrey~Xu Yu} {and}
  \bibinfo{person}{Jiefeng Cheng}.} \bibinfo{year}{2010}\natexlab{}.
\newblock \showarticletitle{Graph Reachability Queries: A Survey}.
\newblock \bibinfo{journal}{\emph{Managing and Mining Graph Data}}
  (\bibinfo{year}{2010}), \bibinfo{pages}{181--215}.
\newblock
\urldef\tempurl%
\url{https://doi.org/10.1007/978-1-4419-6045-0\_6}
\showDOI{\tempurl}


\bibitem[\protect\citeauthoryear{Zeng, Bai, Tan, Feng, Liang, Zhang, and
  Jiang}{Zeng et~al\mbox{.}}{2024}]%
        {zeng2024chainoflayer}
\bibfield{author}{\bibinfo{person}{Qingkai Zeng}, \bibinfo{person}{Yuyang Bai},
  \bibinfo{person}{Zhaoxuan Tan}, \bibinfo{person}{Shangbin Feng},
  \bibinfo{person}{Zhenwen Liang}, \bibinfo{person}{Zhihan Zhang}, {and}
  \bibinfo{person}{Meng Jiang}.} \bibinfo{year}{2024}\natexlab{}.
\newblock \showarticletitle{Chain-of-Layer: Iteratively Prompting Large
  Language Models for Taxonomy Induction from Limited Examples}. In
  \bibinfo{booktitle}{\emph{Proceedings of the 33rd ACM International
  Conference on Information and Knowledge Management (CIKM)}}.
  \bibinfo{pages}{3093--3102}.
\newblock
\urldef\tempurl%
\url{https://doi.org/10.1145/3627673.3679608}
\showDOI{\tempurl}


\bibitem[\protect\citeauthoryear{Zhang, Tao, Chen, Shen, Jiang, Sadler, Vanni,
  and Han}{Zhang et~al\mbox{.}}{2018}]%
        {zhang2018taxogen}
\bibfield{author}{\bibinfo{person}{Chao Zhang}, \bibinfo{person}{Fangbo Tao},
  \bibinfo{person}{Xiusi Chen}, \bibinfo{person}{Jiaming Shen},
  \bibinfo{person}{Meng Jiang}, \bibinfo{person}{Brian Sadler},
  \bibinfo{person}{Michelle Vanni}, {and} \bibinfo{person}{Jiawei Han}.}
  \bibinfo{year}{2018}\natexlab{}.
\newblock \showarticletitle{TaxoGen: Unsupervised Topic Taxonomy Construction
  by Adaptive Term Embedding and Clustering}. In
  \bibinfo{booktitle}{\emph{Proceedings of the 24th ACM SIGKDD International
  Conference on Knowledge Discovery \& Data Mining (KDD)}}.
  \bibinfo{address}{London, UK}, \bibinfo{pages}{2701--2709}.
\newblock
\urldef\tempurl%
\url{https://doi.org/10.1145/3219819.3220064}
\showDOI{\tempurl}


\bibitem[\protect\citeauthoryear{Zhang and Ives}{Zhang and Ives}{2020}]%
        {zhang2020juneau}
\bibfield{author}{\bibinfo{person}{Yi Zhang} {and} \bibinfo{person}{Zachary~G.
  Ives}.} \bibinfo{year}{2020}\natexlab{}.
\newblock \showarticletitle{Finding Related Tables in Data Lakes for
  Interactive Data Science}. In \bibinfo{booktitle}{\emph{Proceedings of the
  2020 ACM SIGMOD International Conference on Management of Data}}.
  \bibinfo{pages}{1951--1966}.
\newblock
\urldef\tempurl%
\url{https://doi.org/10.1145/3318464.3389726}
\showDOI{\tempurl}


\bibitem[\protect\citeauthoryear{Zhang, Shen, Kang, Chen, Jin, and Han}{Zhang
  et~al\mbox{.}}{2025}]%
        {zhang2025chainoffactors}
\bibfield{author}{\bibinfo{person}{Yu Zhang}, \bibinfo{person}{Yanzhen Shen},
  \bibinfo{person}{SeongKu Kang}, \bibinfo{person}{Xiusi Chen},
  \bibinfo{person}{Bowen Jin}, {and} \bibinfo{person}{Jiawei Han}.}
  \bibinfo{year}{2025}\natexlab{}.
\newblock \showarticletitle{Chain-of-Factors Paper-Reviewer Matching}. In
  \bibinfo{booktitle}{\emph{Proceedings of The ACM Web Conference 2025 (WWW
  '25)}}. \bibinfo{publisher}{ACM}, \bibinfo{pages}{1901--1910}.
\newblock
\urldef\tempurl%
\url{https://doi.org/10.1145/3696410.3714708}
\showDOI{\tempurl}


\bibitem[\protect\citeauthoryear{Zhu, Deng, Nargesian, and Miller}{Zhu
  et~al\mbox{.}}{2019}]%
        {zhu2019josie}
\bibfield{author}{\bibinfo{person}{Erkang Zhu}, \bibinfo{person}{Dong Deng},
  \bibinfo{person}{Fatemeh Nargesian}, {and} \bibinfo{person}{Ren{\'e}e~J.
  Miller}.} \bibinfo{year}{2019}\natexlab{}.
\newblock \showarticletitle{{JOSIE}: Overlap Set Similarity Search for Finding
  Joinable Tables in Data Lakes}. In \bibinfo{booktitle}{\emph{Proceedings of
  the 2019 International Conference on Management of Data ({SIGMOD})}}.
  \bibinfo{pages}{847--864}.
\newblock
\urldef\tempurl%
\url{https://doi.org/10.1145/3299869.3300065}
\showDOI{\tempurl}


\bibitem[\protect\citeauthoryear{Zhu, Nargesian, Pu, and Miller}{Zhu
  et~al\mbox{.}}{2016}]%
        {zhu2016lshensemble}
\bibfield{author}{\bibinfo{person}{Erkang Zhu}, \bibinfo{person}{Fatemeh
  Nargesian}, \bibinfo{person}{Ken~Q. Pu}, {and} \bibinfo{person}{Ren{\'e}e~J.
  Miller}.} \bibinfo{year}{2016}\natexlab{}.
\newblock \showarticletitle{{LSH} Ensemble: Internet-Scale Domain Search}.
\newblock \bibinfo{journal}{\emph{Proceedings of the {VLDB} Endowment}}
  \bibinfo{volume}{9}, \bibinfo{number}{12} (\bibinfo{year}{2016}),
  \bibinfo{pages}{1185--1196}.
\newblock
\urldef\tempurl%
\url{https://doi.org/10.14778/2994509.2994534}
\showDOI{\tempurl}


\bibitem[\protect\citeauthoryear{Zhu, Liao, Gu, Huang, Feng, and Qin}{Zhu
  et~al\mbox{.}}{2025}]%
        {zhu2025context}
\bibfield{author}{\bibinfo{person}{Kun Zhu}, \bibinfo{person}{Lizi Liao},
  \bibinfo{person}{Yuxuan Gu}, \bibinfo{person}{Lei Huang},
  \bibinfo{person}{Xiaocheng Feng}, {and} \bibinfo{person}{Bing Qin}.}
  \bibinfo{year}{2025}\natexlab{}.
\newblock \showarticletitle{Context-Aware Hierarchical Taxonomy Generation for
  Scientific Papers via LLM-Guided Multi-Aspect Clustering}. In
  \bibinfo{booktitle}{\emph{Proceedings of the 2025 Conference on Empirical
  Methods in Natural Language Processing}}. \bibinfo{pages}{15616--15634}.
\newblock
\urldef\tempurl%
\url{https://doi.org/10.18653/v1/2025.emnlp-main.788}
\showDOI{\tempurl}


\bibitem[\protect\citeauthoryear{Zobel and Moffat}{Zobel and Moffat}{2006}]%
        {zobel2006inverted}
\bibfield{author}{\bibinfo{person}{Justin Zobel} {and}
  \bibinfo{person}{Alistair Moffat}.} \bibinfo{year}{2006}\natexlab{}.
\newblock \showarticletitle{Inverted Files for Text Search Engines}.
\newblock \bibinfo{journal}{\emph{Comput. Surveys}} \bibinfo{volume}{38},
  \bibinfo{number}{2} (\bibinfo{year}{2006}), \bibinfo{pages}{6:1--6:56}.
\newblock
\urldef\tempurl%
\url{https://doi.org/10.1145/1132956.1132959}
\showDOI{\tempurl}


\end{thebibliography}

\iftechreport
\clearpage
\appendix

\section*{Appendix}

This appendix appears only in the technical report version of the paper and is organized as follows.

\begin{itemize}[leftmargin=*]
\item Appendix~\ref{app:notations} summarizes the key notation used throughout the paper.
\item Appendix~\ref{app:proofs} presents the proofs of the theoretical results.
\item Appendix~\ref{app:details} provides method details omitted from the main text.
\item Appendix~\ref{app:additional_exp} reports evaluation details and experimental results deferred from Section~\ref{sec:exp}.
\item Appendix~\ref{app:related_work} discusses further related work.
\end{itemize}

\section{Notation Table}\label{app:notations}

Table~\ref{tab:notation} summarizes the main notation used throughout Sections~\ref{sec:problem}--\ref{sec:completion}, grouped by the component that introduces each symbol.

\begin{table}[ht]
\centering
\caption{Summary of key notation used in this work.}
\label{tab:notation}
\small
\setlength{\tabcolsep}{6pt}
\renewcommand{\arraystretch}{1.1}
\begin{tabular}{@{}l p{0.68\linewidth}@{}}
\toprule
\textbf{Symbol} & \textbf{Description} \\
\midrule
\multicolumn{2}{@{}l@{}}{\textit{Problem Definition (Section~\ref{sec:problem})}} \\
$\mathcal{P}$, $\mathcal{R}_P$ & Submission corpus and committed review panel. \\
$\mathcal{R}_W$ & Expert lake of external candidate reviewers. \\
$\mathcal{H}_r$ & Publication history of reviewer or expert $r$. \\
$f(p, r)$ & Affinity between submission $p$ and reviewer $r$ (higher is better). \\
$x_{p,r}$ & Indicator that reviewer $r$ is assigned to submission $p$. \\
$K$, $L_r$ & Reviewers per submission, and maximum load of reviewer $r$. \\
$\mathcal{C}$ & Conflict-of-interest pairs excluded from the assignment. \\
$B$, $\mathcal{R}_A$ & Recruitment budget, and augmentation set $\mathcal{R}_A \subseteq \mathcal{R}_W$. \\
\midrule
\multicolumn{2}{@{}l@{}}{\textit{Local Taxonomy Builder (Section~\ref{sec:taxonomy})}} \\
$\mathcal{T}_0$ & Unweighted backbone induced over $\mathcal{D}_{\mathrm{seed}} = \mathcal{P} \cup \mathcal{E}^{\star}$. \\
$\mathcal{T} = (V, E)$ & Calibrated conference-specific local taxonomy. \\
$\mathcal{L}$ & Leaf set of the local taxonomy. \\
$\phi_p$, $\phi_r$ & Leaf distributions of submission $p$ and reviewer $r$ over $\mathcal{L}$. \\
$\mathbf{w} = (w_e)_{e \in E}$ & Calibrated edge weights on $\mathcal{T}$. \\
$S_e$, $m_x(S_e)$ & Leaves of the subtree below edge $e$, and the mass that $x$ places in $S_e$. \\
$d_{\mathcal{T}}(p, r)$ & Tree--Wasserstein distance on $\mathcal{T}$ (Eq.~\eqref{eq:tree-distance}). \\
$\mathcal{J}(\mathbf{w})$ & Edge-weight learning objective over sampled submission pairs $\mathcal{S}$ (Eq.~\eqref{eq:edge_weight_obj}). \\
\midrule
\multicolumn{2}{@{}l@{}}{\textit{Reviewer Evidence Selection (Section~\ref{sec:evidence})}} \\
$\mathcal{U}$, $\mathcal{U}_r$ & Candidate evidence pool of the panel, and its part attributed to reviewer $r$. \\
$\mathcal{E}_r^{\star}$ & Selected evidence subset of reviewer $r$, with $\mathcal{E}_r^{\star} \subseteq \mathcal{H}_r$. \\
$B_r$ & Per-reviewer evidence budget. \\
$\gamma_q \in [0, 1]$ & Representativeness factor of candidate paper $q$. \\
$s(p, q)$ & Quality-weighted proximity of candidate $q$ to submission $p$. \\
$F(x)$ & Submodular evidence-selection objective (Eq.~\eqref{eq:evidence}). \\
\midrule
\multicolumn{2}{@{}l@{}}{\textit{Global Taxonomy Index (Section~\ref{sec:index})}} \\
$\mathcal{A} = \{a_1, \ldots, a_L\}$ & Global vocabulary of canonical expertise atoms. \\
$\mathcal{T}_G = (V_G, E_G)$ & Global atom DAG, where an edge $a_i \to a_j$ means that $a_j$ generalizes $a_i$. \\
$z_{ij}$ & Ancestry indicator, equal to $1$ iff $a_j$ is reachable from $a_i$ in $\mathcal{T}_G$. \\
$\mathbf{w}_r$, $\mathbf{w}_r^{\uparrow}$ & Base and upward-propagated atom profiles of expert $r$. \\
$\alpha_{r, q}$ & Bibliometric scaling factor (recency, authorship, venue). \\
\midrule
\multicolumn{2}{@{}l@{}}{\textit{Global--Local Taxonomy Aligner (Section~\ref{sec:completion})}} \\
$\mathcal{A}_v^{\mathrm{cand}}$ & Top-$k_c$ candidate atoms retrieved for conference node $v$. \\
$\mathcal{A}^{\star}$ & Anchor set selected by the hierarchy-preserving ILP (Eq.~\eqref{eq:map-ilp}). \\
$\mathcal{T}_G^{\star} = (V^{\star}, E^{\star}, \mathbf{w}^{\star})$ & Calibrated, compressed projection of $\mathcal{T}_G$ aligned to $\mathcal{T}$. \\
$\psi_p$, $\psi_r$ & Distributions of $p$ and $r$ over the nodes of $\mathcal{T}_G^{\star}$ (normalized direct atom weights). \\
$\mathbf{h}_{pr}$ & Activation vector, $[\mathbf{h}_{pr}]_e = \big|m_{\psi_p}(S_e) - m_{\psi_r}(S_e)\big|$. \\
$\mathbf{w}^{\star}$ & Calibrated edge weights on $\mathcal{T}_G^{\star}$ (Eq.~\eqref{eq:calibration-lp}). \\
\bottomrule
\end{tabular}
\end{table}

\section{Proofs}
\label{app:proofs}

\subsection{Hardness of Panel Expansion}
\label{app:np-hard-proof}

\begin{theorem}[Hardness of Panel Expansion]
\label{thm:np-hard}
The panel expansion problem in Definition~\ref{def:completion} is NP-hard.
\end{theorem}

\begin{proof}
We provide a formal reduction from Maximum Coverage to the decision version of Definition~\ref{def:completion}, which asks whether some augmentation set $\mathcal{R}_A \subseteq \mathcal{R}_{W}$ with $|\mathcal{R}_A| \leq B$ admits an assignment over $\mathcal{R}_P \cup \mathcal{R}_A$ that satisfies the constraints of Equation~\eqref{eq:assignment} and attains an objective value of at least a target $T$.

\noindent\textbf{Maximum Coverage (decision version).}
Given a universe $U = \{u_1, \ldots, u_n\}$, a collection of subsets $\mathcal{S} = \{S_1, \ldots, S_m\}$ with $S_i \subseteq U$, a budget $B \in \mathbb{Z}_{>0}$, and a target $T \in \mathbb{Z}_{>0}$, decide whether there exists $\mathcal{S}' \subseteq \mathcal{S}$ with $|\mathcal{S}'| \leq B$ such that $|\bigcup_{S \in \mathcal{S}'} S| \geq T$.
Maximum Coverage is known to be NP-complete~\cite{hochbaum1997approximation}.

\noindent\textbf{Reduction.}
Given a Maximum Coverage instance $(U, \mathcal{S}, B, T)$, construct a panel expansion instance as follows:
\begin{itemize}[leftmargin=*,noitemsep]
  \item Submission corpus: $\mathcal{P} = \{p_u : u \in U\}$, with $|\mathcal{P}| = n$.
  \item Initial panel: $\mathcal{R}_P = \emptyset$.
  \item Expert lake: $\mathcal{R}_{W} = \{r_i : i = 1, \ldots, m\}$, where each expert $r_i$ corresponds to subset $S_i$.
  \item Affinity function: $f(p_u, r_i) = 1$ if $u \in S_i$, and $f(p_u, r_i) = 0$ otherwise.
  \item Assignment constraints: $K = 1$ reviewer per submission, load capacity $L_r = n$ for all $r$, empty conflict set $\mathcal{C} = \emptyset$.
  \item Recruitment budget and target: the same $B$ and $T$.
\end{itemize}

\noindent\textbf{Equivalence.}
Under this construction, given any non-empty augmentation set $\mathcal{R}_A \subseteq \mathcal{R}_{W}$ with $|\mathcal{R}_A| \leq B$, the optimal assignment over $\mathcal{R}_A$ assigns each submission $p_u$ to any expert $r_i \in \mathcal{R}_A$ with $u \in S_i$ (yielding score $1$) if such an expert exists, and to an arbitrary expert yielding score $0$ otherwise.
The global assignment objective therefore equals the number of submissions covered:
\[
\sum_{p \in \mathcal{P}} \sum_{r \in \mathcal{R}_A} f(p, r) \cdot x_{p,r} = \left| \bigcup_{r_i \in \mathcal{R}_A} S_i \right|.
\]
An empty $\mathcal{R}_A$ admits no feasible assignment, and the corresponding empty collection covers no element, so both instances reject it because $T \geq 1$.
Consequently, the panel expansion objective achieves value $\geq T$ if and only if the corresponding Maximum Coverage instance admits a solution covering $\geq T$ elements.

The reduction is polynomial: the constructed instance has $|\mathcal{P}| = n$, $|\mathcal{R}_{W}| = m$, and the affinity matrix has $O(nm)$ entries.
Hence Maximum Coverage reduces in polynomial time to the decision version of Definition~\ref{def:completion}, establishing NP-hardness.
\end{proof}

\subsection{Properties of the Evidence Selection Problem}
\label{app:proof-evidence}

This subsection establishes two properties of the evidence selection problem in Equation~\eqref{eq:evidence}: its objective is monotone submodular, and the greedy rule of Equation~\eqref{eq:margin_gain} consequently attains a $1/2$-approximation of the optimum under the per-reviewer budget constraints.
Throughout, we identify an indicator vector $x \in \{0,1\}^{|\mathcal{U}|}$ with the set $\mathcal{E} = \{q \in \mathcal{U} : x_q = 1\}$ of selected candidates, and we write $\prox(p; \mathcal{E})$ and $F(\mathcal{E})$ for $\prox(p; x)$ and $F(x)$.

\begin{proposition}[Submodularity of the Evidence Selection Objective]
\label{prop:submod}
With the concave aggregator $g(t) = \log(1+t)$ and non-negative quality scores $s(p,q) \ge 0$, the evidence selection objective $F(\mathcal{E}) = \sum_{p \in \mathcal{P}} g\big(\prox(p; \mathcal{E})\big)$ in Equation~\eqref{eq:evidence} is a normalized, monotone, submodular set function over the candidate ground set $\mathcal{U}$.
\end{proposition}

\begin{proof}
Fix a submission $p \in \mathcal{P}$ and let $h_p(\mathcal{E}) = g\big(\prox(p; \mathcal{E})\big)$, where $\prox(p; \mathcal{E}) = \sum_{q \in \mathcal{E}} s(p, q)$ by Equation~\eqref{eq:cum-prox}.
Because the scores are non-negative, $\prox(p; \cdot)$ is a non-negative, monotone, and modular set function.

Consider sets $\mathcal{E} \subseteq \mathcal{E}' \subseteq \mathcal{U}$ and a candidate $q \in \mathcal{U} \setminus \mathcal{E}'$, and let $t = \prox(p; \mathcal{E})$, $t' = \prox(p; \mathcal{E}')$, and $\delta = s(p, q)$, so that $0 \le t \le t'$ and $\delta \ge 0$.
Since $g$ is increasing, $h_p(\mathcal{E} \cup \{q\}) - h_p(\mathcal{E}) = g(t + \delta) - g(t) \ge 0$, so $h_p$ is monotone.
Since $g$ is concave, its increment over an interval of fixed length $\delta$ cannot grow as the interval moves to the right,
\begin{equation*}
  g(t + \delta) - g(t) \;\ge\; g(t' + \delta) - g(t'),
\end{equation*}
which is exactly the diminishing-returns condition $h_p(\mathcal{E} \cup \{q\}) - h_p(\mathcal{E}) \ge h_p(\mathcal{E}' \cup \{q\}) - h_p(\mathcal{E}')$. Hence $h_p$ is submodular.

Monotone submodular functions are closed under non-negative linear combinations, so the global objective $F = \sum_{p \in \mathcal{P}} h_p$ is monotone submodular as well. Finally, $F(\emptyset) = \sum_{p \in \mathcal{P}} \log(1+0) = 0$, so $F$ is normalized.
\end{proof}

\begin{proposition}[Greedy $1/2$-Approximation]
\label{prop:approx}
Let $\mathcal{E}^{\mathrm{opt}}$ denote an optimal solution to the evidence selection problem under the per-reviewer budget constraints $|\mathcal{E} \cap \mathcal{U}_r| \le B_r$ for all $r \in \mathcal{R}_P$. The greedy rule of Equation~\eqref{eq:margin_gain} returns a set $\mathcal{E}^\star$ satisfying $F(\mathcal{E}^\star) \ge \tfrac{1}{2}\, F(\mathcal{E}^{\mathrm{opt}})$.
\end{proposition}

\begin{proof}
Recall that $\mathcal{U}_r = \{q \in \mathcal{U} \mid r(q) = r\}$ collects the candidates attributed to reviewer $r$. The candidate pool $\mathcal{U}$ is a disjoint union in which a publication co-authored by several panelists enters once per panelist (Section~\ref{sec:evidence}), so every candidate $q \in \mathcal{U}$ is attributed to exactly one reviewer $r(q)$, and the sets $\{\mathcal{U}_r\}_{r \in \mathcal{R}_P}$ form a disjoint partition of the ground set $\mathcal{U}$.

The constraints restrict feasible subsets $\mathcal{E} \subseteq \mathcal{U}$ to those satisfying $|\mathcal{E} \cap \mathcal{U}_r| \le B_r$ for all $r \in \mathcal{R}_P$. This structure defines the independent sets $\mathcal{I}$ of a \emph{partition matroid} $\mathcal{M} = (\mathcal{U}, \mathcal{I})$~\cite{oxley2011matroid}.

Our goal is therefore to maximize $F(\mathcal{E})$ subject to $\mathcal{E} \in \mathcal{I}$, where $F$ is normalized, monotone, and submodular by Proposition~\ref{prop:submod}.
The rule in Equation~\eqref{eq:margin_gain} is the standard discrete greedy algorithm, which starts from $\mathcal{E} = \emptyset$ and iteratively adds the candidate with the largest marginal gain among those that keep $\mathcal{E}$ independent. A classical theorem by Fisher, Nemhauser, and Wolsey~\cite{fisher1978analysis} proves that this algorithm, applied to a normalized, monotone submodular function under a single matroid constraint, produces a set $\mathcal{E}^\star$ such that
\begin{equation*}
    F(\mathcal{E}^\star) \ge \frac{1}{2} F(\mathcal{E}^{\mathrm{opt}})\,,
\end{equation*}
where $\mathcal{E}^{\mathrm{opt}}$ is an optimal feasible subset.
\end{proof}

Continuous greedy algorithms combined with pipage rounding attain the tight $(1 - 1/e)$-approximation for matroid-constrained submodular maximization~\cite{calinescu2011maximizing}. We use the discrete greedy algorithm instead because it needs only marginal-gain evaluations, whereas the continuous algorithm optimizes a sampled multilinear relaxation. The lazy-greedy implementation of Appendix~\ref{app:impl} makes the same choices as the plain greedy algorithm up to tie-breaking, so the guarantee of Proposition~\ref{prop:approx} also holds for it.

\subsection{Strict Convexity of the Edge-Weight Learning Objective}
\label{app:proof-edge-weight}

\begin{proposition}[Strict Convexity and Uniqueness]
\label{prop:edge-weight-convex}
The edge-weight learning objective $\mathcal{J}(\mathbf{w})$ in Equation~\eqref{eq:edge_weight_obj}, with regularization coefficient $\lambda > 0$, is strictly convex in $\mathbf{w}$. Consequently, minimizing $\mathcal{J}(\mathbf{w})$ over the non-negative orthant $\mathbf{w} \ge \mathbf{0}$ admits a unique global minimum.
\end{proposition}

\begin{proof}
To prove that the objective function $\mathcal{J}(\mathbf{w})$ in Equation~\eqref{eq:edge_weight_obj} is strictly convex, we analyze its Hessian matrix with respect to the edge weight vector $\mathbf{w} \in \mathbb{R}^{|E|}$.

The objective function is defined as
\begin{equation*}
    \mathcal{J}(\mathbf{w}) = \frac{1}{|\mathcal{S}|} \sum_{(p_i, p_j) \in \mathcal{S}} \Big( \mathbf{a}_{ij}^\top \mathbf{w} - \hat{d}_{ij} \Big)^2 + \lambda \|\mathbf{w}\|_2^2\,.
\end{equation*}

The first derivative (gradient) of $\mathcal{J}(\mathbf{w})$ is
\begin{equation*}
    \nabla \mathcal{J}(\mathbf{w}) = \frac{2}{|\mathcal{S}|} \sum_{(p_i, p_j) \in \mathcal{S}} \mathbf{a}_{ij} \Big( \mathbf{a}_{ij}^\top \mathbf{w} - \hat{d}_{ij} \Big) + 2\lambda \mathbf{w}\,.
\end{equation*}

The second derivative (Hessian matrix) $\mathbf{H} \in \mathbb{R}^{|E| \times |E|}$ is
\begin{equation*}
    \mathbf{H} = \nabla^2 \mathcal{J}(\mathbf{w}) = \frac{2}{|\mathcal{S}|} \sum_{(p_i, p_j) \in \mathcal{S}} \mathbf{a}_{ij} \mathbf{a}_{ij}^\top + 2\lambda \mathbf{I}\,,
\end{equation*}
where $\mathbf{I}$ is the identity matrix.

For any non-zero vector $\mathbf{v} \in \mathbb{R}^{|E|}$, we evaluate the quadratic form
\begin{equation*}
    \mathbf{v}^\top \mathbf{H} \mathbf{v} = \frac{2}{|\mathcal{S}|} \sum_{(p_i, p_j) \in \mathcal{S}} \mathbf{v}^\top (\mathbf{a}_{ij} \mathbf{a}_{ij}^\top) \mathbf{v} + 2\lambda \mathbf{v}^\top \mathbf{I} \mathbf{v}\,.
\end{equation*}

Notice that $\mathbf{v}^\top (\mathbf{a}_{ij} \mathbf{a}_{ij}^\top) \mathbf{v} = (\mathbf{a}_{ij}^\top \mathbf{v})^2 \ge 0$. Therefore, the first term is non-negative.
Since the regularization coefficient satisfies $\lambda > 0$ and $\mathbf{v} \neq \mathbf{0}$, the second term is strictly positive:
\begin{equation*}
    2\lambda \mathbf{v}^\top \mathbf{I} \mathbf{v} = 2\lambda \|\mathbf{v}\|_2^2 > 0\,.
\end{equation*}

Consequently, $\mathbf{v}^\top \mathbf{H} \mathbf{v} > 0$ for all $\mathbf{v} \neq \mathbf{0}$, which implies that the Hessian matrix $\mathbf{H}$ is positive definite ($\mathbf{H} \succ 0$).

Because the Hessian of $\mathcal{J}(\mathbf{w})$ is positive definite everywhere on its domain, the objective function is strictly convex. The feasible region $\mathbf{w} \ge \mathbf{0}$ (the non-negative orthant) is closed and convex. Since $\mathcal{J}(\mathbf{w}) \ge \lambda \|\mathbf{w}\|_2^2$, the objective is coercive, so it attains its minimum over this region. Strict convexity rules out a second minimizer, because the midpoint of two distinct minimizers is feasible and would attain a strictly smaller value. The global minimum is therefore unique.
\end{proof}

\section{Method Details}
\label{app:details}

\subsection{Expert Lake Construction}
\label{app:lake}

The expert lake $\mathcal{R}_{W}$ is assembled from two complementary open scholarly sources. The \emph{bibliographic backbone} is obtained from DBLP~\cite{ley2002dblp}, from which author records and their publication histories are crawled. Researchers with fewer than five first-authored publications are filtered out, removing transient authors and leaving a pool of researchers with a demonstrated track record in their respective subfields. Since DBLP records carry metadata (title, authors, venue, year) but lack abstracts, publication-level textual content is subsequently enriched by cross-referencing each retained publication against OpenAlex~\cite{priem2022openalex} and retrieving its abstract via the OpenAlex Works API. Records for which no abstract can be resolved on OpenAlex are dropped, as the downstream keyphrase extraction and canonical atom construction pipeline (Section~\ref{sec:atom-extraction}) requires an abstract to produce reliable expertise atoms.
The resulting expert lake comprises approximately $240{,}000$ researchers and $1{,}300{,}000$ publications with resolved abstracts.

To characterize the lake at the granularity relevant to atom construction, we report the following statistics. Each expert's publication history contains on average 20.2 papers (median 17.5), where papers are counted once per authoring expert. Since publications average 3.7 co-authors, this is consistent with the 1.3M distinct publications in the lake. Keyphrase extraction yields 3.7 atoms per publication on average, and after canonicalization each expert's profile contains 7.5 distinct atoms on average, i.e., roughly a tenfold compression relative to the $\sim$75 raw atom mentions per expert. This compression reflects that the same expertise concept recurs across an expert's papers and is merged into a single canonical atom, empirically confirming the shareability of keyphrase atoms that motivates their use as the vocabulary of the global taxonomy (Section~\ref{sec:atom-extraction}).

\subsection{Bibliometric Scaling Factor \texorpdfstring{$\alpha_{r,q}$}{alpha(r,q)}}
\label{app:scaling-factor}

The scaling factor $\alpha_{r,q} = \exp(-\lambda_t \Delta t_q) \cdot \omega_{\mathrm{auth}}(r, q) \cdot \omega_{\mathrm{venue}}(q)$ introduced in Section~\ref{sec:atom-extraction} combines three multiplicative components. This subsection specifies each component, the data sources used to populate it, and the default constants employed in our experiments.

\noindent\textbf{Recency term $\exp(-\lambda_t \Delta t_q)$.} Here $\Delta t_q$ denotes the age of publication $q$ in years, and $\lambda_t$ is a decay rate chosen so that the half-life $\ln(2)/\lambda_t$ matches the typical turnover horizon of a computer science subfield. We set $\lambda_t = \ln(2)/4 \approx 0.173$, corresponding to a four-year half-life. A paper published eight years before the query therefore contributes one-quarter of its nominal weight, and a twelve-year-old paper contributes one-eighth.

\noindent\textbf{Authorship term $\omega_{\mathrm{auth}}(r, q)$.} Let $k_r \in \{1, \ldots, n_q\}$ denote the position of expert $r$ in the author list of $q$, where $n_q$ is the total number of authors of $q$. The baseline form is a harmonic weighting~\cite{hodge1981credit,hagen2008harmonic},
\begin{equation*}
  \omega_{\mathrm{auth}}^{\mathrm{harm}}(r, q) = \frac{1/k_r}{H_{n_q}},
  \qquad H_{n_q} = \sum_{j=1}^{n_q} \frac{1}{j},
\end{equation*}
which decays smoothly with position and sums to unity across co-authors. When corresponding-author metadata is available for $q$ (e.g., from the \texttt{is\_corresponding} field in OpenAlex or publisher-supplied CrossRef records), we promote flagged authors to full weight:
\begin{equation*}
  \omega_{\mathrm{auth}}(r, q) = \begin{cases} 1 & \text{if } r \text{ is a corresponding author of } q, \\ \omega_{\mathrm{auth}}^{\mathrm{harm}}(r, q) & \text{otherwise.} \end{cases}
\end{equation*}
When corresponding-author metadata is missing, the harmonic form is used for all co-authors. This falls back gracefully for conference papers, whose corresponding-author metadata is often absent in bibliographic sources, and captures the explicit accountability signal when it is present.

The harmonic fallback assumes the contribution-ordered authorship convention prevalent in most of computer science, and its scope should be stated explicitly. In communities that order authors alphabetically, such as theoretical computer science and mathematics, position carries no contribution signal, and in several CS subareas the senior author or PI is conventionally listed last. The corresponding-author promotion already absorbs part of the latter case, since a last-listed PI flagged as corresponding author receives full weight regardless of position. For the former, the authorship term is a modular component of $\alpha_{r,q}$ and admits field-specific instantiations: for alphabetically ordered venues, a uniform weighting $\omega_{\text{auth}}(r,q) = 1/n_q$ is the safer choice, and venue metadata (e.g., the publication's field or the venue's known convention) can select the scheme per publication. Our experiments target computer science conferences, where the harmonic convention with corresponding-author promotion is the empirically dominant pattern.

\noindent\textbf{Venue term $\omega_{\mathrm{venue}}(q)$.} Let $\mathrm{venue}(q)$ denote the publication venue of $q$, resolved against two independent conference ranking lists: the China Computer Federation (CCF) Recommended List (tiers A, B, C) and the CORE Conference Ranking (tiers A*, A, B, C). We first assign per-list weights,
\begin{equation*}
  \omega_{\mathrm{CCF}}(q) \in \{1.0, 0.75, 0.5\}, \qquad \omega_{\mathrm{CORE}}(q) \in \{1.0, 0.85, 0.7, 0.5\},
\end{equation*}
corresponding to the ordered tiers of each list, and default to a floor weight $\omega_{\mathrm{floor}} = 0.25$ when the venue appears in neither list (e.g., arXiv-only preprints, non-indexed workshops, unrefereed technical reports). The final venue weight averages the two list-specific weights when both are defined and falls back to whichever single list covers the venue otherwise,
\begin{equation*}
  \omega_{\mathrm{venue}}(q) = \begin{cases}
    \tfrac{1}{2}\big(\omega_{\mathrm{CCF}}(q) + \omega_{\mathrm{CORE}}(q)\big) & \text{if both lists cover } \mathrm{venue}(q), \\
    \omega_{\mathrm{CCF}}(q) \text{ or } \omega_{\mathrm{CORE}}(q) & \text{if exactly one list covers it,} \\
    \omega_{\mathrm{floor}} & \text{otherwise.}
  \end{cases}
\end{equation*}
Averaging across two independently maintained lists reduces sensitivity to the idiosyncrasies of either ranking and ensures that a venue flagged as top-tier by either community receives a correspondingly high weight.

Together, the three terms convert raw co-authorship into a graded competence signal: $\alpha_{r,q}$ is largest for recent, top-venue publications on which $r$ holds a prominent or corresponding authorship position, and decays smoothly as any of these signals weakens.

\subsection{Detailed Cost Analysis of Incremental Maintenance}
\label{app:maintenance-cost}

This subsection derives the amortized per-event cost
\[
  O\big(|\mathcal{A}_q| \cdot k + |\mathcal{A}_{\mathrm{new}}| \cdot k^2 + d^{\uparrow}\big)
\]
of the three-level maintenance cascade in Section~\ref{sec:maintenance}, where $|\mathcal{A}_q|$ is the number of canonicalized keyphrases extracted from the new publication $q$, $|\mathcal{A}_{\mathrm{new}}|$ is the number of newly created atoms, $k$ is the embedding-neighborhood size, and $d^{\uparrow}$ is the number of ancestors reached by upward propagation.
In terms of Algorithm~\ref{alg:maintenance}, $|\mathcal{A}_q| = |\mathcal{K}_{\mathrm{match}}| + |\mathcal{K}_{\mathrm{new}}|$, and $|\mathcal{A}_{\mathrm{new}}| \le |\mathcal{K}_{\mathrm{new}}|$ counts the novel keyphrases that pass the IDF threshold $\tau_{\mathrm{idf}}$.
We assume an HNSW (or IVF-PQ) index for atom embeddings, hash maps for sparse postings, and a topologically sorted DAG with adjacency lists. Under these standard data structures, all per-atom embedding lookups cost $O(\log L)$ in the vocabulary size $L$, which we treat as a constant dominated by $k$ in practice.

\noindent\textbf{Level 1: Atom-level update (Lines 4--7).}
Each of the $|\mathcal{A}_q|$ extracted keyphrases is canonicalized by retrieving its $k$ embedding nearest neighbors from the dense atom catalog and checking the canonicalization threshold $\tau_{\mathrm{canon}}$, costing $O(k)$ per keyphrase.
Profile and co-occurrence updates for matched atoms are $O(1)$ per atom via hash lookup, and new-atom creation (embedding computation, IDF check, ANN insertion) is $O(\log L)$ amortized per insertion, both dominated by the canonicalization term.
\textbf{Total: $O(|\mathcal{A}_q| \cdot k)$.}

\noindent\textbf{Level 2: Taxonomy structure update (Lines 9--11).}
For each newly created atom $a_{\mathrm{new}}$, candidate edges are scored against its $k$ embedding nearest neighbors at $O(1)$ per edge (since $\mathrm{sim}$, $\mathrm{WP}$, and $\mathrm{PMI}$ are constant-time given precomputed embeddings and co-occurrence vectors), and local transitive reduction examines at most $k^2$ ancestor pairs via the precomputed indicator $z_{ij}$.
Drift-triggered rescorings on existing atoms incur the same $O(k^2)$ cost per drifted atom, but the threshold $\tau_{\mathrm{drift}}$ is set so that each atom triggers a rescoring at most $O(1)$ times over its lifetime, amortizing this contribution to $O(1)$ per event.
\textbf{Total: $O(|\mathcal{A}_{\mathrm{new}}| \cdot k^2)$.}

\noindent\textbf{Level 3: Annotation and auxiliary-layer refresh (Lines 13--16).}
The propagated weight $w_r^{\uparrow}(a_j) = w_{r,j} + \sum_{a_k:\, z_{kj} = 1} w_{r,k}$ is purely additive: a change in a direct weight propagates upward to exactly the ancestors of the affected atom, each receiving a single $O(1)$ update.
We absorb the multi-parent factor of the DAG into $d^{\uparrow}$, interpreting it as the size of the upward-reachable set (bounded by a small constant times the tree depth in practice).
Catalog refreshes are $O(1)$ per affected atom via hash lookup, dominated by the propagation cost.
\textbf{Total: $O(d^{\uparrow})$.}

\noindent\textbf{Combined bound and common case.}
Summing the three levels recovers the stated bound.
In the common case where the new paper introduces no novel atoms ($|\mathcal{A}_{\mathrm{new}}| = 0$) and no existing atom has drifted, the Level 2 term vanishes and the cost reduces to $O(|\mathcal{A}_q| + d^{\uparrow})$ under the constant-$k$ convention, essentially linear in the number of keyphrases of a single paper.
For the operations analyzed above, no term in this expression depends on the lake size $|\mathcal{R}_W|$ or on the number of papers already indexed. Keeping node posting lists sorted and the periodic consolidation below add costs outside this per-event bound.

\noindent\textbf{Periodic consolidation.}
The amortized analysis assumes that small drifts do not accumulate to the point where the DAG becomes structurally suboptimal.
We trigger a periodic global re-validation pass (rescoring all edges, recomputing transitive reduction) every $T$ events, with $T$ chosen so that the amortized cost of the global pass is absorbed into the Level 2 term.
This ensures the asymptotic bound holds in expectation rather than only on a per-event basis.

\subsection{Hierarchy-Preserving Mapping ILP}
\label{app:mapping-ilp}

We formalize the anchor-selection step of Section~\ref{sec:structural-alignment} as an integer program and then linearize it into the integer linear program (ILP) solved on Line~5 of Algorithm~\ref{alg:aligner}. Let $y_{v,i} \in \{0, 1\}$ be a binary indicator that takes value $1$ if conference node $v \in V$ is mapped to global atom $a_i$ and $0$ otherwise, and let $\mathcal{A}_v^{\mathrm{cand}}$ denote the candidate pool of top-$k_c$ atoms retrieved for $v$ by centroid cosine similarity. Finally, let $z_{ij} \in \{0, 1\}$ be the precomputed global ancestry indicator of Section~\ref{sec:global-taxonomy}, equal to $1$ if atom $a_j$ is reachable from atom $a_i$ along the hypernym edges of $\mathcal{T}_G$ and $0$ otherwise.

The hierarchy-preserving mapping is then obtained by solving
\begin{equation}
\label{eq:map-ilp}
\begin{aligned}
  \max_{y} \;\; & \sum_{v \in V} \sum_{a_i \in \mathcal{A}_v^{\mathrm{cand}}} y_{v,i} \cos(\mathbf{c}_v, \mathbf{e}_{a_i}) \\
  & \qquad {} - \lambda \!\!\! \sum_{(v,v') \in E} \sum_{i, j} y_{v,i}\, y_{v',j}\, (1 - z_{ji}), \\
  \text{s.t.} \;\; & \sum_{a_i \in \mathcal{A}_v^{\mathrm{cand}}} y_{v,i} = 1, \quad \forall\, v \in V, \\
  & y_{v,i} \in \{0, 1\}, \quad \forall\, v \in V,\ a_i \in \mathcal{A}_v^{\mathrm{cand}},
\end{aligned}
\end{equation}
where, for each local edge $(v, v')$ with parent $v$ and child $v'$, the indices $i$ and $j$ range over the candidate pools $\mathcal{A}_v^{\mathrm{cand}}$ and $\mathcal{A}_{v'}^{\mathrm{cand}}$.

The objective encodes the two criteria identified in Section~\ref{sec:structural-alignment}. The first term rewards semantic fit: for every conference node $v$, the chosen anchor $a_i$ contributes its centroid--atom cosine similarity $\cos(\mathbf{c}_v, \mathbf{e}_{a_i})$, drawing $v$ toward atoms that are topically close in embedding space. The second term imposes a structural penalty: for every local parent--child edge $(v, v') \in E$, any pair of anchors $(a_i, a_j)$ assigned to the two endpoints is penalized by the coefficient $(1 - z_{ji})$, which is zero whenever $a_j$ is a descendant of $a_i$ in $\mathcal{T}_G$ and one otherwise. The penalty weight $\lambda > 0$, which is the input $\lambda$ of Algorithm~\ref{alg:aligner} and is unrelated to the regularization coefficient of Equation~\eqref{eq:edge_weight_obj}, sets the trade-off between the two criteria, with larger $\lambda$ prioritizing hierarchical faithfulness over per-node semantic fit. The equality constraint $\sum_{a_i} y_{v,i} = 1$ enforces that each conference node is mapped to exactly one global atom, and the integrality constraint preserves the discrete assignment semantics.

The presence of the bilinear product $y_{v,i} y_{v',j}$ in the structural-penalty term makes Equation~\eqref{eq:map-ilp} a quadratic integer program. We linearize it using the standard product-to-binary substitution: for each quadruple $(v, v', i, j)$ with $(v, v') \in E$ and $z_{ji} = 0$, which are the only products that carry a penalty, introduce an auxiliary binary variable $u_{v,v',i,j} \in \{0, 1\}$ together with the three linking constraints $u_{v,v',i,j} \leq y_{v,i}$, $u_{v,v',i,j} \leq y_{v',j}$, and $u_{v,v',i,j} \geq y_{v,i} + y_{v',j} - 1$, which together enforce $u_{v,v',i,j} = y_{v,i} \cdot y_{v',j}$ at any feasible integer solution. Substituting $u_{v,v',i,j}$ for the product yields a standard integer linear program, which is solved efficiently by off-the-shelf MIP solvers~\cite{gurobi,cplex} on all problem sizes encountered in our experiments. The resulting anchor set $\mathcal{A}^{\star} = \bigcup_v \{a_i : y_{v,i} = 1\}$ is passed to the Steiner-tree extraction step of Algorithm~\ref{alg:aligner}.

\section{Additional Experiments}
\label{app:additional_exp}

\subsection{Dataset Details}
\label{app:datasets}

\noindent\textbf{Established benchmarks.}
We use four widely adopted benchmark datasets: NIPS06, SIGIR07, SciRepEval, and KDD20, where SciRepEval is a composite benchmark derived from enriched versions of NIPS06 and ICIP16. Obtaining ground-truth labels for reviewer matching is challenging due to the strict confidentiality of the peer-review process, so none of these benchmarks is collected from real assignment logs. Instead, a small candidate reviewer pool is curated for each paper, and human annotators assess the suitability of each candidate on three relevance levels: strongly relevant, weakly relevant, and irrelevant.

\noindent\textbf{\neurips construction.}
To simulate a real paper-assignment workflow (Study~1 in Section~\ref{sec:human-eval}), we construct a new benchmark, \neurips. Because relevance labels are costly to collect, existing conference benchmarks typically adopt a retrieval-style protocol: a small set of accepted papers from a conference serves as query papers, and only a few candidate reviewers per query paper are annotated for relevance. These benchmarks therefore cover only a few submission--reviewer pairs per query rather than the full submission-by-reviewer matrix, and they provide no conflict-of-interest set. Built from the publicly available NeurIPS 2025 acceptance and authorship data, \neurips contains $5{,}286$ accepted papers and a reviewer pool of $11{,}235$ candidates. We instantiate its conflict set from the relations recorded on or auto-detected by OpenReview, including co-authorship, shared institutional affiliation, and advisor--advisee relations.

The candidate reviewer pool is constructed from the NeurIPS 2025 author community using the reciprocal-reviewer protocol common in major conferences: all accepted NeurIPS 2025 papers are retrieved via the OpenReview API, all authors with at least five prior publications are selected as candidates, and any papers published after the NeurIPS 2025 submission deadline are excluded from each reviewer's publication history to prevent temporal data leakage. Dataset statistics for all five datasets are reported in Table~\ref{tab:datasets}.

\subsection{Additional Implementation Details}
\label{app:impl}

\noindent\textbf{Hyperparameter Setting.} The taxonomy backbone is induced by prompting DeepSeek-V3.2 (non-thinking mode) to iteratively partition and name topic nodes, up to a maximum depth $D = 4$ (Section~\ref{sec:exp:setup}). The evidence-selection submodular program (Equation~\eqref{eq:evidence}) is solved under a per-reviewer budget of $B_r = 15$ papers by the lazy-greedy algorithm~\cite{lin2011submodular}, which exploits submodularity to avoid unnecessary marginal-gain re-evaluations. The adaptive inverse temperature $\beta_u$ of the soft routing (Section~\ref{sec:core}) uses the stabilizer $\epsilon = 10^{-3}$, and the edge-weight learning objective (Equation~\eqref{eq:edge_weight_obj}) is solved as a non-negative least-squares problem using SciPy's \texttt{nnls} routine with $\lambda = 0.01$.
Candidate evidence publications are restricted by a recency threshold of $Y = 5$ years, and publications appearing after the conference submission deadline are excluded from each reviewer's profile to prevent temporal leakage.

\noindent\textbf{Baselines.} Baseline implementations follow the original papers. TPMS~\cite{charlin2013tpms} is re-implemented from public reference code. SciBERT~\cite{beltagy2019scibert}, SPECTER~\cite{cohan2020specter}, and SPECTER2~\cite{singh2023scirepeval} use the official HuggingFace checkpoints. CoF~\cite{zhang2025chainoffactors} is run with the three-embedding configuration specified by its authors. LLM Judge queries DeepSeek-V3.2 once per submission--reviewer pair and rates the pair on the 0--10 scale described in Section~\ref{sec:exp:affinity}.

\subsection{Impact of Taxonomy Induction Backbone}
\label{app:abl-backbone}

The local taxonomy builder is agnostic to the method used to induce the unweighted backbone $\mathcal{T}_0$ (Section~\ref{sec:core}). To isolate the contribution of the backbone choice, we substitute four representative alternatives into Line~4 of Algorithm~\ref{alg:affinity} while leaving every other component (evidence selection, soft routing, edge-weight calibration, tree--Wasserstein scoring) unchanged. The variants span the dominant families of hierarchical induction:
\begin{itemize}[leftmargin=*]
    \item \textbf{\sysHAC} (agglomerative hierarchical clustering)~\cite{murtagh2012algorithms}. Bottom-up agglomeration with cosine distance and average linkage, merging the most similar clusters until the root is formed.
    \item \textbf{\sysHDB} (HDBSCAN)~\cite{campello2015hdbscan}. Density-based hierarchical clustering that recovers a condensed cluster tree without requiring a fixed branching factor or depth.
    \item \textbf{\sysvMF} (recursive von Mises--Fisher)~\cite{banerjee2005vmf}. Top-down recursive partitioning with directional concentration tailored to normalized embeddings.
    \item \textbf{\sysLLM} (LLM-guided induction with GPT-4o-mini). Taxonomy induced by prompting GPT-4o-mini to iteratively build and refine a hierarchical structure, following the multidimensional alignment approach of TaxoAdapt~\cite{kargupta2025taxoadapt}.
    \item \textbf{\sys} (default, LLM-guided induction with DeepSeek-V3.2). Taxonomy induced by prompting DeepSeek-V3.2 (non-thinking mode) to iteratively partition and name topic nodes, up to a maximum depth $D = 4$ (Section~\ref{sec:exp:setup}).
\end{itemize}

Table~\ref{tab:abl-backbone} reports the results. Both LLM-guided backbones outperform the three clustering-based alternatives under both Soft and Hard metrics, reflecting the semantic reasoning capabilities of LLMs in producing coherent leaf clusters that align with fine-grained expertise. \sysLLM attains the highest scores, slightly above the default (39.03 vs.\ 38.00 Soft P@5 and 18.13 vs.\ 17.40 Hard P@5), but its GPT-4o-mini-based induction incurs \$14.50 in API charges and $2.7$ hours of wall-clock time for the taxonomy building phase alone on \kdd, which has only 174 submissions. By comparison, the default \sys\ spends \$5.3 in API fees on all four benchmarks together (Section~\ref{sec:exp:affinity}).

Among the clustering-based alternatives, \sysvMF is the strongest but trails the default on all four metrics. \sysHDB yields the lowest Hard precision, as its density-driven cluster boundaries leave a non-trivial fraction of documents in noise regions, which produces overly diffuse leaf distributions. \sysHAC yields the lowest Soft precision, consistent with the well-known fragmentation of bottom-up agglomeration when scaled to many documents.

\begin{table}[t]
\centering
\caption{Impact of the taxonomy induction backbone on \kdd. Best results are in bold.}
\label{tab:abl-backbone}
\small
\setlength{\tabcolsep}{5pt}
\begin{tabular}{l cc cc}
\toprule
\multirow{2}{*}{\textbf{Variant}} & \multicolumn{2}{c}{\textbf{Soft}} & \multicolumn{2}{c}{\textbf{Hard}} \\
\cmidrule(lr){2-3} \cmidrule(lr){4-5}
 & \textbf{P@5} & \textbf{P@10} & \textbf{P@5} & \textbf{P@10} \\
\midrule
\sysHAC        & 34.42 & 25.80 & 15.50 & 12.22 \\
\sysHDB        & 35.16 & 26.23 & 15.28 & 12.00 \\
\sysvMF        & 36.30 & 27.09 & 16.02 & 12.54 \\
\sys (Default) & 38.00 & 28.53 & 17.40 & 13.56 \\
\sysLLM        & \textbf{39.03} & \textbf{29.56} & \textbf{18.13} & \textbf{14.21} \\
\bottomrule
\end{tabular}
\end{table}

\subsection{Sensitivity to Taxonomy Depth}
\label{app:hyper}

We study the sensitivity of \sys\ on \kdd to the maximum taxonomy depth $D$, the recursion depth used in taxonomy induction, varying $D \in \{1, 2, 3, 4, 5, 6, 7\}$, which controls the granularity of the induced local hierarchy. As shown in Figure~\ref{fig:hyperparameter}, very shallow taxonomies ($D \leq 2$) severely underfit: topic nodes are overly coarse, collapsing distinct subfields and weakening the discriminability of the tree--Wasserstein alignment. Increasing the depth improves matching quality up to $D = 4$ by enabling finer routing and a more informative transport geometry. Beyond that point, overly deep taxonomies ($D \geq 5$) over-fragment the corpus: documents disperse into sparse, noisy leaf assignments that harm Hard precision and unnecessarily inflate taxonomy construction cost. We therefore set $D = 4$ as the default.

\begin{figure}[t]
    \centering
    \includegraphics[width=0.75\linewidth, trim={48 141 22 3}, clip]{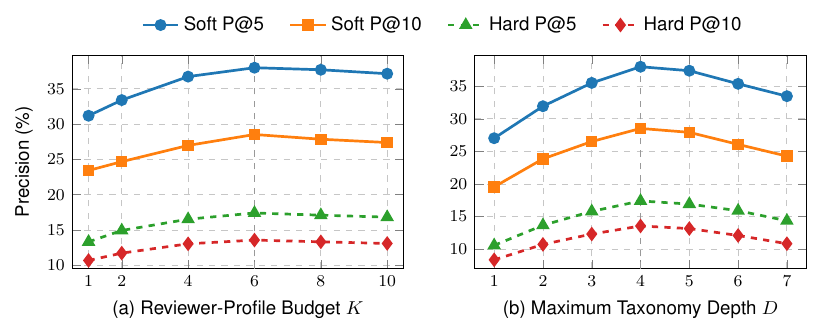}\\[2pt]
    \includegraphics[width=0.48\linewidth, trim={203 17 0 21}, clip]{figures/hyperparameter_sensitivity.pdf}
    \caption{Sensitivity of \sys\ to the maximum taxonomy depth $D$ on \kdd: Soft and Hard Precision@5 and Precision@10 (\%) for $D = 1, \ldots, 7$.}
    \Description{Line chart of Soft and Hard Precision@5 and Precision@10 on KDD20 as the maximum taxonomy depth D grows from 1 to 7. All four curves rise until D = 4 and decline afterwards.}
    \label{fig:hyperparameter}
\end{figure}

\subsection{Impact of Bibliometric Weighting in the Local Representation}
\label{app:bib-ablation}

To validate the design choice of Section~\ref{sec:core}, we evaluate a variant TaxoReview\textsubscript{Bib} that replaces the uniform mixture of Equation~\eqref{eq:reviewer-dist} with an $\alpha_{r,q}$-weighted mixture, $\phi_r(\ell) \propto \sum_{q \in \mathcal{E}^\star_r} \alpha_{r,q}\,\phi_q(\ell)$, reusing the bibliometric factor of Section~\ref{sec:atom-extraction} unchanged. On KDD20, TaxoReview\textsubscript{Bib} attains 36.78/27.23 Soft P@5/P@10 and 16.84/13.11 Hard P@5/P@10, consistently below the default's 38.00/28.53 and 17.40/13.56 (Table~\ref{tab:ablation}). The degradation is consistent with the analysis in Section~\ref{sec:core}: the evidence set is already curated for representativeness and recency, so the additional scaling concentrates mass on a few highly weighted papers and distorts the topical balance that the submodular selection establishes.

\subsection{Human Evaluation Protocol Details}
\label{app:human-eval}

This subsection details the protocol of the two studies in Section~\ref{sec:human-eval}, whose results are reported in Table~\ref{tab:human-eval}.
\emph{Annotators.} We recruited 18 qualified researchers, each with at least two publications in the areas covered by the sampled submissions.
\emph{Interface and blinding.} Judgments were collected through a web application presenting one submission (title and abstract) together with the pooled, de-duplicated union of candidates produced by all methods in randomized order, with method identities hidden. For each candidate, the interface displays the publication history, and the annotator selects one of the three relevance levels or abstains via a ``cannot judge'' option, with abstained pairs excluded from all counts.
\emph{Sampling.} Study~1 sampled 298 submissions from NeurIPS25, yielding 1,254 labeled pairs. Study~2 sampled 292 queries evenly across the three query types of Section~\ref{sec:retrieval} and judged the top-$B$ candidates per method with $B = 5$, yielding 1,323 labeled pairs.
\emph{Conflict detection for lake experts.} Co-authorship is resolved against DBLP records and shared affiliation against OpenAlex institution metadata via author-name matching. This bibliographic detection is best-effort, and residual conflicts invisible to publication metadata are covered by self-declaration in deployment (Section~\ref{sec:retrieval}).
\emph{Metrics.} Soft/Hard precision report the fraction of judged candidates rated at least weakly / strongly relevant, and Avg.\ Score is the mean graded relevance on the 0--2 scale over the same judged pairs. Expressed as fractions, Avg.\ Score equals the sum of the Soft and Hard precisions. $\tau_b$ is the mean per-query Kendall correlation between a method's affinity scores and the human labels over the pooled judged candidates, where queries whose judged pool shows no label variation are excluded from the mean.
\emph{Released data.} All submissions, candidate pools, anonymized labels, and the annotation interface materials are available in our repository~\cite{taxoreview_repo}.

\section{Extended Related Work}\label{app:related_work}
\noindent\textbf{Data Discovery.}
Data discovery has long been a topic in data management, particularly in the data lake context~\cite{nargesian2019datalake,fernandez2018aurum}, where a large, heterogeneous, continuously growing repository must be organized so that items relevant to a query can be located efficiently. Our framework for panel expansion mirrors this process, which is why we term $\mathcal{R}_W$ an \emph{expert lake}: reviewers enter the repository without a unifying schema, and the Global Taxonomy Index imposes structure offline so that coverage-driven queries can be served online. Existing work falls into two categories.
\emph{Structured data discovery} operates over relational artifacts in the lake and retrieves tables, columns, or values that match a query table: joinable-table search via overlap set similarity~\cite{zhu2019josie,zhu2016lshensemble}, unionable-table search via schema or value-domain similarity~\cite{nargesian2018tus,bogatu2020dataset}, semantically enriched union search via contextual column embeddings or external knowledge bases~\cite{fan2023starmie,khatiwada2023santos}, and related-table discovery for downstream analytics~\cite{zhang2020juneau}.
\emph{Unstructured data discovery} operates over text documents, images, or other non-tabular items and retrieves them via flat similarity in a learned embedding space, usually served by an HNSW or IVF-PQ index~\cite{malkov2020hnsw,fu2019nsg,jegou2011pq}. Cross-modal lakes that combine tables and documents under a unified similarity space fall into this family as well~\cite{eltabakh2023crossmodal}.
\sys\ aims to bring data discovery to a third setting in which the items to be discovered are \emph{experts} rather than tables or documents, the repository is organized as a DAG of hypernym--hyponym subsumptions over canonical expertise atoms rather than as a flat collection, and relevance is measured by the hierarchy-aware tree--Wasserstein distance of Section~\ref{sec:taxonomy} rather than flat set overlap or cosine similarity. This specialization preserves the candidate pruning that motivates data-lake indexing while serving queries whose item type, structure, and geometry are specific to paper--reviewer assignment.

\noindent\textbf{Similarity Indexing and Top-$k$ Retrieval.}
At its core, panel expansion is a top-$k$ retrieval problem over a structured repository, and its efficiency hinges on the choice of index. The database and information retrieval literature offers three broad families. \emph{Flat sparse indexes} organize items over a discrete token vocabulary: inverted indexes~\cite{zobel2006inverted} map each token to a posting list of items, and threshold-style aggregation over ranked sources~\cite{fagin2001ta,ilyas2008topk} turns sparse probes into provably sublinear top-$k$ retrieval under monotone scoring. \emph{Flat dense indexes} organize items as vectors in a learned embedding space and answer similarity queries via approximate nearest neighbor (ANN) search. Graph-based structures such as HNSW~\cite{malkov2020hnsw}, NSG~\cite{fu2019nsg}, and disk-resident DiskANN~\cite{subramanya2019diskann}, together with quantization-based methods~\cite{jegou2011pq}, deliver sublinear lookup over million- to billion-scale vector collections under well-characterized quality--throughput trade-offs~\cite{wang2021ann}. \emph{Hierarchical indexes} impose a tree or DAG over items so that queries can be answered at multiple granularities. Examples include ontology- and taxonomy-indexed retrieval in semantic search and the tree-based optimal-transport indexes that enable sublinear nearest neighbor search under Wasserstein geometry~\cite{indyk2003fast,le2019tree,backurs2020flowtree}. Each family is designed around a different item representation (discrete tokens, dense vectors, or hierarchically organized entities) and a correspondingly different similarity geometry (set overlap, Euclidean, or tree metric).

The Global Taxonomy Index of Section~\ref{sec:index} sits in the hierarchical family but is shaped by two requirements that distinguish it from prior tree-structured indexes. First, the lake is large and continuously growing, so the index is deliberately kept \emph{lightweight} and \emph{incrementally updatable}, scaling with canonical expertise atoms rather than full paper texts and absorbing new publications without routine full rebuilds. Second, and more importantly, it provides a solid foundation for retrieval under a faithful proxy of the \emph{conference-specific tree--Wasserstein metric}.

\fi

\end{document}